\documentclass[tightenlines,floatfix,11pt]{article}

\usepackage{epsfig}
\usepackage{amsfonts}
\usepackage{array}
\usepackage{amsmath}
\usepackage{slashed}
\usepackage{pifont}
\usepackage{amssymb}
\usepackage{dsfont}
\usepackage{hyperref}
\usepackage{comment}
\usepackage[capitalise]{cleveref}

\usepackage{cite}
\usepackage{hhline}

\usepackage{multirow}
\usepackage{adjustbox}
\usepackage{float}
\renewcommand{\arraystretch}{1.5}

\usepackage[dvipsnames,table]{xcolor}

\newcommand{\al}{\alpha}
\newcommand{\bt}{\beta}
\newcommand{\g}{\gamma}

\newcommand{\dt}{\delta}

\newcommand{\la}{\lambda}

\newcommand{\Or}{\mathcal O}

\newcommand{\vL}{\ensuremath{\mathcal{L}}}

\newcommand{\ga}{\gamma}

\newcommand{\dslash}[1]{#1 \llap{/\kern-0.5pt}}
\newcommand{\Dslash}[1]{#1 \llap{/\kern+1.5pt}}
\newcommand{\DDslash}[1]{#1 \llap{/\kern+2.3pt}}
\newcommand{\dslashh}[1]{#1 \llap{/\kern+1pt}}

\newcommand{\beq}{\begin{equation}}
\newcommand{\eeq}{\end{equation}}

\newcommand{\bea}{\begin{eqnarray}}
\newcommand{\eea}{\end{eqnarray}}
\newcommand{\bma}{\begin{pmatrix}}
\newcommand{\ema}{\end{pmatrix}}
\newcommand{\nn}{\nonumber}

\newcommand{\nocontentsline}[3]{}
\newcommand{\tocless}[2]{\bgroup\let\addcontentsline=\nocontentsline#1{#2}\egroup}

\newcommand{\eR}{\epsilon_R}

\newcommand{\eP}{\epsilon_P}

\newcommand{\cmark}{{\color{darkgreen}\ding{51}}}%
\newcommand{\xmark}{{\color{darkred}\ding{55}}}%

\newcommand{\nnw}{\nonumber\\[1ex]}

\newcommand{\hc}{\mathrm{h.c.}}

\newcommand{\XXint}[3]{{\setbox0=\hbox{$#1{#2#3}{\int}$}
\vcenter{\hbox{$#2#3$ }}\kern-.65\wd0}}

\newcommand{\remark}[1]{}

\definecolor{darkgreen}{rgb}{0,0.5,0}
\definecolor{darkblue}{rgb}{0,0,0.5}
\definecolor{darkred}{rgb}{0.5,0,0}
\definecolor{beige}{rgb}{0.7,0.4,0.3}

\graphicspath{ {./Figures/} }

\newcommand{\ccg}{\cellcolor{gray!15}}
\newcommand{\ccgg}{\cellcolor{blue!15}}

\newcommand{\RNum}[1]{\uppercase\expandafter{\romannumeral #1\relax}}

\numberwithin{equation}{section}

\allowdisplaybreaks[1]

\begin{document}
\begin{titlepage}

\begin{flushright}
LA-UR-23-31027
\end{flushright}

\vspace{1cm}

\begin{center}
{\LARGE  \bf  
Anomalies in global SMEFT analyses: \\[1ex]
a case study of first-row CKM unitarity}
\vspace{1cm}

{\large \bf  Vincenzo Cirigliano,$^a$ Wouter Dekens,$^{a}$
  Jordy de  Vries,$^{b,c}$\\[1mm] 
  Emanuele Mereghetti,$^d$ Tom Tong$^{e}$}
\vspace{0.5cm}

\vspace{0.25cm}

{\large 
$^a$ 
{\it 
Institute for Nuclear Theory, University of Washington, Seattle WA 98195-1550, USA}}\\
{\large 
$^b$ 
{\it Institute for Theoretical Physics Amsterdam and Delta Institute for Theoretical Physics,University of Amsterdam, Science Park 904, 1098 XH Amsterdam, \\The Netherlands}}\\
{\large 
$^c$ 
{\it Nikhef, Theory Group, Science Park 105, 1098 XG, Amsterdam, The Netherlands}}\\
{\large 
$^d$ 
{\it Theoretical Division, Los Alamos National Laboratory, Los Alamos, NM 87545, USA}}\\
{\large 
$^e$ 
{\it Center for Particle Physics Siegen, University of Siegen, 57068 Siegen, Germany}}

\end{center}

\begin{abstract}

Recent developments in the Standard Model analysis of semileptonic charged-current processes involving light quarks have revealed $\sim 3\sigma$ tensions in Cabibbo universality tests involving meson,  neutron, and nuclear beta decays. 
In this paper, we explore beyond the Standard Model explanations of this so-called Cabibbo Angle Anomaly in the framework of the Standard Model Effective Field Theory (SMEFT), including not only low-energy charged current processes (`L'), but also electroweak precision observables (`EW') and Drell-Yan collider processes (`C') that probe the same underlying physics across a broad range of energy scales. 
The resulting `CLEW' framework not only allows one to test explanations of the Cabibbo Angle Anomaly, but is set up to provide near model-independent  analyses with minimal assumptions on the flavor structure of the SMEFT operators.   
Besides the global analysis, we consider a large number of simpler scenarios, each with a subset of SMEFT operators, and investigate how much they improve upon the Standard Model fit. We find that the most favored scenarios, as judged by the Akaike Information Criterion, are those that involve right-handed charged currents. Additional interactions, namely oblique operators, terms modifying the Fermi constant, and operators involving right-handed neutral currents, play a role if the CDF determination of the $W$ mass is included in the analysis.

\end{abstract}

\vfill
\end{titlepage}

\tableofcontents
\newpage

\section{Introduction} 

	Precision studies of semileptonic charged-current (CC)  interactions involving light quarks, mediating decays of mesons, neutrons, and nuclei, provide stringent tests of the Standard Model (SM) and its possible extensions.   
	Prominent examples are the precision tests of lepton universality and quark-lepton (Cabibbo) universality~\cite{Cabibbo:1963yz}. 
	The latter is encoded in the SM by the first-row unitarity relation for the Cabibbo-Kobayashi-Maskawa (CKM)~\cite{Cabibbo:1963yz,Kobayashi:1973fv} matrix: 
	\beq\label{CAA}
	\Delta_{\rm CKM} \equiv |V_{ud}|^2 + |V_{us}|^2 + |V_{ub}|^2  - 1 = 0~. 
	\eeq
           $\beta$ decays and kaon decays allow one to extract $V_{ud}$ and $V_{us}$ with fractional uncertainties of 0.03\% and 0.22\% respectively \cite{ParticleDataGroup:2022pth}, 
          leading to an uncertainty of $\sim 5 \times 10^{-4}$ in $|V_{ud}|^2 + |V_{us}|^2$, thus making $|V_{ub}|^2 \simeq 1.5 \times 10^{-5}$ largely irrelevant for this test.  
	 In the last few years, mainly due to the evolution of the theoretical input on radiative corrections in $\beta$ decays \cite{Seng:2018yzq,Seng:2018qru, Czarnecki:2019mwq,Shiells:2020fqp,Hayen:2020cxh,Hardy2020Oct} and on the $K \to \pi$ vector matrix element in lattice QCD \cite{Carrasco:2016kpy,FermilabLattice:2018zqv,FlavourLatticeAveragingGroupFLAG:2021npn},  several tensions have emerged with the SM interpretation of semileptonic decays and Cabibbo universality. These are collectively dubbed the Cabibbo Angle Anomaly (CAA). 
	 The two main features of the CAA can be summarized as follows: first, the best-fit value of $\Delta_{\rm CKM} = -1.48 (53) \times 10^{-3}$, as obtained, for example, in~\cite{Cirigliano:2022yyo}, deviates from zero at the 3$\sigma$ level.   
         Second,  the extractions of $V_{us}$ from $K \to \pi \ell \nu$ and $V_{us}/V_{ud}$ from $\Gamma(K \to \mu \nu)/\Gamma (\pi \to   \mu \nu)$ 
         are by themselves inconsistent with CKM unitarity at the 3$\sigma$ level.

	This state of affairs has generated significant activity on two fronts. On the one hand, the community continues to scrutinize the theoretical aspects of the SM analysis, which involves non-perturbative input for radiative corrections and more generally for hadronic and nuclear matrix elements~\cite{Feng:2020zdc,Seng:2020wjq,Ma:2021azh,Yoo:2023gln,Cirigliano:2023fnz,Seng:2022cnq,Seng:2023cvt,Cirigliano:2022yyo,1100705,Seng:2022epj,Seng:2022inj,Ma:2023kfr,Seng:2023cgl}. On the other hand, there is interest in looking for new physics scenarios that might explain the CAA, should it persist~\cite{{Belfatto:2019swo,Grossman:2019bzp,Crivellin:2020lzu,Kirk:2020wdk,Crivellin:2020ebi,Alok:2021ydy,Crivellin:2021bkd,Crivellin:2022rhw,Belfatto:2021jhf,Belfatto:2023tbv}}. On the latter front, previous work has been rooted both in specific extensions of the SM and in effective field theory (EFT) approaches. In turn, EFT approaches have followed different strategies. Motivated by phenomenological considerations or specific classes of BSM scenarios, several groups have analyzed the CAA within a subset of the dimension-six SMEFT operators that can affect semileptonic decays~\cite{Grossman:2019bzp,Crivellin:2020lzu,Kirk:2020wdk,Crivellin:2020ebi,Alok:2021ydy,Crivellin:2022rhw}, with a global analysis still missing. In a different approach, Refs.~\cite{Gonzalez-Alonso:2016etj,Falkowski:2017pss} performed a global analysis of low-energy charged-current data (including $\tau$ decays in Ref.~\cite{Cirigliano:2021yto}) using the full semileptonic operator basis in the EFT valid below the weak scale (LEFT). The LEFT analysis by construction loses the correlation of charged-current precision measurements with other electroweak precision observables, which is explicit when using SMEFT operators that are invariant under $SU(2) \times U(1)$.  

	In this work, we fill the gap described above by presenting a global SMEFT analysis of the possible BSM origins of the CAA. This forces us to look at the CAA in conjunction with other observables that have sensitivity to (a subset of the) SMEFT operators that affect semileptonic charged-current decays. It turns out that a minimal analysis requires including two additional sets of observables: 
         (i) traditional electroweak precision observables (EWPO), which are affected by vertex corrections and four-lepton operators (shifting the Fermi constant $G_F$) also appearing in $\beta$ decays; 
         (ii) LHC charged-current and neutral-current (NC) Drell-Yan (DY) processes, namely $pp \to \ell \bar \nu_\ell + X$ and $pp \to \ell \bar \ell + X$, which are affected by semileptonic four-fermion operators and vertex corrections also appearing in $\beta$ decays. 
        To achieve a self-consistent analysis, we should consider the complete set of SMEFT operators that affect low-energy CC processes, EWPO, and Drell-Yan processes, which is larger than the set that affects only $\beta$ decays.  
        Note that by considering the EWPO we are also forced to confront the so-called $W$ mass anomaly~\cite{CDF:2022hxs}~\footnote{In light of tensions in the W mass measurements, we have performed the analysis both with and without the CDF 2022 result~\cite{CDF:2022hxs}.}. 
        
	Given the considerations above, it is evident that the CAA and EWPO are interconnected. Thus, a comprehensive analysis accounting for all constraints should encompass observables from collider processes (`C'), low-energy (`L') charged-current processes, and electroweak precision observables (`EW'). We will refer to this as a `CLEW' analysis. The relevance of the CLEW analysis framework can be appreciated by noting the following:  
	\begin{itemize}
	\item If one focuses on BSM explanations of the CAA and only considers low-energy CC observables, it is possible to select BSM scenarios that are incompatible with EWPO and LHC Drell-Yan measurements. Only a global CLEW analysis down-selects explanations of the CAA which are compatible with constraints from weak and TeV scales. 
        \item Similarly, it has been shown that explanations of the $W$ mass anomaly that focus only on EWPO can select SMEFT parameters that aggravate the CAA~\cite{Cirigliano:2022qdm,Bagnaschi:2022whn}.
	\end{itemize}
 
	Although the CLEW framework was originally motivated by explaining anomalies in low-energy charged-current processes, it is set up to perform nearly global SMEFT fits to the precision observables that do not involve flavor-changing neutral currents (FCNC) and CP violation. This framework can be further extended in the future to include other precision measurements, such as low-energy neutral-current phenomena (e.g.\ atomic parity violation or parity-violating electron scattering) and semileptonic $\tau$ decays. We currently do not include these additional observables because they either have lower precision or constrain combinations of SMEFT operators that are orthogonal to the ones affecting low-energy charged-current processes (see Section~\ref{sec:flavors-global} for details).

        In SMEFT analyses, one has to confront the proliferation of parameters when considering the most general flavor structure of the Wilson coefficients. This is usually dealt with by imposing flavor symmetries. 
        We start by presenting our results for a `warm-up' scenario in which we impose the $U(3)^5$ flavor symmetry on the SMEFT Wilson coefficients.  
	This is instructive because it closely parallels existing analyses. However, as we will discuss, this assumption is quite restrictive and can lead to inconsistencies with low-energy observables if they are not explicitly included in the analysis.  
	We then present a general `flavor-assumption-independent' analysis, which is highly desirable because it avoids any model dependence. 
	We achieve this by focusing on a certain subset of Wilson coefficients that contribute to the CLEW observables, but (mostly) leave FCNC processes unaffected. 
	The main challenge in setting up such an analysis is the need to decouple CC observables from FCNC observables (EWPO and FCNC are relatively easily decoupled) by suitable choices of independent Wilson coefficients. 
	Although exact decoupling is not possible, we will describe a plausible scenario and possible ways to test it in the future, with extended fits that include FCNC observables. 

Based on the identified set of SMEFT operators and Wilson coefficients, we then analyze the CAA systematically. 
We present a global fit with up to 37 independent operators and use the Akaike Information Criterion (AIC) to investigate whether fits with fewer operators perform better. We then identify, without making flavor assumptions, which set of SMEFT operators provides the optimal resolution to the CAA.
	
	The paper is organized as follows. In Section~\ref{sect:framework} we set up the framework of our analysis, identifying the operators and observables that we consider. In Section~\ref{sect:strategy} we describe our strategy to deal with flavor structure and set up three classes of analyses in addition to the SM analysis. 
    In Section~\ref{tools} we summarize the statistical tools used in our work. We present our results in Sections~\ref{resultsSM}-\ref{sec:global_37} providing a discussion of the main features driving the various fits. Apart from discussing the nearly-global analysis, we also consider a number of simpler scenarios that involve subsets of SMEFT operators and investigate which one leads to the most favored solution of the CAA.
    In Section~\ref{sect:falsifying}, we explore the potential of future measurements and theoretical developments to probe the nonzero couplings, which the statistical analysis identifies as the simplest explanation for the CAA. We offer our conclusions and outlook in Section~\ref{sect:conclusion}, while technical details are provided in the Appendices. We collect the results of the `flavor-assumption-independent' fit in the Supplemental Material.

\section{Analysis framework}
\label{sect:framework}

\subsection{Standard Model Effective Field Theory}
\label{sect:smeft}

Assuming that BSM physics appears at a scale $\Lambda$ well above the electroweak scale, $\Lambda\gg v$, its effects can be captured by an EFT. If the BSM dynamics is weakly coupled, the resulting TeV-scale effective Lagrangian linearly realizes the electroweak symmetry $SU(2) \times U(1)$ and contains an SM-like $SU(2)$ Higgs doublet. The relevant EFT is the SMEFT~\cite{Buchmuller:1985jz,Grzadkowski:2010es}, which extends the SM with operators of canonical dimension $d>4$, suppressed by powers of $\Lambda^{4-d}$.
The first BSM operator appears at dimension five~\cite{Weinberg:1979sa} and gives rise to neutrino Majorana masses.   
The leading contributions to the observables of interest in this work arise from dimension-six operators $Q_i$, which are described by the following effective Lagrangian
\bea\label{eq:Lag}
\vL = \vL_{\rm SM}+\sum_{i} C_i   Q_i\,,
\eea
where the Wilson coefficients, $C_i$, have mass dimension $-2$. 
There are 2499 operators in SMEFT at dimension six~\cite{Alonso:2013hga},   
and we adopt the widely used \textit{Warsaw basis}~\cite{Grzadkowski:2010es}. 
As discussed in the Introduction, our analysis includes only the operators that affect low-energy CC (semi)leptonic processes, EWPO, and Drell-Yan at the LHC. 
We list the relevant operators in Table~\ref{tab:operators1} along with the classes of observables to which they contribute, making it clear that a joint analysis of these three classes of observables is required for consistency. 

Our notation is such that $l^T = (\nu_L, e_L)$ and $q^T = (u_L, d_L)$ stand for left-handed lepton and quark $SU(2)$ doublets, while $u = u_R$, $d = d_R$, and $e = e_R$ are the right-handed up-type, down-type, and charged-lepton fields. We use $p,r,s,t$ for generation indices and work in a basis in which the electron and down-quark Yukawa matrices are diagonal. This implies that the fields $d_{L,R}$, $e_{L,R}$ correspond to the mass eigenstates, while for the up-type quarks we have $u_L = V^\dagger u_L^{\rm mass}$, where $V$ is the CKM matrix~\footnote{We will not be concerned with neutrino mass effects in the current work, implying we do not distinguish between neutrino mass and flavor eigenstates.}. For further details of our notation, we refer to Appendix~\ref{app:SMEFT}.

In this work, we will mainly be concerned with the SMEFT Lagrangian at tree level and only consider loop effects to include sizable QCD corrections at leading-log accuracy. 
This affects only the operators in the $(\bar LR)(\bar LR)$ and $(\bar LR)(\bar RL)$ classes in Table~\ref{tab:operators1}. We will evaluate these coefficients at a renormalization scale of $\mu=1$ TeV when presenting the results.
We use the Lagrangian in Eq.~\eqref{eq:Lag} to make predictions for observables at or above the electroweak scale, such as the $W$ mass, Z-pole observables, and Drell-Yan production cross sections. In the EWPO analysis, we choose our input parameters as the Fermi constant $G_F$, extracted from muon decay, the $Z$ mass, $m_Z$, and the fine-structure constant, $\alpha_{em}$. 

To predict the low-energy charged current processes in terms of the Wilson coefficients $C_i$ of Eq.~\eqref{eq:Lag}, we switch to the LEFT~\cite{Jenkins:2017jig}, the low-energy EFT valid below the weak scale.
This is formally done by integrating out the heavy SM fields and matching the operators in Eq.~\eqref{eq:Lag} to an $SU(3)_c\times U(1)_{em}$-invariant Lagrangian relevant for kaon, pion, and $\bt$ decays. 
The matching to this Lagrangian, and the translation to conventions often used in the literature, are discussed in Appendix~\ref{app:epstransl}.

\begin{table}[t!]
	\begin{center}
		\begin{minipage}[t]{.8\textwidth}
			\renewcommand{\arraystretch}{1.5}
			\begin{tabular}[t]{|c|c|c|c|c|}\hline
				\multicolumn{2}{|c|}{\textbf{Operators}}& L & EW & C \\\hline
				\multicolumn{2}{|c|}{\boldmath$H^4 D^2$} &&&\\
				\hline
				$Q_{H D}$   & $\ \left(H^\dag D^\mu H\right)^* \left(H^\dag D_\mu H\right)$&\multicolumn{3}{c|}{parameter shift ($m_Z$)}  \\\hline
				\multicolumn{2}{|c|}{\boldmath$X^2H^2$}&&& \\
				\hline
				$Q_{H WB}$     & $ H^\dag \tau^I H\, W^I_{\mu\nu} B^{\mu\nu}$ &\multicolumn{3}{c|}{parameter shift ($\sin\theta_W$)}\\\hline
				
				\multicolumn{2}{|c|}{\boldmath$\psi^2H^2 D$}&&& \\
				\hline
				$Q_{H l}^{(1)}$      & $(H^\dag i\overleftrightarrow{D}_\mu H)(\bar l_p \gamma^\mu l_r)$&\xmark &\cmark &\cmark\\
				$Q_{H l}^{(3)}$      & $(H^\dag i\overleftrightarrow{D}^I_\mu H)(\bar l_p \tau^I \gamma^\mu l_r)$&\cmark &\cmark &\cmark\\
				$Q_{H e}$            & $(H^\dag i\overleftrightarrow{D}_\mu H)(\bar e_p \gamma^\mu e_r)$&\xmark &\cmark &\cmark\\
				$Q_{H q}^{(1)}$      & $(H^\dag i\overleftrightarrow{D}_\mu H)(\bar q_p \gamma^\mu q_r)$&\xmark &\cmark &\cmark\\
				$Q_{H q}^{(3)}$      & $(H^\dag i\overleftrightarrow{D}^I_\mu H)(\bar q_p \tau^I \gamma^\mu q_r)$&\cmark &\cmark &\cmark\\
				$Q_{H u}$            & $(H^\dag i\overleftrightarrow{D}_\mu H)(\bar u_p \gamma^\mu u_r)$&\xmark &\cmark &\cmark\\
				$Q_{H d}$            & $(H^\dag i\overleftrightarrow{D}_\mu H)(\bar d_p \gamma^\mu d_r)$&\xmark &\cmark &\cmark\\
				$Q_{H u d}$ + h.c.   & $i(\widetilde H ^\dag D_\mu H)(\bar u_p \gamma^\mu d_r)$ & \cmark & \xmark &\cmark
				\\\hline
    \multicolumn{2}{|c|}
    {\boldmath$(\bar L L)(\bar L L)$} &&&\\
				\hline
				$Q_{ll}$        & $(\bar l_p \gamma^\mu l_r)(\bar l_s \gamma_\mu l_t)$ &\multicolumn{3}{c|}{parameter shift ($G_F^{}$)}\\\hline
				$Q_{lq}^{(1)}$                & $(\bar l_p \gamma^\mu l_r)(\bar q_s \gamma_\mu q_t)$ &\xmark & \xmark &\cmark\\
				$Q_{lq}^{(3)}$                & $(\bar l_p \gamma^\mu \tau^I l_r)(\bar q_s \gamma_\mu \tau^I q_t)$ &\cmark & \xmark &\cmark
				\\\hline
				\multicolumn{2}{|c|}{\boldmath$(\bar LR)(\bar RL)+\hc$} &&&\\
				\hline
				$Q_{ledq}$ & $(\bar l_p^j e_r)(\bar d_s q_{tj})$ &\cmark & \xmark &\cmark\\\hline
				
				\multicolumn{2}{|c|}{\boldmath$(\bar LR)(\bar L R)+\hc$} &&&\\
				\hline
				$Q_{lequ}^{(1)}$ & $(\bar l_p^j e_r) \epsilon_{jk} (\bar q_s^k u_t)$ &\cmark & \xmark &\cmark\\
				$Q_{lequ}^{(3)}$ & $(\bar l_p^j \sigma_{\mu\nu} e_r) \epsilon_{jk} (\bar q_s^k \sigma^{\mu\nu} u_t)$&\cmark & \xmark &\cmark\\\hline
			\end{tabular}
		\end{minipage}
	\end{center}
	\caption{The dimension-six SMEFT operators  (in the Warsaw basis~\cite{Grzadkowski:2010es}) that are relevant for our analysis, with subscripts $p, r, s, t$ indicating weak-eigenstate generation indices. The last three columns indicate which observables the operators contribute to. `L' stands for the neutron, nuclear, and meson decays discussed in Appendices~\ref{app:beta} and~\ref{app:meson}; `EW' stands for the electroweak precision observables of Appendix~\ref{app:EWPO}; `C' stands for the $pp \to \ell\ell$ and $pp \to \ell\nu$ processes discussed in Appendix~\ref{app:LHC}.}
 \label{tab:operators1}
\end{table}

\subsection{Observables}
\label{Observables}
The choice of SMEFT operators included in the fits is dictated by the observables that we want to analyze. 
The observables fall into three classes: low-energy CC (semi)leptonic processes (L), electroweak-precision observables (EW), and collider probes such as Drell-Yan at the LHC (C). 

\subsubsection{Low-energy charged-current observables} 
\label{sec:betaObs}

We include semileptonic processes involving electrons and muons mediated by light quark ($u,d,s$) charged-current interactions. Relevant observables involve measurements of neutron and nuclear $\beta$ decays as well as pion and kaon decays. 
\\\\
{\bf Neutron and nuclear $\bt$ decays:}  \\
We closely follow the analysis of Ref.~\cite{Falkowski:2020pma} and take into account the neutron lifetime, $0^+ \to 0^+$ transitions, and mirror decays.
Within the SM, these measurements determine $V_{ud}$, which, when combined with the extraction of $V_{us}$ discussed below, currently leads to a deviation from CKM unitarity. In SMEFT, these measurements are sensitive to new vector and scalar interactions. Apart from the decay rates, we consider correlation measurements, such as those between neutrino and electron momenta $\sim \vec p_e\cdot \vec p_\nu$, described by the coefficient $a$, or between nuclear spin and electron momentum $\sim \vec p_e\cdot \vec J$, captured by the coefficient $A$. These measurements allow one to extract the ratio of vector-to-axial couplings $g_A/g_V$, and are sensitive to different combinations of SMEFT interactions.

The $\bt$-decay observables considered are listed in Appendix~\ref{app:beta}. We include them in the fit using a $\chi^2$ function provided by the authors of Ref.~\cite{Falkowski:2020pma}. An important subset of them, especially with respect to CAA, are the $\mathcal{F}t$ values of superallowed transitions. For a transition, $i$, these values are proportional to the inverse of the decay rate, $1/\Gamma_i$, with transition-dependent phase-space factors and radiative corrections taken out. Their theoretical predictions can be written as
\bea\label{eq:ft}
{\cal F }t_i = \frac{4 \pi^3 \log 2}{M_F^2 m_e^5}\left[C_V^2 \pm 2 C_V C_S\gamma_i\Big\langle \frac{m_e}{E_e}\Big\rangle_i\right]^{-1}\,,
\eea
where the upper (lower) sign corresponds to a $\beta^-$ ($\beta^+$) decay and the Fermi matrix element is $M_F^2=2$ for $0^+\to0^+$. Here 
\bea
C_V&=&\sqrt{2}G_F V_{ud}g_V\sqrt{1+\Delta_R^V}(1+\epsilon_L^{ed}+\epsilon_R^{ed})\,,\qquad C_S = \sqrt{2} G_F V_{ud}g_S \epsilon_S^{ed}\,,
\eea
and $\gamma_i =\sqrt{1-\alpha^2 Z_i^2}$, with $Z$ the atomic number of the final-state nucleus. The Wilson coefficients are given in terms of the $\epsilon_i$ that vanish in the SM, but are generally nonzero in SMEFT, see Appendix~\ref{app:epstransl}. The second term in square brackets arises from the so-called $b$ Fierz interference term, where $\langle m_e/E_e\rangle_i$ denotes the averaged ratio $m_e/E_e$. This term vanishes in the SM, but receives corrections from non-standard scalar interactions in SMEFT.
$C_V$ and $C_S$ depend on the nucleon vector and scalar charges, $g_V$ and $g_S$, and short-distance radiative corrections, $\Delta_R^V$, see Appendix~\ref{app:beta}. The latter have recently been the subject of several re-evaluations~\cite{Seng:2018qru,Seng:2018yzq,Czarnecki:2019mwq,Shiells:2020fqp,Hayen:2020cxh,Cirigliano:2023fnz,Ma:2023kfr}, and we use the value from Ref.~\cite{Seng:2018yzq}.

To take into account the theoretical uncertainties associated with the transition-dependent corrections, we follow the treatment in Ref.~\cite{Falkowski:2020pma}. Uncertainties enter Eq.~\eqref{eq:ft} through $\mathcal Ft_i = (1+\delta_R')(1+\delta_{NS}^V-\delta_C^V) ft_i$, with $ft_i$ denoting the uncorrected values of $ft$. Here $\delta_R'$ represents long-distance radiative corrections, $\delta_{NS}^V$ captures nuclear-structure-dependent effects, and $\delta_C^V$ is an isospin-breaking correction. The theoretical uncertainties related to these corrections are then taken into account through
\bea\label{eq:ft2}
\mathcal Ft_i^{\rm th.} = {\cal F}t_i^{\rm expt.} (1+\eta_1 \Delta \delta_R^{\prime \, i}+\eta_2\Delta \delta_{NS,A}+\eta_3\Delta\delta^i_{NS,E})\,,
\eea
where $\eta_{1,2,3}$ are treated as Gaussian parameters with a $1\sigma$ confidence interval of $\eta_i\in [-1,1]$. The uncertainty $\Delta \delta_R^{\prime \, i}$ is taken to be one third of the $Z^2\alpha^3$ term in $ \delta_R^{\prime \, i}$~\cite{Towner:2007np,Hardy:2014qxa,Gonzalez-Alonso:2018omy}, while $\Delta \delta_{NS,A}=3.3\cdot 10^{-4}$~\cite{Hardy:2020qwl}, and $\Delta\delta^i_{NS,E} = 0.8 \cdot 10^{-4}Q/{\rm MeV}$~\cite{Gorchtein:2018fxl}, with $Q$ the transition energy. 
In practice, the uncertainties related to $\eta_1$ turn out to be negligible~\cite{Falkowski:2020pma}. For the `experimental' values, ${\cal F}t_i^{\rm expt.}$, we use the results of Ref.~\cite{Hardy:2020qwl}, while we employ the theoretical expressions of Ref.~\cite{Falkowski:2020pma} for the left-hand side of Eq.~\eqref{eq:ft2}.
As we will see below, the CAA will often manifest itself through nonzero values of $\Delta_R^V$ and $\eta_{2,3}$. In such cases, the theory predictions are unable to accommodate the measured rates of $0^+\to0^+$ without significantly altering the corrections due to $\delta_i$. 
\\\\
{\bf Kaon and pion decays:}\\
We closely follow Ref.~\cite{Gonzalez-Alonso:2016etj} and organize the two-body decays into a decay rate, $\Gamma(K\to \mu\nu_\mu)$, and three ratios, $R_\pi=\Gamma(\pi\to e\nu_e)/\Gamma(\pi\to \mu\nu_\mu)$, $R_K=\Gamma(K\to e\nu_e)/\Gamma(K\to \mu\nu_\mu)$, and $R_\mu=\Gamma(K\to \mu\nu_\mu)/\Gamma(\pi\to \mu\nu_\mu)$. 
In the SM, the ratio $R_\mu$ allows for a precise determination of $V_{us}/V_{ud}$. Going beyond the SM, all two-body decays receive contributions from axial and pseudoscalar SMEFT operators. 

The two-body rate for a pseudoscalar meson $P$ is given by
\bea
\Gamma_{P\ell 2}=\frac{G_F^2|\tilde V_{uD}^\ell|^2 f_{P^\pm}^2}{8\pi} m_{P^\pm} m_\ell^2\left(1-\frac{m_\ell^2}{m_{P^\pm}^2}\right)^2(1+\delta_{\rm em}^{P\ell})(1+\Delta_{\ell2}^P)\,,
\eea
where $m_{P^\pm}$ and $f_{P^\pm}$ are, respectively, the meson mass and decay constant, $\delta_{\rm em}^{P\ell}$ captures radiative corrections, and 
\bea
\tilde V_{uD}^{\ell} &=& \left(1+\epsilon_L^{\ell D}+\epsilon_R^{\ell D}-\epsilon_L^\mu\right)V_{uD}\,,\nn\\
\Delta_{\ell2}^P &=& - 4\eR^{\ell D} - \frac{2m_{P^\pm}^2}{m_\ell (m_u + m_D)}\eP^{\ell D} + \mathcal{O}(v^4/\Lambda^4)\,,
\eea
where $D={d,s}$ for $P={\pi,K}$ and the coefficients $\epsilon_i$ denote the contributions of the SMEFT operators, see Appendix~\ref{app:epstransl} for details.
We see that $\Gamma(\pi\to \ell\nu_\ell)$ and $\Gamma(K\to \ell\nu_\ell)$ stringently constrain pseudoscalar interactions whose contributions are enhanced by a factor of $(m_{P^\pm}^2)/(m_\ell (m_u + m_D))$, where $\ell=\{e,\,\mu\}$.

For the three-body decays, we include the rates of $\pi^+\to\pi^0 e\nu_e$, $K\to \pi e\nu_e$, and $K\to \pi \mu\nu_\mu$. The latter two kaon modes determine $V_{us}$ in the SM, while all these three-body decays are sensitive to vector, scalar, and tensor SMEFT interactions. For the total rates we have
\bea
\Gamma_{K \ell 3} = \frac{G_F^2 m_K^5}{192\pi^3} C_K S_{\rm ew}|\tilde V_{us}^\ell|^2 f_+(0)^2 I_K^\ell(1+\delta^c+\bar\delta^{c\ell}_{\rm em})^2\,,
\eea
where $C_K = 1/2$ or 1 for charged or neutral kaons. $S_{\rm ew}$, $\delta^c$, and $\bar\delta^{c\ell}_{\rm em}$ denote short-distance, isospin, and radiative corrections.
Finally, $f_+(0)$ is the $K\to \pi$ form factor at zero momentum, while $I_K$ is the phase-space integral that depends on the BSM scalar and tensor interactions.
In addition, we use measurements of the shape of the spectrum in these $K_{l3}$ decays to constrain the scalar and tensor operators. 
The experimental and theoretical inputs required are collected in Appendix~\ref{app:meson}.

\subsubsection{Electroweak precision observables}

We include the `traditional' observables measured at the $Z$ pole~\cite{ALEPH:2005ab}. These include the decay widths, asymmetries, and the hadronic cross section obtained from $e^+ e^-\to Z\to\bar qq$, as well as the $W$ mass. As noted earlier, the recent CDF determination of $m_W$~\cite{CDF:2022hxs} deviates significantly from the SM and the average of other measurements~\cite{ParticleDataGroup:2022pth}. Therefore, in our analyses we will consider two scenarios, one assuming the CDF determination and one where we use the world average value in PDG. In addition, we include several measurements at hadron colliders of processes at electroweak-scale energies, such as the branching ratios of the $W$ boson and asymmetries in $pp \to \bar \ell \ell$~\cite{Balkin:2022glu,Efrati:2015eaa}.

Some of the traditional EW observables are measured at the sub-permille level and provide precise determinations of SM parameters, like $m_Z$, as well as stringent tests of the SM predictions. 
Due to their sensitivity to BSM physics,
the study of EWPO in the SMEFT has received considerable attention~\cite{Han:2004az,Ciuchini:2013pca,Falkowski:2014tna,Berthier:2015oma,deBlas:2021wap,deBlas:2022hdk,Breso-Pla:2021qoe,Bruggisser:2022rhb}.
Recent analyses have highlighted the important interplay between EWPO, Higgs observables, and diboson production at the LHC~\cite{Ellis:2018gqa,Ellis:2020unq,Almeida:2021asy,Fan:2022yly},
the role of dimension-eight SMEFT operators~\cite{Corbett:2021eux,Corbett:2023qtg}, and the implications of quark flavor assumptions~\cite{Balkin:2022glu,Efrati:2015eaa,Bellafronte:2023amz}.
Beyond the SM, the traditional EW observables receive mainly corrections from operators that contribute to the fermion-$Z$ couplings, i.e.\ operators in the $\psi^2 H^2D$ class in Table~\ref{tab:operators1}. The additional observables from hadron colliders are helpful in closing free directions that would otherwise appear when considering SMEFT scenarios that do not assume flavor universality.
The observables used, as well as the relevant SM predictions, are collected in Appendix~\ref{app:EWPO} and Table~\ref{tab:EWPO}.

\subsubsection{Collider probes} 
The measurement of Drell-Yan tails at the LHC has also been shown to be an effective probe of BSM interactions, allowing us to probe semileptonic effective operators with different quark and lepton flavors~\cite{Cirigliano:2012ab,Greljo:2017vvb,Alioli:2018ljm,Torre:2020aiz}. These constraints are complementary to those derived from low-energy flavor-physics observables. To incorporate Drell-Yan  observables into our analysis, we implement the Mathematica package \texttt{HighPT}~\cite{Allwicher:2022mcg,Allwicher:2022gkm}, which includes the cross sections for Drell-Yan dilepton ($pp\to \bar\ell\ell$) and mono-lepton ($pp\to \bar\ell\nu$) searches by ATLAS and CMS, see Appendix~\ref{app:LHC} for details. \texttt{HighPT} includes the SMEFT contributions to the cross sections at LO in QCD, with detector effects simulated with \texttt{Pythia8} and \texttt{Delphes3}. Although NLO QCD effects have been shown to be important, reaching up to 30\% at high invariant mass~\cite{Alioli:2018ljm}, the LO bounds are sufficient to provide a good estimate of the LHC sensitivity.
The SM and SMEFT DY predictions are affected by theoretical uncertainties, induced, for example, by errors on the parton distributions or the omission of QCD corrections. Since the DY sensitivity relies on large deviations from the SM shape, rather than on a high-precision comparison with the SM prediction, and the uncertainty in high-transverse or invariant-mass bins is dominated by experimental uncertainties, we neglect these theory errors in our analysis.

In principle, the dimension-six operators can induce corrections to the cross section that scale as $\sim |C_i|^2\sim v^4/\Lambda^4$. Such terms formally appear at the same order as genuine dimension-eight operators, which are not included in this work. Therefore, we only take into account SMEFT effects up to $\Or(v^2/\Lambda^2)$ for most of this work. Although not fully consistent, we will make an exception in Section~\ref{22fit} to estimate the sensitivity of the LHC to (pseudo)scalar operators, whose contributions vanish at $\Or(1/\Lambda^2)$. In the future, it will be interesting to extend the study to $\Or(1/\Lambda^{4})$ in a consistent manner~\cite{Boughezal:2021tih,Kim:2022amu,Boughezal:2022nof}.

\section{Analysis strategy}
\label{sect:strategy}

While the SMEFT framework provides a powerful tool to systematically and model-independently perform a global analysis of particle physics experiments involving a broad range of energies, there are technical difficulties towards a practical implementation. The main complication is the large number of independent operators, for example, the 2499 baryon-number-conserving operators at dimension six. 
Even after restricting the operator structures to a subset of the full basis as done in Table~\ref{tab:operators1}, which focuses on the CLEW observables, including all generation indices leads to an unmanageable number of Wilson coefficients. 
This problem is often circumvented by making certain flavor and CP assumptions on the operator basis. For example, the fit of EWPO can be greatly simplified if one assumes a global $U(3)^5$ flavor symmetry. This assumption implies that only 10 operators affect EWPO at tree level~\cite{deBlas:2022hdk} (more operators appear if loop corrections are implemented~\cite{Bellafronte:2023amz}). 
Similarly, CP-invariance is often assumed as well. 

These assumptions are not fully satisfactory as they essentially reintroduce model dependence into the analysis. 
The downside of using a strong flavor assumption is that it can miss interesting BSM explanations of anomalies. For example, as we shall see, the CAA can be nicely explained by right-handed charged current operators in up-down and up-strange transitions~\cite{Bernard:2007cf,Grossman:2019bzp,Cirigliano2022Aug}. However, the corresponding SMEFT operators are forbidden under $U(3)^5$ and are strongly suppressed within minimal flavor violation (MFV)~\cite{DAmbrosio:2002vsn}. Entire classes of BSM models, such as left-right symmetric models, are automatically discarded~\cite{Dekens:2021bro}, demonstrating that the analysis is no longer model independent. 

Here, our goal will be to mitigate this loss of model independence, which is introduced when flavor assumptions are made, as much as possible. 
To achieve this, the first task is to identify the most general set of Wilson coefficients, denoted by $C_{n}$, that affect the CLEW observables discussed in Section~\ref{Observables} and Table~\ref{tab:operators1}. 
As we shall see, some of these couplings, denoted by $\tilde{C}_k$, unavoidably affect observables not included in our analysis as well, such as FCNC. A truly global analysis would thus need to include measurements of FCNCs explicitly, along with the CLEW observables discussed here. 
However, we will argue that due to the strength of FCNC constraints on the $\tilde{C}_k$ coefficients and the fact that the contributions of $\tilde{C}_k$ to the CLEW observables are in many cases suppressed by powers of the Cabibbo angle $\lambda \sim 0.2$, 
the global fit can be approximated by setting $\tilde{C}_k=0$ in our analysis. This makes use of
the approximate decoupling of the global analysis into smaller fits, in particular, into flavor-changing and flavor-conserving sectors. This corresponds to an approximate factorization of the likelihood function. Consequently, we expect that the constraints on $C_n$ orthogonal to $\tilde C_k$ remain largely unchanged in a truly global fit. 

In the following, we identify the Wilson coefficients relevant for the $U(3)^5$ and flavor-assumption-independent analyses, and present the associated fitting scenarios.

\subsection{$U(3)^5$}\label{sec:U35}

To compare with the existing literature and highlight possible drawbacks, we first consider a scenario based on the symmetry group of flavor universality, namely $U(3)_l\times U(3)_e\times U(3)_q\times U(3)_u\times U(3)_d$. This flavor symmetry imposes strong constraints on the operators in Table~\ref{tab:operators1}. 
For the couplings that contribute to EWPO, neglecting terms proportional to the SM Yukawa couplings, one finds the following relations
\bea\label{eq:U35coeffs}
C_{Hl}^{(1,3)}\propto \mathbf{1}\,,\qquad C_{He}^{}\propto \mathbf{1}
\qquad 
C_{Hq}^{(1,3)}\propto \mathbf{1}\,,\qquad C_{Hu}^{}\propto \mathbf{1}\,,\qquad C_{Hd}^{}\propto \mathbf{1}\,.
\eea
For operators that involve right-handed fields and contribute to low-energy CC observables, the flavor symmetry implies
\begin{equation}
    C_{Hud}=C_{ledq}= C_{lequ}^{(1,3)}=0\,.
\end{equation}
Semileptonic four-fermion operators involving left-handed fields induce 9 terms, all with the same Wilson coefficient
\begin{equation}\label{eq:U35coeffs2}
    C^{(3)}_{lq} \propto \mathbf{1}_{\ell} \times \mathbf{1}_{q}\,. 
\end{equation}
In the purely leptonic sector, $Q_{ll}$ involves two bilinears with the lepton field $L$. In this case, we can write two invariant structures
\begin{equation}
    \left[C_{ll}\right]_{ij \ell m} \propto  C_0 \delta^{i j} \delta^{\ell m} + \hat{C}_{ll}\delta^{i m} \delta^{\ell k}\,.
\end{equation}
$C_0$ does not induce charged-current decays of the muon or $\tau$ lepton, and, in particular, does not affect $G_F$.  
$\hat C_{ll} = [C_{ll}]_{1221} = [C_{ll}]_{1331} =[C_{ll}]_{2332}$ affects muon decay and the extraction of $G_F$. 
In the case of $U(3)^5$, the corrections to $\mu \rightarrow e \bar{\nu}_e \nu_\mu$,
$\tau \rightarrow e \bar{\nu}_e \nu_\tau$, and $\tau \rightarrow \mu \bar{\nu}_\mu \nu_\tau$ are the same, such that the effect of $\hat{C}_{ll}$ on the lepton decays becomes invisible if $G_F$ is used as input parameter. 
Finally, purely bosonic operators, $Q_{HD}$ and $Q_{HWB}$ also affect the EWPO. However, it turns out that these contributions always appear in fixed combinations with the vertex corrections in Eq.~\eqref{eq:U35coeffs}. They therefore do not increase the effective number of fit parameters, see Section~\ref{resultsU3} for details.

In summary, for this scenario we will consider a set of 9 Wilson coefficients involving the couplings in Eq.~\eqref{eq:U35coeffs} and \eqref{eq:U35coeffs2} together with $\hat C_{ll}$.

\subsection{Flavor-symmetry independent analysis} 
\label{sec:flavors-global}

The first step towards a flavor-assumption-independent analysis is to identify the most general set of Wilson coefficients that affect the CLEW observables, allowing for a general flavor structure, starting from Table~\ref{tab:operators1}. 
To this end, we need to pick a basis, and we find it convenient to phrase the discussion in terms of the quark-mass basis, which is reached from the weak basis by simply replacing $u_L \to V^\dagger u_L$, where $V$ is the CKM matrix.
In addition, it will be useful to identify the couplings that give rise to neutral-current interactions. In the quark mass basis, the induced couplings of up- and down-type quarks to the $Z$ boson are given by the following linear combinations of Wilson coefficients 
\bea\label{eq:CHqud}
C_{Hq}^{(u)} &=& V\left[ C_{Hq}^{(1)}-C_{Hq}^{(3)}\right]V^\dagger\,,\qquad C_{Hq}^{(d)} =  C_{Hq}^{(1)}+C_{Hq}^{(3)}\,.
\eea
Similarly, the semileptonic Wilson coefficients appear as
\bea
\label{eq:Clq}
C_{lq}^{(u)} &=& V\left[ C_{lq}^{(1)}-C_{lq}^{(3)}\right]V^\dagger\,,\qquad C_{lq}^{(d)} =  C_{lq}^{(1)}+C_{lq}^{(3)}\,,\qquad \bar C^{(1,3)}_{lequ} = VC^{(1,3)}_{lequ}\,,
\eea
where $C_{lq}^{(u)}$ and $\bar C^{(1,3)}_{lequ}$ represent the neutral-current couplings between up-type quarks and charged leptons, while $C_{lq}^{(d)}$ and $C_{ledq}$ control the neutral currents between down-type quarks and charged leptons.

The contributions of these new Wilson coefficients to CLEW observables can be read off in a straightforward way in the case of EWPO and neutral-current Drell-Yan, which involve the diagonal entries of the above couplings. 
On the other hand, $d,s\to u$ transitions in CC processes, such as $\beta$-decay and CC Drell-Yan, 
are sensitive to the combinations ($D=d,s$) \begin{eqnarray}
\left[V C_{Hq}^{(d)}-C_{Hq}^{(u)}V\right]_{uD}\,,\qquad
\left[V C_{lq}^{(d)}-C_{lq}^{(u)}V\right]_{\ell\ell uD}\,,\nonumber\\
\left[V C_{le dq}^\dagger\right]_{\ell \ell 11}\,,\qquad
\left[V C_{le dq}^\dagger\right]_{\ell \ell 22}\,,\qquad
\left[\left(\bar C^{(1,3)}_{le qu}\right)^\dagger V\right]_{\ell \ell 11}\,.
\end{eqnarray}
These combinations depend on off-diagonal Wilson coefficients, thus inducing `cross-talk' with FCNC observables not included in our analysis. 
Now we discuss the impact of these probes in greater detail.\\

Contributions of left-handed interactions to CC processes are proportional to 
\bea\label{eq:epsLflav}
V_{uD}\epsilon_L^{D\ell} &=& \frac{v^2}{2}\left[V C_{Hq}^{(d)}-C_{Hq}^{(u)}V\right]_{uD}-\frac{v^2}{2}\left[V C_{lq}^{(d)}-C_{lq}^{(u)}V\right]_{\ell\ell uD}\,.
\eea
Neglecting terms that are suppressed by more than one power of $\la\equiv V_{us}$, only a few off-diagonal couplings are expected to contribute without CKM suppression,
\bea \label{eq:lowEflavor}
\epsilon_L^{d\ell} &\simeq& \frac{v^2}{2}\left[
C_{\substack{Hq\\11}}^{(d)}-C_{\substack{Hq\\11}}^{(u)}+\frac{V_{us}}{V_{ud}}C_{\substack{Hq\\21}}^{(d)}-\frac{V_{cd}}{V_{ud}}C_{\substack{Hq\\12}}^{(u)}\right]-(C_{Hq}\to C_{lq})
\,,\nn\\
\epsilon_L^{s\ell} &\simeq& \frac{v^2}{2}\left[
\frac{V_{ud}}{V_{us}}C_{\substack{Hq\\12}}^{(d)}- \frac{V_{cs}}{V_{us}}C_{\substack{Hq\\12}}^{(u)}+C_{\substack{Hq\\22}}^{(d)}-C_{\substack{Hq\\11}}^{(u)}-\frac{V_{ts}}{V_{us}}C_{\substack{Hq\\31}}^{(u)}\right]-(C_{Hq}\to C_{lq})
\,.\eea
The off-diagonal couplings are stringently constrained by the decays of pseudoscalar mesons to charged leptons. In particular, $[C_{Hq}^{(d)}]_{12,21}$ will have a large impact on kaon decays such as $K\to \pi \bar \ell \ell$, while $[C_{Hq}^{(u)}]_{12}$ affects the analogous $D$ meson decays. We discuss the resulting constraints in more detail in Appendix~\ref{app:flavor}, where the main conclusion is that the off-diagonal elements appearing in Eq.~\eqref{eq:lowEflavor} are expected to have a minimal impact on low-energy CC observables because of stringent FCNC constraints. Similarly, $[C_{Hq}^{(u)}]_{31}$ in Eq.~\eqref{eq:lowEflavor} is constrained by top-quark decays and its contribution to $\epsilon^{s\ell}_L$ is suppressed by a factor of $V_{ts}/V_{us}$, so that it cannot significantly affect the low-energy CC observables. These arguments justify a basis that includes only the diagonal couplings of $C_{\substack{Hq}}^{(u,d)}$.

Very similar arguments hold for the coefficients $[\bar C^{(1,3)}_{lequ}]^\dagger$ and $C_{ledq}^\dagger$, whose contributions to the effective low-energy couplings, $\epsilon_{S,P,T}$, have a flavor structure analogous to the $C_{Hq}^{(u)}$ and $C_{Hq}^{(d)}$ terms in Eq.~\eqref{eq:epsLflav}, respectively.

The discussion of $C_{lq}^{(u,d)}$ is somewhat more involved. Although their appearance in Eq.~\eqref{eq:epsLflav} and~\eqref{eq:Clq} is analogous to the case of $C_{Hq}^{(u,d)}$ couplings, they differ due to the fact that these operators also induce couplings to neutrinos of the form $(\bar\nu\nu)( \bar q q)$. The flavor structures that govern neutrino interactions with up- and down-type quarks are given by
\bea\label{eq:clqnu}
C_{lq}^{(\nu-u)} = VC_{lq}^{(d)}V^\dagger \,,\qquad C_{lq}^{(\nu-d)} = V^\dagger C_{lq}^{(u)}V \,.
\eea
Thus, off-diagonal entries of $C_{lq}^{(u,d)}$ can be stringently constrained by meson decays to charged leptons and neutrinos. The latter are discussed in more detail in Appendix~\ref{app:flavor}, where we find that $[C_{lq}^{(\nu-d)}]_{12}$ is very stringently constrained by $K\to \pi  \nu \bar \nu$,
practically forcing $[C_{lq}^{(u)}]_{\ell \ell 12} \simeq 0$ and $[C_{lq}^{(u)}]_{\ell \ell 11} \simeq [C_{lq}^{(u)}]_{\ell \ell 22}$.\\

The inclusion of low-energy CC processes in the CLEW analysis also affects the linearly independent combinations of the purely bosonic Wilson coefficients $C_{HWB}$ and $C_{HD}$ (directly associated with the Peskin-Takeuchi oblique parameters $S$ and $T$~\cite{Peskin:1991sw}) that can be constrained by the data. 
While ten operators affect EWPO under $U(3)^5$, only eight linear combinations are actually constrained. We denote these combinations by $\hat{C}_i$, see Eq.~\eqref{eq:BSM_primary}, which indicates that $C_{HF}^{(3)}$, with $F=\{l,q\}$, appear in a linear combination with $C_{HWB}$ and $C_{HD}$, while $C_{HF}^{(1)}$ and $C_{Hf}$, with $f=\{u,d,e\}$, appear in combination with $C_ {HD}$.
Since the low-energy CC observables (L) are affected by $C_{HF}^{(3)}$ but not by $C_{HWB}$ and $C_{HD}$ (see Table~\ref{tab:operators1}), they break the degeneracy between ${C}_{HF}^{(3)}$ and $C_{HWB,HD}$ present in the EW fits. 
The LEW data thus constrain one linear combination of $C_{HD}$ and $C_{HWB}$, which we call $C_{ST}$, while leaving the orthogonal direction, $C_{TS}$, unconstrained. 

The combinations $C_{ST,TS}$ take the form
\begin{align}
\bma C_{ST}\\ C_{TS}\ema \ \equiv \frac{1}{\sqrt{c_w^2+16s_w^2}}\bma
4s_w & c_w\\
-c_w & 4s_w
\ema \bma  C_{HWB} \\ C_{HD}\ema \,,
\end{align}
where $s_w = \sin{\theta_w}$ and $c_w = \cos{\theta_w}$, with $\theta_w$ being the weak mixing angle.
$C_{HF}^{(1)}$ and $C_{Hf}$ do not affect the low-energy CC, but do influence the LHC observables by entering the Drell-Yan process. Their corrections to the Drell-Yan cross section exhibit the same energy dependence as the SM background. In contrast, the SMEFT 4-fermion operators give corrections with a stronger energy dependence and dominate the high-energy tail of Drell-Yan processes. Thus, although some constraints are exerted on $C_{HF}^{(1)}$ and $C_{Hf}$ by the LHC observables, they are too weak to substantially lift the degeneracy with $C_{HD}$. Thus, only the linear combination $C_{ST}$ is well constrained by our CLEW data sets. The orthogonal combination, $C_{TS}$, on the other hand, is poorly constrained and we do not include it.

For each class of operator in Table~\ref{tab:operators1} we can now summarize which Wilson coefficients should be included in a general analysis of low-energy CC and EWPO data:
\begin{itemize}
\item $\psi^2 H^2 D$\\
For $Q_{Hl}^{(1,3)}$ and $Q_{He,Hd,Hu}$ we take into account the diagonal entries that appear in EWPO, including couplings to all flavors apart from the top. $Q_{Hud}$ does not induce neutral currents or effects in EWPO, and we include the terms that contribute to low-energy CC processes, namely the $11$ and $12$ components. 
We include the diagonal entries of the $C_{Hq}^{(u,d)}$ coefficients that enter in EWPO or low-energy observables, again excluding the top coupling.
\item $(\bar LL)(\bar LL)$\\
Here, the relevant couplings include $[C_{ll}]_{1221}$ that enters through its effect on muon decay. We include the diagonal entries of the $C_{lq}^{(u,d)}$ couplings that appear in low-energy CC measurements, namely $[C^{(u,d)}_{lq}]_{\ell\ell 11}$ and $[C^{(d)}_{lq}]_{\ell\ell 22}$. In addition, we impose the constraint that these coefficients do not induce large off-diagonal entries of $C_{lq}^{(\nu-d)}$, which implies $[C^{(u)}_{lq}]_{\ell\ell pr}\propto \delta_{pr} [C^{(u)}_{lq}]_{\ell\ell pr}$. Although this does not affect low-energy observables, the existence of couplings to the heavier generations does have an effect on collider observables.
\item $(\bar LR)(\bar RL)$\\
Taking into account the diagonal terms that enter the low-energy measurements leads to the inclusion of $[C_{ledq}]_{\ell\ell11}$ and $[C_{ledq}]_{\ell\ell22}$. 
\item $(\bar LR)(\bar LR)$\\
The same analysis tells us that we should include $[\bar C_{lequ}^{(1,3)}]_{\ell\ell 11}$.
\item Finally, we include the linear combination $C_{ST}$ of the two purely bosonic interactions $C_{HD}$ and $C_{HWB}$. 
\end{itemize}
We summarize the Wilson coefficients involved in the left panel of Table~\ref{tab:globalWC}.\\

The above set of operators are chosen because they provide corrections to the EWPO and CC processes while not being constrained to negligible levels by FCNC constraints. Four-fermion operators give rise to contributions to Drell-Yan processes that grow with energy. They are therefore strongly constrained by measurements of the high-invariant or transverse-mass tails of the NC and CC Drell-Yan distributions. CC Drell-Yan is primarily sensitive to four-fermion operators that affect $\beta$ decays at low energy, namely
$[V C_{lq}^{(d)}-C_{lq}^{(u)}V]_{\ell\ell ud}$,
$[V C_{lq}^{(d)}-C_{lq}^{(u)}V]_{\ell\ell us}$,
$\left[C_{le dq}\right]_{\ell \ell 11}$,
$\left[C_{le dq}\right]_{\ell \ell 22}$,
and $[C^{(1,3)}_{lequ}]_{\ell \ell 11}$. Operators with heavy $c$ and $b$ quarks do contribute to the Drell-Yan cross section, but their effects are suppressed by the heavy-quark PDFs and can, in the first approximation, be neglected. The above derived operator basis is then complete for the analysis of CC Drell-Yan processes. 

Through gauge invariance, the semileptonic CC four-fermion operators in principle also contribute to NC Drell-Yan which is included in our analysis. Strictly speaking, these measurements are affected by additional operators $C_{ld}$, $C_{lu}$,
$C_{ed}$, $C_{eu}$, and $C_{qe}$ with their respective flavor indices. Including these interactions would also require the extension of the set of observables since these operators are probed by processes such as parity-violating electron-proton scattering. Here, we do not discuss these operators further because they do not modify the EWPO and low-energy CC processes.
Nevertheless, explicitly including these operators would be an interesting extension of the current framework, which we leave for future work.

\begin{table}\small
$$\begin{array}{l|l}
\centering
{\rm Global \ analysis} & {\rm Indices}\\\hline
 C_{\substack{Hl\\pr}}^{(1,3)}\,,\quad C_{\substack{He\\pr}}^{}\, & pr \in \{ee,\,\mu\mu,\,\tau\tau\}\\
C_{\substack{Hq\\pr}}^{(d)}\,,\quad  C_{\substack{Hd\\pr}}^{} & pr \in \{11,22,33\}\\
C_{\substack {Hq\\pr}}^{(u)}\,,\quad C_{\substack{Hu\\pr}}^{}& pr \in \{11,22\}\\
C_{\substack{Hud\\pr}} & pr\in\{11,12\}\,\\
 C^{(d)}_{\substack{lq \\ \ell\ell pr}}\,, \quad C_{\substack{ledq\\ \ell\ell pr}} & \ell\in\{e,\mu\}\,,\quad pr\in \{11,22\}
\\
C^{(u)}_{\substack{lq \\ \ell\ell 11}}\,, \quad \bar C^{(1,3)}_{\substack {lequ\\\ell\ell 11}} & \ell\in \{e,\mu\}\\
C_{ST} & \\
C_{\substack{ll\\2112}}\, & 
\end{array}\qquad
\begin{array}{l|l}
\centering
{\rm Low~energy~CC~analysis} & {\rm Indices}\\\hline
 C_{\substack{Hl\\pr}}^{(3)}\, & pr \in \{ee,\,\mu\mu\}\\
C_{\substack{Hq\\pr}}^{(d)} & pr \in \{11,22\}\\
C_{\substack {Hq\\ 11}}^{(u)}& \\
C_{\substack{Hud\\pr}} & pr\in\{11,12\}\,\\
 C^{(d)}_{\substack{lq \\ \ell\ell pr}}\,, \quad C_{\substack{ledq\\ \ell\ell pr}} & \ell\in\{e,\mu\}\,,\quad pr\in \{11,22\}
\\
C^{(u)}_{\substack{lq \\ \ell\ell 11}}\,, \quad \bar C^{(1,3)}_{\substack {lequ\\\ell\ell 11}} & \ell\in \{e,\mu\}\\
& \\
C_{\substack{ll\\2112}}\, & 
\\
\end{array}
$$
\caption{Left panel: The 37 Wilson coefficients that are relevant to the global analysis, including low-energy charged-current observables, 
EWPO, and CC Drell-Yan as described in Section~\ref{sec:flavors-global}. 
Right panel: The subset of 22 Wilson coefficients that contribute to low-energy charged-current observables as described in Section~\ref{sec:flavors-global}.}
\label{tab:globalWC}
\end{table}

\subsection{Three classes of analyses}
\label{3classes}
Having discussed the relevant operators, we will perform fits in three different scenarios:
\begin{enumerate}
    \item {\bf $U(3)^5$ fit:} We start with a scenario assuming $U(3)^5$ flavor symmetry, neglecting terms involving the SM Yukawa couplings. 
     This scenario highlights the close connection between EWPO, low-energy CC observables, and high-energy Drell-Yan processes. We will show that fits considering only a single class of observables, as often done in the literature,  can lead to a poor description for observables in the other classes, and are thus inconsistent. 
     
    \item {\bf Flavor-symmetry-independent intermediate fit:} A scenario involving only the 22 Wilson coefficients that affect low-energy CC processes and the CAA (see the right panel of Table~\ref{tab:globalWC}), but which nevertheless includes the EWPO and Drell-Yan data. Although not self-consistent, we will see that this scenario provides a good description of the data, in particular of the CAA, as long as the CDF measurement of $m_W$ is not considered. 
    
    \item  {\bf Flavor-symmetry-independent global fit:} A global fit in which all operators in the left panel of Table~\ref{tab:globalWC} are included.
    \end{enumerate}

In addition, we will consider several simpler scenarios, involving subsets of the operators, which will help build some intuition for the larger, more general analysis.
In all cases, we will study the role of different sets of observables. 
We will label each of these analyses by the observables considered and the Wilson coefficients involved in the scenario. L denotes the low-energy CC processes related to nuclear, neutron, and meson decay; EW stands for traditionally defined electroweak precision observables (EWPO); and C (collider) stands for the Drell-Yan processes described above. For instance, the fit L22 would describe a fit of the 22 operators introduced in Scenario 2 to the low-energy CC observables only, whereas CLEW22 would correspond to the same operators, but now also including EWPO and Drell-Yan processes. For fits that include EW, we will often consider two separate cases depending on whether we include or exclude the CDF measurement of $m_W$.

\section{Statistical tools}\label{tools}

In the next section, we present results for a large number of fits. We will use $\chi^2$ minimization to perform these analyses and employ the Akaike Information Criterion (AIC)~\cite{1100705} to compare the quality of various fits. 
The method of $\chi^2$ minimization is a statistical tool commonly used to fit theoretical models to empirical data. It measures the sum of the squared differences between the observed data and the model predictions, weighted by their respective uncertainties. By minimizing the function $\chi^2$, we obtain the best-fit parameters that most closely align the theoretical model with the experimental observations.

A $\chi^2$-fit is performed to a given set of experimental measurements $E = \{E_1,\,E_2,\,...\}$ with a covariance matrix $V$, where each $E_i$ is the experimental value of an observable. We denote the theoretical predictions for those observables as $O = \{O_1,\,O_2,\,...\}$, where each $O_i$ is a function of the model parameters. When fitting to the SM, the parameters are $\lambda$, hadronic and nuclear matrix elements appearing in low-energy observables (such as the ratio of Fermi and Gamow-Teller matrix elements in mirror transitions, see Appendix~\ref{app:beta}),
and the parameters that capture theoretical uncertainties related to corrections in $\beta$ decays, which are discussed below Eq.~\eqref{eq:ft2}. 
In SMEFT, $O_i$ also depend on the Wilson coefficients considered in the analysis. We construct our $\chi^2$ function by
\begin{align}
    \chi^2 = (O - E)^T V^{-1} (O - E) \,,
\end{align}
where $V^{-1}$ is the inverse of the covariance matrix, which is related to the variances of the observables and their correlation matrix through, $V_{ii} = \sigma_{i}^2$ and corr$_{ij} = \sigma_i^{-1} V_{ij}\sigma_j^{-1}$, respectively. 
Whenever they are available, we add theoretical calculations of hadronic and nuclear matrix elements to the $\chi^2$ in the same way, effectively treating them as experimental measurements. Similarly, $\eta_{1,2,3}$, which appear in $\beta$ decays, are also treated as Gaussian parameters, see Appendix~\ref{app:beta} for details.

Minimizing the $\chi^2$ function gives the best-fit values of the model parameters. We will take the minimum of the $\chi^2$ of the SM as our benchmark and compare it with that of a specific set of Wilson coefficients in SMEFT. The difference of their minimal $\chi^2$ values,
\begin{align}
    \Delta \chi^2 = \chi^2_{\text{SM}} - \chi^2_{\text{SMEFT}} \,,
\end{align}
indicates the improvement on the fit provided by SMEFT. Note that this is only a useful comparison when both $\chi^2_{\rm SM}$ and $\chi^2_{\rm SMEFT}$ involve the same set of observables. If the two fits include different observables, then the two minimal $\chi^2$ values cannot be compared directly.

Generally speaking, the more parameters a model includes, such as Wilson coefficients, the lower the minimal $\chi^2$, since the extra parameters provide more degrees of freedom to shift the theoretical predictions towards the experimental values. To compare the quality of two fits including the same set of observables but different parameters, instead of using the $\chi^2$ per degree of freedom criterion, we apply the AIC~\cite{1100705}.

AIC is often used in model selection. It is a measure that balances the goodness of fit of a model against its complexity. The AIC is computed using the number of parameters in the model and the maximum likelihood of the model. In our case, the AIC of a fit can be presented as
\begin{align}
    \text{AIC} = \left( \text{Minimal} \ \chi^2 \right) + 2 \times \left( \text{Number of the parameters} \right) \,,
\label{eq:AIC}
\end{align}
where the parameters here refer to the considered Wilson coefficients and $\lambda$. Depending on the observables that are considered, the parameters may also include several matrix elements and parameters that describe theoretical uncertainties. The lower the AIC, the better the model is considered to describe the data, so the difference ${\rm AIC}_{i} -{\rm AIC}_j$ is a measure of the improvement of model $j$ over $i$. 
As more parameters (such as Wilson coefficients) are added to a model, it may better fit the data, but it also becomes more complex and runs the risk of overfitting. 
The AIC balances these considerations by penalizing models for introducing more parameters without a corresponding decrease in $\Delta \chi^2$. 
Used in conjunction with the $\chi^2$ method, AIC tends to find a model that not only fits the experimental data well but also avoids overcomplication, thus improving its predictive accuracy.

In our analysis, we will often be interested in comparing the SM fit with SMEFT scenarios. We will thus take the SM as the reference model, and define differences in AIC as  
\begin{equation}\label{DeltaAICSM}
\Delta {\rm AIC}_i={\rm AIC_{SM}}-{\rm AIC}_i~,
\end{equation}
where $i$ denotes a particular SMEFT model. Here and in what follows, we will define a `SMEFT model' as the SMEFT with a subset of Wilson coefficients turned on, with the remaining couplings set to zero. 
An advantage of the definition in Eq.~\eqref{DeltaAICSM} is that the SM parameter $\lambda$, the matrix elements, and the parameters that describe the theoretical uncertainties drop out in the calculation of $\Delta {\rm AIC}_i$. 
Notice that, with our definition, $\Delta {\rm AIC}_i > 0$ implies that a model performs better than the SM.

In addition to the overall quality of the fit, as measured by the AIC, we are also interested in the best-fit values, uncertainties, and correlations of the Wilson coefficients and the other fit parameters. We use the fact that the observables are linear in the Wilson coefficients to rewrite 
\begin{align}
    \chi^2 =\chi^2_{\rm min} +(C - \mu )^T \hat V^{-1} (C - \mu) \,,
\end{align}
where $C$ and $\mu$ are vectors of model parameters (Wilson coefficients) and their best-fit values, while $\hat V$ is the covariance matrix of the model parameters 
\begin{align}
 (\hat V^{-1})_{ij} = \frac{1}{2}\frac{\partial^2 \chi^2}{\partial C_i \partial C_j}\Big|_{C=\mu}\,.
\end{align} 
The uncertainties of $C_i$ can now be read from $\hat \sigma_i^2 = \hat V_{ii}$, while the correlations between the model parameters are given by $\widehat{\rm corr}_{ij} = \hat \sigma_i^{-1} \hat V_{ij}\hat \sigma_j^{-1}$. 

Because $\hat V^{-1}$ is symmetric, it can be diagonalized by an orthogonal matrix, which implies that there are linear combinations of $C_i$, corresponding to the eigenvectors of $\hat V^{-1}$, that are not correlated. This diagonalization allows one to see which combinations of Wilson coefficients the observables are most sensitive to and which combinations are most favored to be nonzero. This diagonalization can be performed through $R^T\hat V^{-1}R = \hat v$, with $\hat v$ a diagonal matrix and $R$ an orthogonal matrix. The uncorrelated combinations of Wilson coefficients and the best-fit values are given by $C' = R^T C$ and $\mu' = R^T \mu$. Here, $R$ is determined by the eigenvectors of $\hat V^{-1}$, while the uncertainties of $C'_i$ are determined by the eigenvalues of $\hat V^{-1}$. 

\subsection{Model averaging} 

When considering a large number of scenarios involving different sets of operators, it becomes interesting to see how (un)favored nonzero values of a particular Wilson coefficient are across different fits.
One robust method to tackle this is model averaging using the AIC. Here, we follow Burnham and Anderson~\cite{burnham2003model}, who advocated for the assignment of weights to individual models based on their AIC scores, thus allowing the derivation of model-averaged estimates for the parameters. 
The weight of a model is defined by
\begin{align}\label{eq:weight}
w_i = \frac{e^{ \frac{1}{2} \Delta {\rm AIC}_i}}{\sum_j e^{\frac{1}{2} \Delta {\rm AIC}_j}} \,.
\end{align}
Notice that the weights $w_i$ do not depend on the fact that we used the SM as a reference in the definition of $ \Delta {\rm AIC}_j$. 
The quality of the models that involve a given SMEFT coefficient, $\theta$, will be gauged by considering the sum of the weights of all the models in which $\theta$ is turned on,
\begin{align}
    W_\theta = \sum_i w_i I_\theta(i)\,,
    \label{eq:Wt}
\end{align}
where $I_\theta(i) = 1$ if the model, $i$, contains the parameter $\theta$, and $I_\theta(i) = 0$ if it does not. For the most significant coefficients, with $W_\theta =\Or(1)$, we will be interested in providing a model-averaged central value and an estimator of the standard deviation~\cite{burnham2003model},
\begin{align}
    \bar\theta = \frac{\sum w_i  I_\theta(i) \theta_i}{W_\theta},  \qquad \sigma_\theta = \frac{\sum w_i  I_\theta(i) \sqrt{ \sigma_{\theta_i}^2 + (\theta_i - \bar\theta)^2} }{W_\theta} \,,
\label{eq:ma}
\end{align}
where $\theta_i$ and $\sigma_{\theta_i}$ are the best-fit value and uncertainty of the Wilson coefficient $\theta$ in model~$i$, respectively. 
Wilson coefficients which have $W_\theta\ll 1$ are only present in models of poor quality, making their model-averaged values less important, and we will not consider them.

\section{SM analysis}
\label{resultsSM}
We begin our analysis by studying the Standard Model and the status of the CAA.
The numbers of measurements included in the different data sets are shown in Table~\ref{tab:num_obs}. For the collider data, this represents the total number of bins extracted from the package \texttt{HighPT}. Our most comprehensive data set, CLEW, contains 239 measurements in total.

The matrix elements affecting the low-energy fit are summarized in Table~\ref{tab:hadronic}. They do not affect the number of degrees of freedom as we add a term in $\chi^2$ for each, effectively treating the theoretical central value and uncertainty as a `measurement'. 
The parameters $\eta_{1,2,3}$, which capture theoretical uncertainties in superallowed $\beta$ decays, are treated in a similar manner. When counting degrees of freedom, in addition to the number of observables and the fit parameters of a model, we need to include six ratios of Fermi and Gamow-Teller matrix elements. They enter in mirror $\beta$ decays and are fit at the same time as the SM and SMEFT coefficients. 
\begin{table*}[t]
    \centering
    \begin{tabular}{|c|c|c|c|c|c|}
 \hline
 Data sets & EW & L & C & LEW & CLEW \\
 \hline
 Number of measurements & 41 & 53 & 145 & 94 & 239 \\
 \hline
 Degrees of freedom for SM & 41 & 46 & 145 & 87 & 232 \\
 \hline
    \end{tabular}
    \caption{Number of measurements included in different data sets and the number of degrees of freedom relevant for the $\chi^2$ analysis of the SM.}
    \label{tab:num_obs}
\end{table*}

We begin with the low-energy CC observables, for which the relevant degrees of freedom are $V_{us}\equiv\lambda$ as well as the six ratios of nuclear matrix elements which enter in mirror $\beta$ decays. The resulting value for $\lambda$ and the minimum $\chi^2$ are given by,
\begin{align}
\label{eq:chi2SM_L}
    \lambda = 0.22497\pm0.00027\,,\qquad \chi^2_{\rm min} =53\,, \qquad \chi^2/{\rm d.o.f.} = 1.1 \,,
\end{align}
obtained when fitting the SM to the L observables. The fact that this fit gives a good $\chi^2/{\rm d.o.f.}$ is in part due to the fact that the CAA is diluted by a large number of other observables, which mostly show good agreement with the SM. However, the subset of observables related to the extraction of $V_{ud}$ and $V_{us}$ still shows significant tension with the SM as indicated in Eq.~\eqref{CAA} and the discussions in e.g.\ Refs.~\cite{Crivellin:2020lzu,Cirigliano:2022yyo}.

\begin{figure}[t!]
\centering
\includegraphics[width=0.8\textwidth]{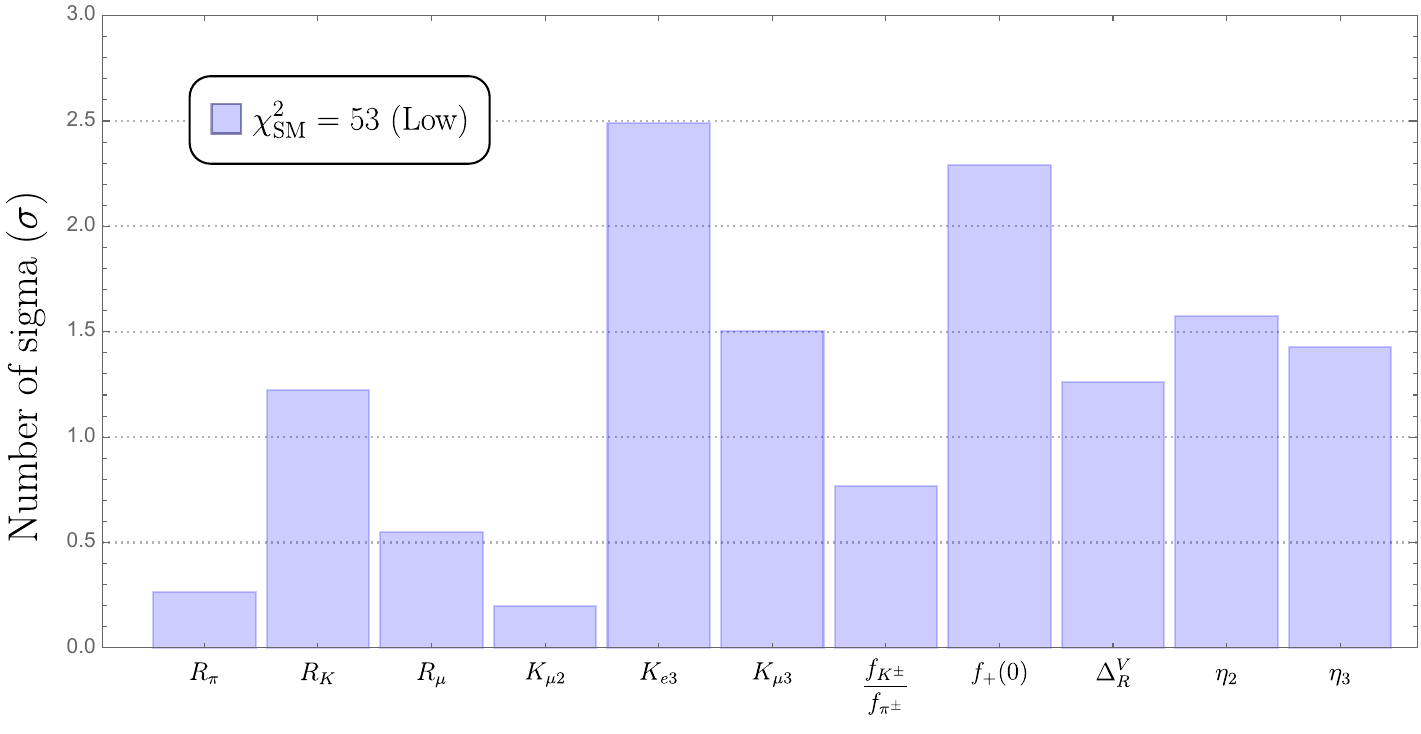}
\caption{Chart of several low-energy CC observables, relevant to the CAA, showing the difference between the SM prediction at the best-fit point and the experimental results in units of standard deviation. For the matrix elements ($f_{K^\pm}/f_{\pi^\pm}$, $f_+(0)$ and $\Delta_R^V$) and nuisance parameters ($\eta_{2,3}$), the chart shows the deviation from the central value of the theoretical prediction.}
\label{fig:SM_L}
\end{figure}

In our SM analysis, the unitarity of the CKM matrix is fixed and $V_{ud}$ and $V_{us}$ are not independent. The CAA as presented in Eq.~\eqref{CAA} is then manifested in Fig.~\ref{fig:SM_L} where we show the difference (in units of standard deviations) between the SM predictions at the best-fit point and the experimental values for several observables. We also show the deviation of $f_{+}(0)$,
$f_K/f_\pi$, and $\Delta_R^V$ and of the nuisance parameters $\eta_{2,3}$ from their preferred theoretical values. In the case of meson decays, the CAA appears mainly in the $K_{\ell 3}$ decays, as the corresponding decay rates and the relevant form factor, $f_+(0)$, deviate from their experimental/theoretical values by 2.5 standard deviations. 

For neutron and nuclear $\beta$ decays, the CAA mostly shows up through deviations in the radiative corrections, $\Delta_R^V$, and the parameters $\eta_{2,3}$, which parameterize theoretical uncertainties related to nuclear-structure dependent corrections. The observables of interest, the $\mathcal{F}t$ values for the $0^+\to0^+$ transitions, 
typically show deviations of $\lesssim 1\sigma$ and
are not shown in Fig.~\ref{fig:SM_L}. This means that the fit aims to align $\mathcal{F}t$ values closely with experimental determinations by selecting matrix elements and parameters $\eta_{2,3}$ that deviate from their predicted values.

\begin{figure}[t!]
\centering
\includegraphics[width=.8\textwidth]{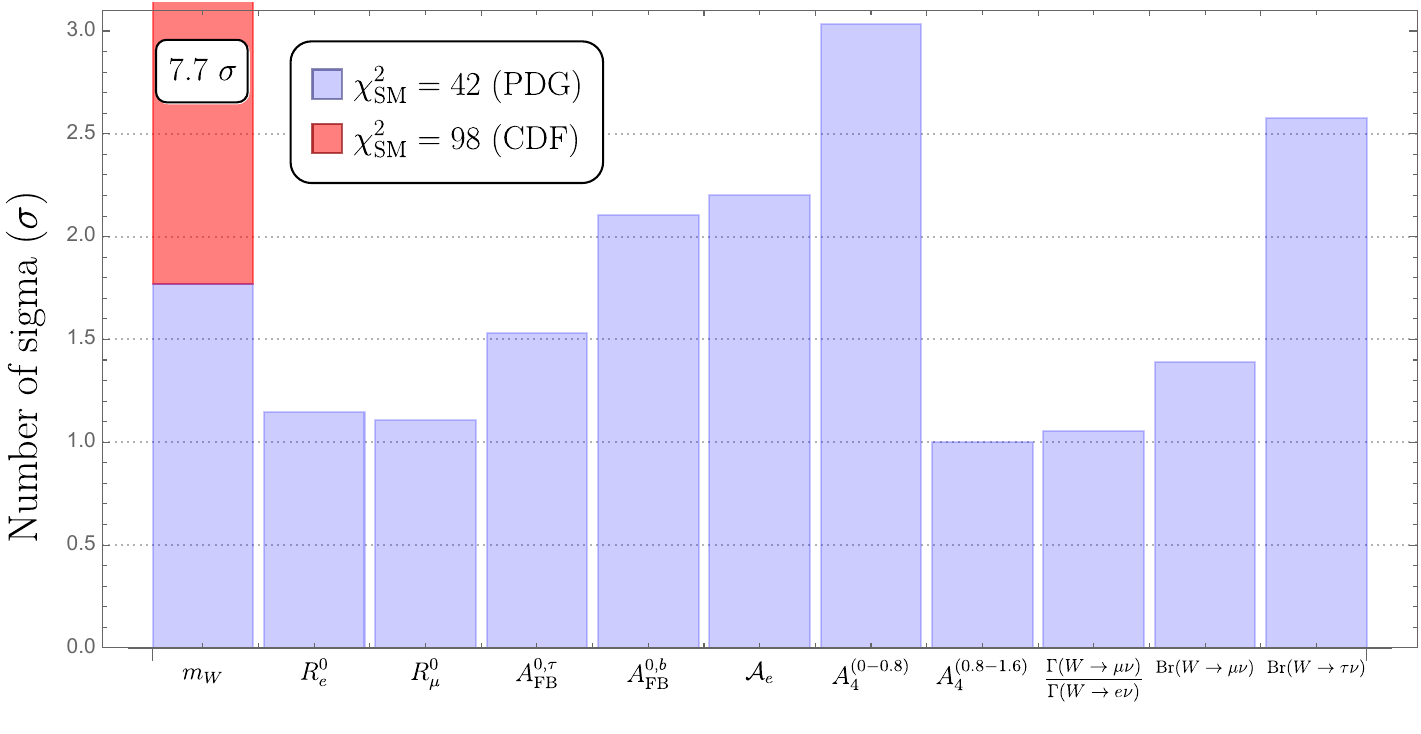}
\caption{Observables 
that deviate by more than $1\sigma$ from their experimental values at the best-fit point of the EW fit. The red bar shows the $7.7\sigma$ deviation associated with the CDF value of the W mass, while the blue bars implement the PDG value. }
\label{fig:SM_EW}
\end{figure}

Moving on to the class of EW observables, we find a minimal $\chi^2$ of 
\begin{align}\label{eq:chi2SM_EW}
     \chi^2_{\rm min} &=42.2\,, & \chi^2/{\rm d.o.f.} &= 1.1\,, & & (m_W^{\rm PDG})\,, \nonumber \\  \chi^2_{\rm min} &=97.9\,, & \chi^2/{\rm d.o.f.} &= 2.4\,, & & (m_W^{\rm CDF})\,,
\end{align}
depending on which measurement of $m_W$ is used.
We again show the discrepancy between the fit and experiment in Fig.~\ref{fig:SM_EW}, now for all observables with a deviation greater than $1\sigma$. Clearly, the largest discrepancy arises in the case of $m_W^{\rm CDF}$, while pulls of a few $\sigma$ appear for several ratios of decay widths and asymmetries. The most 
significant ones appear for asymmetries measured at LEP (e.g.\ $A_{\rm FB}^{0,b}$ and ${\cal A}_e$), measurements near the $Z$ pole at the LHC (one of the bins of the asymmetry $A_4$), and the branching ratio of $W\to \tau \nu$.

Finally, we consider the SM fit to the collider observables, which consist of the different CC and NC Drell-Yan processes, $pp\to \bar \ell \ell$ and $pp\to \ell \nu$. In turn, each of these channels involves a relatively large number of invariant and transverse mass bins. The minimal $\chi^2$ therefore tends to be fairly large, and we find
\begin{align}
\label{eq:chi2SM_C}
     \chi^2_{\rm min} =104\,, \qquad \chi^2/{\rm d.o.f.} = 1.1\,.
\end{align}
The result of fitting multiple sets of observables, such as LEW or CLEW, corresponds to simply combining individual fits, which implies that the $\chi^2_{\rm min}$ of the CLEW fit is given by the sum of Eqs.~\eqref{eq:chi2SM_L}, \eqref{eq:chi2SM_EW}, and \eqref{eq:chi2SM_C}. This is due to the fact that only the fit to low-energy data involves free parameters in the form of $\lambda$, matrix elements, and parameters that describe theoretical uncertainties, while the EWPO and collider data are, to a good approximation, independent of these variables.

\section{SMEFT analysis with $U(3)^5$ flavor assumption}
\label{resultsU3}

We start by considering a BSM scenario in which we impose a $U(3)^5$ flavor symmetry on the SMEFT coefficients. Ref.~\cite{deBlas:2022hdk} investigated the impact of the measurement of the CDF W mass on the EWPO fit under these assumptions. The EWPO depend on eight combinations of Wilson coefficients~\cite{Falkowski:2014tna}, namely $\hat{C}_{H l}^{(1,3)}$, $\hat{C}_{H q}^{(1,3)}$, $\hat{C}_{H e}$, $\hat{C}_{H u}$, $\hat{C}_{H d}$, and $\hat C_{ll}$. As mentioned in Section~\ref{sec:U35}, the hat-notation is used to identify the linear combinations that cannot be separated using EWPO alone:
\begin{align}
\hat C^{(3)}_{H F} &= C_{H F}^{(3)} + \frac{c_w}{s_w} C_{HWB} + \frac{c_w^2}{4 s_w^2} C_{HD}\,, \nnw
\hat C^{(1)}_{HF} &= C_{HF}^{(1)} - \frac{Y_F}{2} C_{HD}\,, \nnw
\hat C_{Hf} &= C_{Hf} - \frac{Y_f}{2} C_{HD}\,,
\label{eq:BSM_primary}
\end{align}
for $F=\{l,q\}$ and $f=\{u,d,e\}$ and $Y_{F,f}$ denotes the corresponding weak hypercharge. We follow~\cite{Cirigliano:2022qdm} and define
\begin{equation}
 C_{\Delta}= 2 \left[ \hat C_{H q}^{(3)} - \hat C_{H l}^{(3)} + \hat C_{ll} \right]\,,
\end{equation}
where $\hat C_{ll} = [C_{ll}]_{1221}$. Defining $C_{\Delta}$ will be useful, as it is the linear combination of Wilson coefficients that appears in the EWPO that contributes to deviations from CKM unitarity. Therefore, we will use this relation to trade $\hat C_{ll}$ for $C_{\Delta}$. 
The SMEFT corrections to the $W$ mass can be expressed in terms of these operators as \cite{Berthier:2015oma,Bjorn:2016zlr}
\begin{align}
	\frac{\delta m_W^2}{m_W^2} &= 
	v^2 \ \frac{s_w c_w}{s_w^2 - c_w^2} 
	\left[ 2 \, C_{H WB} + \frac{c_w}{2 s_w} \, C_{H D} + 
	\frac{s_w}{c_w} \left( 2 \, C_{H l}^{(3)} - \hat C_{ll} \right) \right] \nnw
        &= v^2 \ \frac{s_w^2}{s_w^2 - c_w^2} \left( 2 \, \hat C_{H l}^{(3)} - \hat C_{ll} \right) = v^2 \ \frac{s_w^2}{s_w^2 - c_w^2} \left( \hat C_{H l}^{(3)} + \hat C_{H q}^{(3)} - \frac{1}{2} C_\Delta \right) \,.
\label{eq:mW}
\end{align}
The expression of $s_w$ in terms of the input parameters $G_F$, $m_Z$, and $\alpha_{em}$ is given in Eq.~\eqref{eq:sw}. Finally, under the assumption of $U(3)^5$ flavor symmetry, the violation of CKM unitarity is described by 
\beq\label{DeltaCKM}
\Delta_\mathrm{CKM} = |\tilde V_{ud}|^2+  |\tilde V_{us}|^2 -1 = v^2 \left( C_\Delta -2 C_{lq}^{(3)}\right)\,.
\eeq
Here, $\tilde V_{ij}$ are the effective CKM elements that are probed in low-energy measurements of $\beta$ and $K$ decays, while $\tilde V_{ub}$ can be neglected at the current level of precision.
$C_{\Delta}$ entirely captures the contribution to $\Delta_{\rm CKM}$ of the operators that enter EWPO, whereas $C_{lq}^{(3)}$ does not play a role in EWPO and is therefore traditionally not included.

\begin{table*}[t]
    \centering
    \begin{tabular}{|c|c|c|c|c|}
 \hline
 & EW & LEW$_{\rm 1}$ 	& LEW$_{\rm 2}$  & CLEW	\\ 
 \hline
$\hat{C}_{H l}^{(1)}$ & $ -0.0091 \pm 0.011 $ &$ -0.016 \pm 0.011$  & $ -0.0091 \pm 0.011 $ &            $ -0.016 \pm 0.011 $             \\ 
$\hat{C}_{H l}^{(3)}$ & $ -0.057 \pm 0.015 $  & $-0.046\pm 0.014  $  &  $ -0.057 \pm 0.015 $  &		 $ -0.046 \pm 0.014 $  	 \\   
$\hat{C}_{H e}$       & $ -0.024 \pm 0.0086 $  &  $-0.027 \pm 0.0085 $
& $-0.024 \pm 0.0086 $  &  	$ -0.027 \pm 0.0085 $    	
\\  
$\hat{C}_{H q}^{(1)}$ & $ -0.029\pm 0.043 $  &$  -0.045\pm 0.042  $ & $ -0.029\pm 0.043 $  &		$ -0.044\pm 0.042 $  	\\   
$\hat{C}_{H q}^{(3)}$ & $ -0.095 \pm 0.032 $ &$ -0.041\pm 0.014 $ &$ -0.095 \pm 0.032 $ &		  $ -0.040\pm0.014$		\\    
$\hat{C}_{H u}$       & $ -0.0046 \pm 0.12 $    & $-0.12 \pm 0.098 $ & $ -0.0046 \pm 0.12 $    &	  $ -0.13 \pm 0.098$    		       \\  
$\hat{C}_{H d}$       & $ -0.55 \pm 0.25 $        &  $-0.33  \pm 0.22 $ &    $ -0.55 \pm 0.25 $        &  	  $ -0.33\pm0.22$         	 \\  
$C_{\Delta}$          & $  -0.15 \pm 0.068 $             &  $ -0.030 \pm 0.0083 $ &    $  -0.15 \pm 0.068 $      & 	  $  -0.029 \pm 0.0083 $    \\
$C_{lq}^{(3)}$          &  --                             & --              &    $ -0.063 \pm 0.034$ & 		  $ 0.00029\pm 0.00058$ 			
\\ \hline \hline 
$\Delta \chi^2$  & 66 & 73 & 77 & 73 \\
 $\chi^2$/d.o.f. & 1.0 & 1.0 & 0.95 & 1.1 \\
$\Delta$AIC     & 50 & 57 & 59 & 55 \\  
\hline
    \end{tabular}
    \caption{$U(3)^5$ fit results while including the CDF $m_W$ measurement. Coefficients are given in units of TeV$^{-2}$. In the EW and LEW$_{\rm 1}$ fits, the $U(3)^5$ invariant coefficient $C_{lq}^{(3)}$ is set to zero, while it is included as a fit parameter in LEW$_{\rm 2}$ and CLEW. The bottom three rows report several quantities that measure the quality of the fits.}
\label{tab:SMEFT2}
\end{table*}

\begin{figure}[t]
	\centering
	\includegraphics[width=1\textwidth]{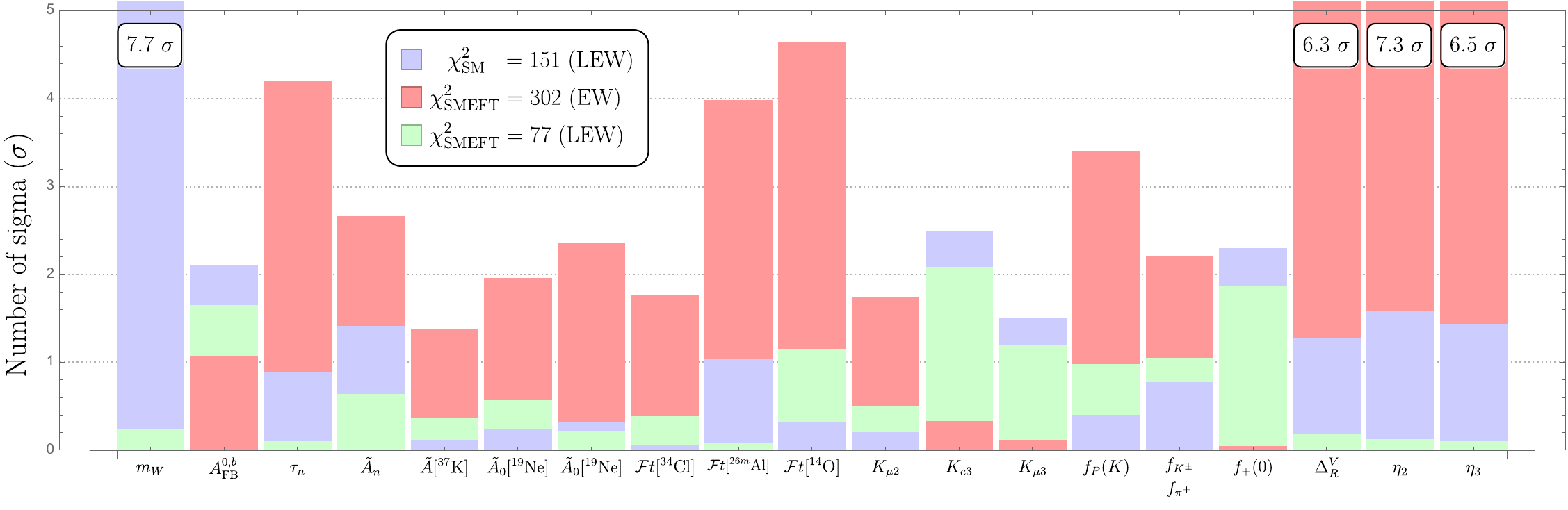}
 \vspace{-0.5cm}
	\caption{
 Comparison of three scenarios in the description of EWPO and low-energy CC observables, \textit{including} the CDF measurement of $m_W$ marked by a deviation of $7.7\sigma$. In blue, we denote the results of a SM fit to the LEW data. The red bars correspond to a SMEFT fit to just the EW data, under the flavor assumption of $U(3)^5$. The last three red bars, $\Delta_R^V$, $\eta_2$, and $\eta_3$, have off-the-charts deviations greater than $6\sigma$. The green bars correspond to a $U(3)^5$ SMEFT fit to the LEW data.}
	\label{fig:SuperU35}
\end{figure}

\subsection{$U(3)^5$ results \textit{including} the CDF W mass}
We begin our analysis by fitting the 8 linear combinations of Wilson coefficients to the set of EWPO defined in Appendix~\ref{app:EWPO} including, in particular, the 2022 CDF measurement of the W mass. We list the best-fit values and the 1$\sigma$ ranges in the EW column of Table~\ref{tab:SMEFT2} labeled with EW. Within uncertainties, the results agree with Refs.~\cite{deBlas:2022hdk,Cirigliano:2022qdm}. As noted in~\cite{Cirigliano:2022qdm}, the fit value of $C_\Delta$ based on EWPO is nonzero at a significant level and corresponds to a violation of CKM unitarity.
Plugging $C_{\Delta}$ in Eq.~\eqref{DeltaCKM}, we find
\beq\label{DeltaCKM2}
\left. \Delta_\mathrm{CKM} \right|_{\rm EW} = -0.009 \pm 0.004\,,
\eeq
indicating a percent-level deviation from CKM unitarity, significantly larger than allowed by current experimental determinations~\cite{Cirigliano:2022qdm}. 

We can study the impact of the Wilson coefficients in the EW column of Table~\ref{tab:SMEFT2} on each of the low-energy CC observables. We do so by setting the Wilson coefficients at their central values, while floating the relevant matrix elements, the parameters that describe the uncertainties of the theory, and $\lambda$. In Fig.~\ref{fig:SuperU35} we compare this scenario (in red) with the SM (in blue) and the SMEFT (in green), fit to both the EWPO and low-energy CC observables (LEW). We only show observables that change by more than $1\sigma$ from the SM fit. The red bars show that, while the $U(3)^5$ EW fit resolves the $m_W$ anomaly, it leads to a very poor description of a wide range of low-energy observables when they are not explicitly included in the analysis. 

To get a sensible result, we must include low-energy CC observables in the fit. This leads to the fit results in the second column of Table~\ref{tab:SMEFT2}. The inclusion of low-energy observables significantly reduces the value of $|C_\Delta|$, which also leads to lower values for $|\hat C_{H l}^{(3)}|$ and $|\hat C_{H q}^{(3)}|$. This confirms the conclusion of Ref.~\cite{Cirigliano:2022qdm} that the $m_W$-anomaly, in general, cannot be studied using EWPO alone. The modified fit leads to a much improved description of the low-energy CC observables as can be seen by comparing the red and green bars in Fig.~\ref{fig:SuperU35}. 

It can be argued that the above conclusion is too strong. Even within the $U(3)^5$ flavor symmetry it is possible to decouple the EWPO from the low-energy CC observables by including $C_{lq}^{(3)}$ in the fit, as can be seen from Eq.~\eqref{DeltaCKM}. We demonstrate this in the third column of Table~\ref{tab:SMEFT2} where we observe in the EW and LEW columns that a nonzero value of $C_{lq}^{(3)}$ can absorb violations of CKM unitarity, while leaving the other Wilson coefficients unchanged. 

Although this may seem to be a reasonable resolution of both the $m_W$ anomaly and the CAA, significant values of $C_{lq}^{(3)}$ modify the Drell-Yan processes measured at the LHC. To test whether this leads to relevant deviations of the high-$p_T$ tails of Drell-Yan processes, we include the observables in Table~\ref{tab:searches}. The fit results\footnote{We have checked that adding additional $U(3)^5$-invariant four-fermion operators that only affect Drell-Yan processes but not the LEW observables, does not change this conclusion.} are given in the CLEW column of Table~\ref{tab:SMEFT2} and we see that the LHC observables essentially force $|C_{lq}^{(3)}| \leq 10^{-3}\,{\rm TeV}^{-2}$, far too small to compensate for any significant low-energy effects induced by the other Wilson coefficients.
The resulting CLEW values in Table~\ref{tab:SMEFT2} then exactly agree with the values of the LEW$_{1}$ column in the same table.

The quality of the fits are also shown at the bottom of Table~\ref{tab:SMEFT2}. 
The $U(3)^5$ fit to the LEW observables (corresponding to the LEW$_{\rm 1}$ column) gives $\Delta \chi^2 = 73$ and $\chi_{\mathrm{SMEFT}}^2/\mathrm{d.o.f.}=1.0$, which implies a very good fit and an improvement over the SM, $\Delta\mathrm{AIC} = 57$. The CLEW fit that includes $C_{lq}^{(3)}$ gives a slightly worse $\Delta\mathrm{AIC} = 55$ due to the addition of a fit parameter. 

The large difference between the SM and $U(3)^5$ fits is mainly driven by the CDF determination of $m_W$. This is reflected in the nonzero values of several Wilson coefficients, with three of them at 3 to 4$\sigma$: $\hat{C}_{H l}^{(3)}$, $\hat{C}_{H q}^{(3)}$, and $\hat{C}_{H e}$. 
Their values in the first column (EW) of Table~\ref{tab:SMEFT2} can be understood from the relation
\begin{align}
\hat C_{H l}^{(3)} + \hat C_{H q}^{(3)} &= \frac{1}{v^2} \left( 1 - \frac{c_w^2}{s_w^2} \right) \frac{\delta m_W^2}{m_W^2} + \frac{1}{2}C_\Delta  \simeq  -0.15 \; {\rm TeV}^{-2},
\end{align}
which agrees well with the best-fit point 
\begin{equation}
\hat C_{H l}^{(3)} + \hat C_{H q}^{(3)} =
    -(0.152 \pm 0.047) \; {\rm TeV}^{-2} \,.
\end{equation}
The value of $\hat C_{H e}$ can be understood from the SMEFT correction to the partial width of the $Z$ boson that decays into right-handed electrons~\cite{Kribs:2020jgn} 
\begin{align}
\frac{\delta \Gamma(Z \to e_R \, \bar{e}_R)}{\Gamma(Z \to e_R \, \bar{e}_R)} &= \frac{v^2}{c_w^2 - s_w^2} \left[ \frac{2c_w}{s_w}\,C_{H WB} + \frac{1}{2}\,{C_{H D}} + \left( 2 \, C_{H l}^{(3)} - \hat C_{ll} \right) + \frac{s_w^2 - c_w^2}{s_w^2}\, C_{H e} \right] \nnw
&= -\frac{v^2}{s_w^2 - c_w^2} \left( 2 \, \hat C_{H l}^{(3)} - \hat C_{ll} \right) - \frac{v^2}{s_w^2} \hat C_{H e} \nnw
&= -\frac{1}{s_w^2} \left( \frac{\delta m_W^2}{m_W^2} + v^2 \, \hat C_{H e} \right) \,,
\label{eq:Zee}
\end{align}
where $\Gamma(Z \to e_R \, \bar{e}_R)$ is derived from the total width $R_e^0$ and the forward-backward asymmetry factor ${\cal A}_{e}$. Therefore,
\begin{align}
\hat C_{He} = - \frac{1}{v^2}  \left[ \frac{\delta m_W^2}{m_W^2} + s_w^2 \frac{\delta \Gamma(Z \to e_R \, \bar{e}_R)}{\Gamma(Z \to e_R \, \bar{e}_R)} \right] \simeq -0.026  \; {\rm TeV}^{-2} \,,
\end{align}
which roughly agrees with the best-fit value of $\hat C_{He} = -(0.024 \pm 0.0086)$ TeV$^{-2}$ in the first column (EW) of Table~\ref{tab:SMEFT2}.

Finally, $\hat C_{Hd} = -(0.54 \pm 0.25)$ TeV$^{-2}$ in Table~\ref{tab:SMEFT2PDG} is approximately 2$\sigma$ from zero (this is true even without the CDF measurement of $m_W$, see below), which is caused by the discrepancy in the forward-backward asymmetry of the bottom quark, $A_{FB}^{0,b}$ (see Fig.~\ref{fig:SM_EW}). This leads to~\cite{Kribs:2020jgn} 
\begin{align}\label{AFB_Chd}
\hat C_{Hd} = -\frac{1}{3v^2}  \left[ \frac{\delta m_W^2}{m_W^2} + s_w^2 \frac{\delta \Gamma(Z \to b_R \, \bar{b}_R)}{\Gamma(Z \to b_R \, \bar{b}_R)} \right] \simeq -0.55 \; {\rm TeV}^{-2}\,.
\end{align}

\subsection{$U(3)^5$ results \textit{without} the CDF W mass}\label{U3PDG}

\begin{table*}[!t]
    \centering
    \begin{tabular}{|c|c|c|}
 \hline
 & EW & LEW	\\ 
 \hline
$\hat{C}_{H l}^{(1)}$ & $ 0.0026 \pm 0.011 $ & $ -0.0014 \pm 0.011 $ \\
$\hat{C}_{H l}^{(3)}$ & $ -0.019 \pm 0.016 $ & $ -0.011 \pm 0.015 $ \\
$\hat{C}_{H e}$       & $ -0.0011 \pm 0.0092 $ & $ -0.0027 \pm 0.0091 $ \\
$\hat{C}_{H q}^{(1)}$ & $ -0.033 \pm 0.043 $ & $ -0.043 \pm 0.042 $\\
$\hat{C}_{H q}^{(3)}$ & $ -0.056 \pm 0.033 $ & $ -0.022 \pm 0.014 $\\
$\hat{C}_{H u}$       & $ -0.02 \pm 0.12 $ & $ -0.095 \pm 0.098 $ \\
$\hat{C}_{H d}$       & $ -0.54 \pm 0.25 $ & $-0.41  \pm 0.22 $ \\
$C_{\Delta}$          & $ -0.11 \pm 0.069 $ & $ -0.029 \pm 0.0083 $ \\ \hline
    \end{tabular}
    \caption{$U(3)^5$ fit excluding the $m_W$ CDF measurement. We have set
    $C^{(3)}_{lq}=0$ as required from the LHC Drell-Yan measurements. The coefficients are given in units of TeV$^{-2}$.}
    \label{tab:SMEFT2PDG}
\end{table*}

\begin{figure}[t]
	\centering
	\includegraphics[width=1\textwidth]{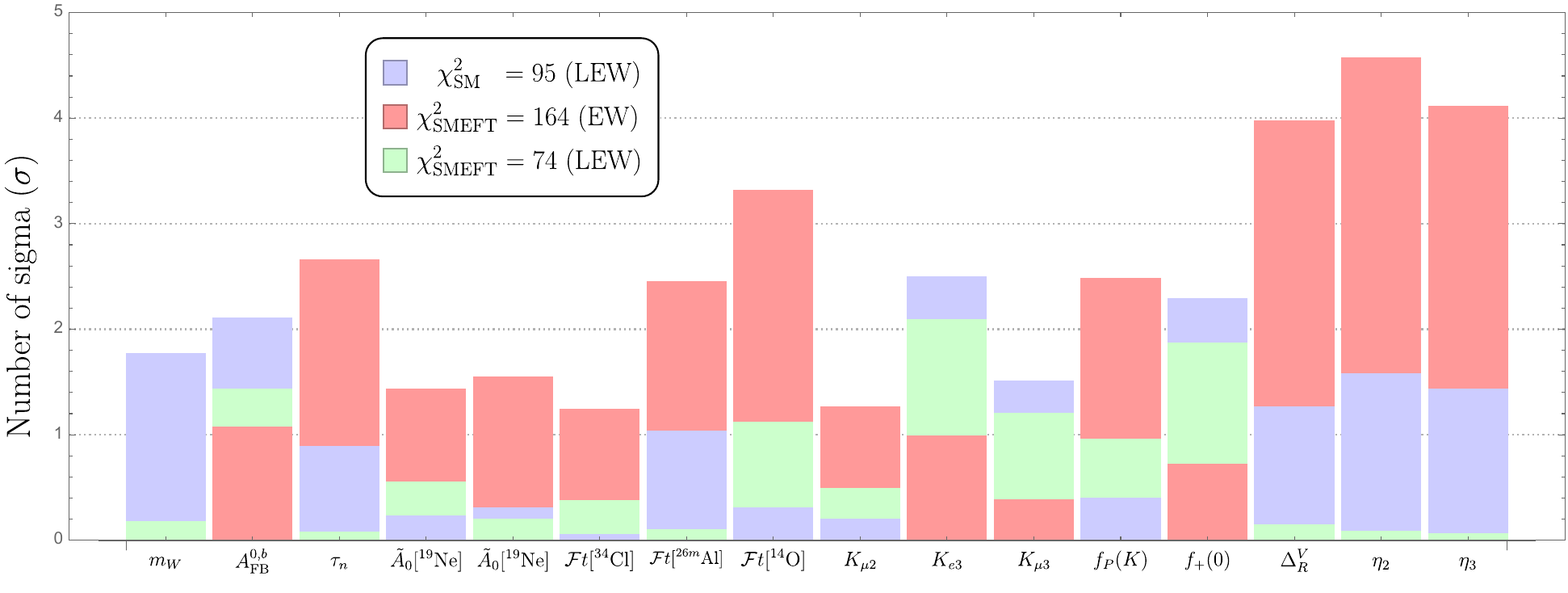}
 \vspace{-0.5cm}
	\caption{Same as Fig.~\ref{fig:SuperU35}, assuming the PDG value of $m_W$.}
	\label{fig:SuperU35PDG}
\end{figure}

We briefly investigate the impact on the above results if we use the world average value of the W mass in PDG, $m_W = 80.377 \pm 0.012\,$GeV. The SM fit to the LEW data results in the blue bars of Fig.~\ref{fig:SuperU35PDG} corresponding to a total $\chi_{\mathrm{SM}}^2= 95$ ($\chi_{\mathrm{SM}}^2/\mathrm{d.o.f.}=1$) with an information criterion of $\mathrm{AIC}_{\mathrm{SM}} = 97$. By excluding CDF $m_W$, the total $\chi^2$ has been reduced by 56 (see Section~\ref{resultsSM}), but there is still tension remaining due to the CAA. 

We then performed a SMEFT fit to the EWPO (see the EW column in Table~\ref{tab:SMEFT2PDG}) and predicted the corresponding low-energy CC observables. Similar to the previous section, this results in a fit that alleviates minor discrepancies in the EWPO (such as $m_W$), but it provides an unsatisfactory description of the low-energy CC processes. Instead, if we perform a SMEFT LEW fit, we obtain the LEW column of Table~\ref{tab:SMEFT2PDG} corresponding to the green bars in Fig.~\ref{fig:SuperU35PDG}. This fit gives $\chi_{\mathrm{SMEFT}}^2= 74$ ($\Delta \chi^2 = 21$ and $\chi_{\mathrm{SMEFT}}^2/\mathrm{d.o.f.}=0.9$). It has a better $\Delta\mathrm{AIC} = 7$ than the SM fit and addresses the CAA through $C_\Delta$, which is nonzero at the $3\sigma$ level. From Fig.~\ref{fig:SuperU35PDG}, we see that $C_\Delta$ reduces the tension in superallowed $\beta$ decays, but cannot accommodate $K_{\ell 3}$ at the same time.
Compared to the fit using $m_W$(PDG), the result for $C_{\Delta}$ remains the same. However, the 3$\sigma$ deviations in $\hat{C}_{H l}^{(3)}$, $\hat{C}_{H e}$, $\hat{C}_{H q}^{(3)}$ disappeared. This is not surprising, as all were driven by the value of $m_W$ measured by CDF. 

\subsection{Intermediate conclusions}

The first conclusion is that within the $U(3)^5$ scenarios it is not possible to decouple the EWPO from the low- and high-energy CC observables. These observables depend on an overlapping set of Wilson coefficients and have a similar sensitivity to BSM physics. We have shown that fitting the SMEFT Wilson coefficients to the EWPO only, irrespective of whether the CDF measurement of $m_W$ is included or not, generally leads to unacceptably large BSM effects in low-energy $\beta$ and meson decay processes. Furthermore, due to the pronounced sensitivity of CC Drell-Yan, semileptonic four-fermion operators cannot offset these effects. We are forced to combine the sets of observables. Given this perspective, the conventional set of EWPO, as discussed in the literature, is no longer adequate. We recommend consistently incorporating both low- and high-energy CC observables. Similar conclusions were reached in Refs.~\cite{Cirigliano:2022qdm, Blennow:2022yfm,Bagnaschi:2022whn,ThomasArun:2023wbd}.

That being said, the inclusion of CDF $m_W$ obviously affects the fit results. Taken at face value, it clearly shows that the SM provides a poor fit. We find that a $U(3)^5$ scenario can simultaneously account for the $m_W$ anomaly and part of the CAA. This fit performs significantly better than the SM with $\Delta\mathrm{AIC} = 57$ and requires four Wilson coefficients that are 3-4$\sigma$ away from zero, which provides a clue for model building in $U(3)^5$ scenarios. We stress that different values of these coefficients will be obtained 
if the EWPO observables are considered in isolation, which leads to severe problems in the description in the low-energy CC observables, as shown in Fig.~\ref{fig:SuperU35}. 

Excluding the CDF measurement, the picture is less clear. While the $U(3)^5$ LEW fit can partially accommodate the CAA -- as evidenced by the enhanced descriptions of neutron, nuclear, and meson decays -- this improvement is offset by the inclusion of additional fit parameters. When the dust settles, the AIC of the $U(3)^5$ fit is still better ($\Delta\mathrm{AIC} = 7$) than the SM fit due to a partial resolution of the CAA.

\section{Flavor-independent intermediate fit and the CAA in SMEFT}
\label{22fit}

In this section, we consider scenario 2 of Section~\ref{3classes} which focuses on the CAA and \textit{excludes} the CDF $m_W$. We have argued that an analysis that does not rely on theoretical assumptions regarding the flavor structure of BSM physics involves 37 independent SMEFT operators, 22 of which contribute to the low-energy CC processes. In the following, we start by discussing scenarios including the 22 Wilson coefficients that affect the low-energy observables. Although we will perform a global fit to the CLEW observables of all 22 Wilson coefficients, the results are not straightforward to interpret. We will first explore several more focused fits, considering only a subset of operators. This approach will assist in dissecting the results and offer guidance for model building.

\subsection{Right-handed operators}

\begin{figure}[t]
\centering
\includegraphics[scale=0.7]{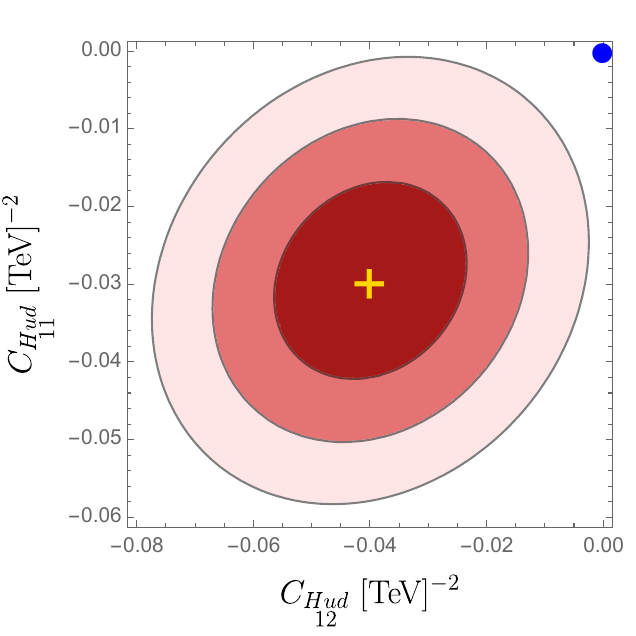}
\caption{Results of the L2(RH) fit to low-energy CC observables involving two Wilson coefficients $[C_{Hud}]_{11}$ and $[C_{Hud}]_{12}$, given in units of TeV$^{-2}$. Three contours illustrate the likelihoods of 1, 2, and $3\sigma$. The best-fit point is marked by a yellow cross, whereas the origin (the SM point) is marked by a blue dot.}
\label{fig:L2}
\end{figure}

The analysis of the CKM anomaly performed in Ref.~\cite{Cirigliano2022Aug} indicated that right-handed (RH) charged-current interactions could provide a viable explanation for the CAA. These interactions are induced by the SMEFT operator $Q_{Hud}$, which is forbidden under $U(3)^5$ and strongly suppressed in MFV scenarios. We consider two independent Wilson coefficients $\left[C_{Hud}\right]_{11,12}$ and, when fitting to low-energy CC observables, we refer to this fit as L2(RH). The results can be found in the L2(RH) column of Table~\ref{tab:L8}, showing nonzero values of the RH up-down and up-strange interactions at more than 3$\sigma$, with a small correlation between the two couplings, see Fig.~\ref{fig:L2} for details.

The corresponding eigenvectors are given by
\bea
0.28 \,C_{\substack{Hud\\11}}+0.96 \,C_{\substack{Hud\\12}} &=& -0.011\times (4.2\pm1)\,{\rm TeV}^{-2}\,,\nn\\
0.96 \,C_{\substack{Hud\\11}}-0.28 \,C_{\substack{Hud\\12}} &=& -0.0076\times (2.1\pm1)\,{\rm TeV}^{-2}\,,
\eea
and thus, respectively, $4.2\sigma$ and $2.1\sigma$ away from zero. 

\begin{table*}[t!]
	\centering
	\begin{tabular}{|c|c|c|c|}
  \hline
  & L2(RH) & L6(SPS) & L8 \\
  \hline
  $[C_{Hud}]_{11}$ & $-0.030 \pm 0.008$ & -- & $-0.058 \pm 0.079$  \\
  $[C_{Hud}]_{12}$ &
  $-0.040 \pm 0.011$ & -- & $0.080 \pm 0.35$ \\
  $[C_{ledq}]_{1111}$ &  -- & $-0.014 \pm 0.006$ & $0.0010 \pm 0.0075$ \\ 
  $[C_{ledq}]_{1122}$ & -- & $-0.014 \pm 0.006$ & $ 0.0009 \pm 0.0075$ \\
  $[\bar{C}_{lequ}^{(1)}]_{1111}$ & -- & $-0.014 \pm 0.006$ & $ 0.0010 \pm 0.0075$ \\ 
  $[C_{ledq}]_{2211}$ & -- & $ 0.0062 \pm 0.0042$ & $0.017 \pm 0.039$ \\ 
  $[C_{ledq}]_{2222}$ & -- & $0.0006 \pm 0.0045$&  $  -0.0096 \pm 0.036$ \\
  $[\bar{C}_{lequ}^{(1)}]_{2211}$& -- & $0.0054 \pm 0.0043$ & $  0.014 \pm 0.036$ \\
  \hline
	\end{tabular}
 \caption{Central values and $1\sigma$ uncertainties for the Wilson coefficients in the L2 (RH), L6 (SPS) and L8 (RS) fits, given in units of TeV$^{-2}$.}
\label{tab:L8}
\end{table*}

\begin{figure}[t]
\centering
\includegraphics[scale=0.34]{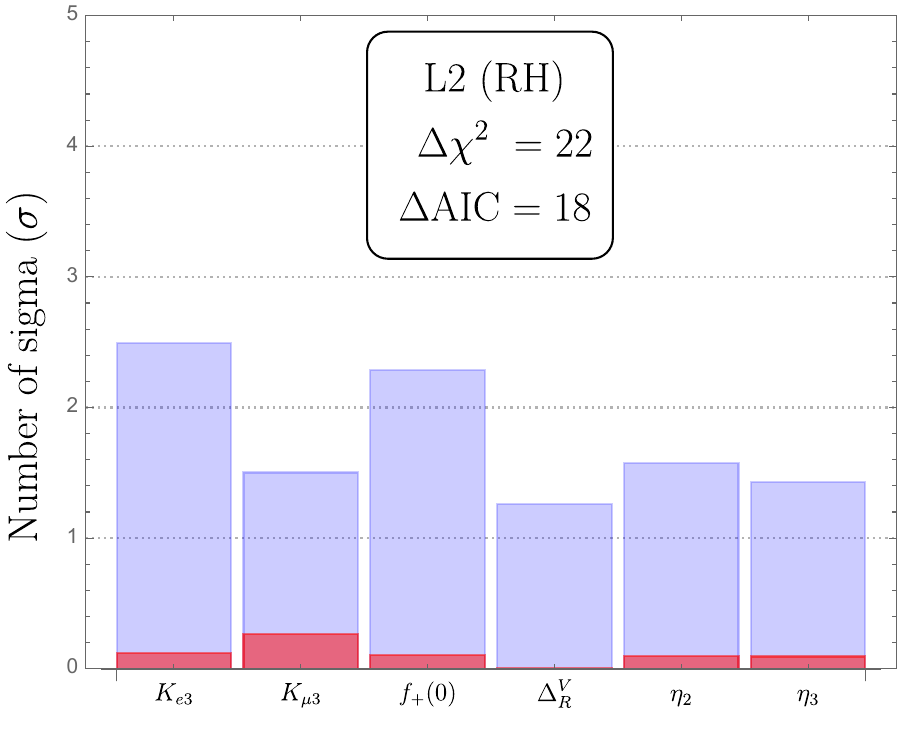}
\includegraphics[scale=0.34]{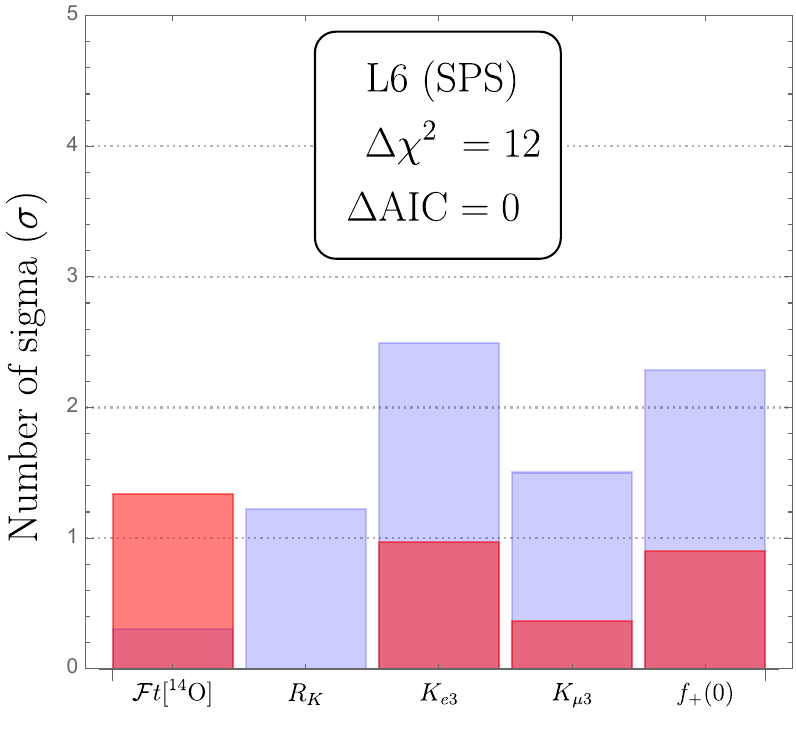}
\includegraphics[scale=0.34]{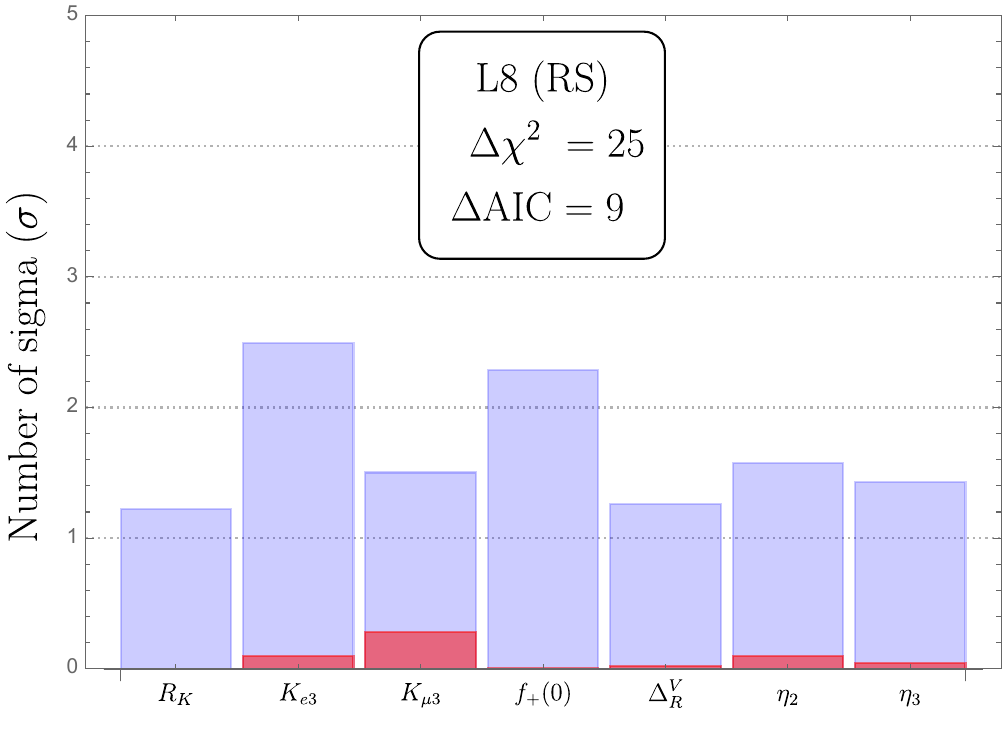}
\caption{Difference between the theoretical predictions at the best-fit point and the experimental results or the theoretical central values, for observables and matrix elements relevant to low-energy CC data. The blue bars represent the SM results, while the red bars correspond to three SMEFT fits. Only SMEFT bars that differ by more than $1\sigma$ from the SM are shown. Left: L2(RH). Center: L6(SPS). Right: L8(RS).}
\label{fig:ObsL2L6}
\end{figure}

Compared to the SM fit, the minimal $\chi^2$ decreased from 52 to 30, an improvement of $\Delta \chi^2=22$, which implies $\Delta \text{AIC(RH2)}=18$. In the left panel of Fig.~\ref{fig:ObsL2L6}, we display the observables that show improvement over the SM fit. Specifically, the RH currents align the three-body kaon decays as well as nuclear $\beta$ decays with observations, as evidenced by the $\Delta_R^V$ values and the $\eta_{2,3}$ parameters. 

\subsection{Scalar/Pseudoscalar operators}

Scalar/pseudoscalar currents also influence the charged-current processes that determine $V_{ud}$ and $V_{us}$. As shown in Appendix~\ref{app:epstransl}, there are six relevant Wilson coefficients that enter at tree level. We label the fit to them as L6(SPS) and present the results in Table~\ref{tab:L8} and the middle panel of Fig.~\ref{fig:ObsL2L6}.

There are strong correlations between the different contributions. In the case of semileptonic couplings to electrons, the pseudoscalar operators give contributions to $R_\pi$ and $R_K$ (defined as $R_{\pi,K}=\Gamma_{\pi,K\to e\nu_e}/\Gamma_{\pi,K\to \mu\nu_\mu}$) that are enhanced by $m^2_{\pi, K}/[m_e (m_u + m_{d,s})]$ with respect to the SM. As a consequence, the combinations
$([C_{ledq}]_{1111}- [\bar{C}^{(1)}_{lequ}]_{1111})$
and
$([C_{ledq}]_{1122}- [\bar{C}^{(1)}_{lequ}]_{1122})$ are severely constrained. The scalar combination $([C_{ledq}]_{1111} + [\bar{C}^{(1)}_{lequ}]_{1111})$
affects $0^+ \to 0^+$ transitions and must be nonzero to address the CAA. For the couplings to muons, the contributions to $R_\pi$ and $R_K$ are only enhanced by $1/m_\mu$, so the correlations are weaker, while the scalar combination
$([C_{ledq}]_{2222} + [\bar{C}^{(1)}_{lequ}]_{2211})$ affects $K_{\mu 3}$. 
This discussion can be neatly summarized by studying the eigenvectors of the fit:
\begin{center}
$
\begin{array}{cc|cccccc}
 \sigma\, ({\rm TeV}^{-2})& \mu/\sigma & {C}_{\substack{ledq\\1111}} & {C}_{\substack{ledq\\1122}} & {C}_{\substack{ledq\\2211}} & {C}_{\substack{ledq\\2222}} & \bar{C}_{\substack{lequ\\1111}}^{(1)} & \bar{C}_{\substack{lequ\\2211}}^{(1)} \\
 \hline
 2.7 \cdot 10^{-6}&0.2 & -0.62 & -0.15 & 0 & 0 & 0.77 & 0 \\
 5.8 \cdot 10^{-6}&1.2 & 0.53 & -0.8 & 0 & 0 & 0.28 & 0 \\
 1.5 \cdot 10^{-3}&0.9 & 0 & 0 & 0.79 & -0.2 & 0 & -0.58 \\
 1.7 \cdot 10^{-3}&1.8 & 0 & 0 & 0.23 & -0.78 & 0 & 0.58 \\
 6.7 \cdot 10^{-3}&1.1 & 0 & 0 & 0.57 & 0.59 & 0 & 0.57 \\
 1.1 \cdot 10^{-2}&2.3 & -0.58 & -0.58 & 0 & 0 & -0.58 & 0  \\
\end{array}
$
\end{center}
where $mu$ and $\sigma$ denote the best-fit values and uncertainties of the eigenvectors.
The eigenvectors reflect the correlations argued above, but also show that the picture is more complicated than having a single eigenvector that is clearly nonzero. 

The main differences between the L2(RH) and L6(SPS) fits can be seen from the two left panels of Fig.~\ref{fig:ObsL2L6}. L2(RH) helps to resolve discrepancies in the three-body kaon decay processes, specifically $K_{e3}$ and $K_{\mu3}$. Additionally, it brings the $\eta_{2,3}$ parameters associated with $0^+\to0^+$ transitions closer to their predicted central values. L6(SPS) can also alleviate, but to a lesser degree, the tensions in the three-body kaon decays, while also removing the small tension in $R_K=\Gamma(K\to e\nu_e)/\Gamma(K\to \mu\nu_\mu)$. However, this introduces extra tension in the superallowed $\beta$ decay of ${}^{14}$O and pushes the $\eta_{2,3}$ parameters further away from their central values. Taken together, the fit of L6(SPS) yields a $\Delta \mathrm{AIC} \approx 0$, making its performance on par with the SM, but significantly inferior to L2(RH).
\\
\\
\textbf{Combining RH + SPS operators.}
We now combine the two RH operators and the six pseudoscalar ones to perform an L8(RS) fit. The results show $\Delta \chi^2 = 25$, which is slightly higher than L2(RH), at the price of six additional Wilson coefficients and thus a worse $\Delta\mathrm{AIC}=9$. The relevant observables are shown in the right panel of Fig.~\ref{fig:ObsL2L6}. We observe that L8(RS) closely mirrors L2(RH), with scalar/pseudoscalar interactions offering a slight improvement of $R_K$.

\begin{figure}[t!]
\centering
\includegraphics[width=0.95\textwidth]{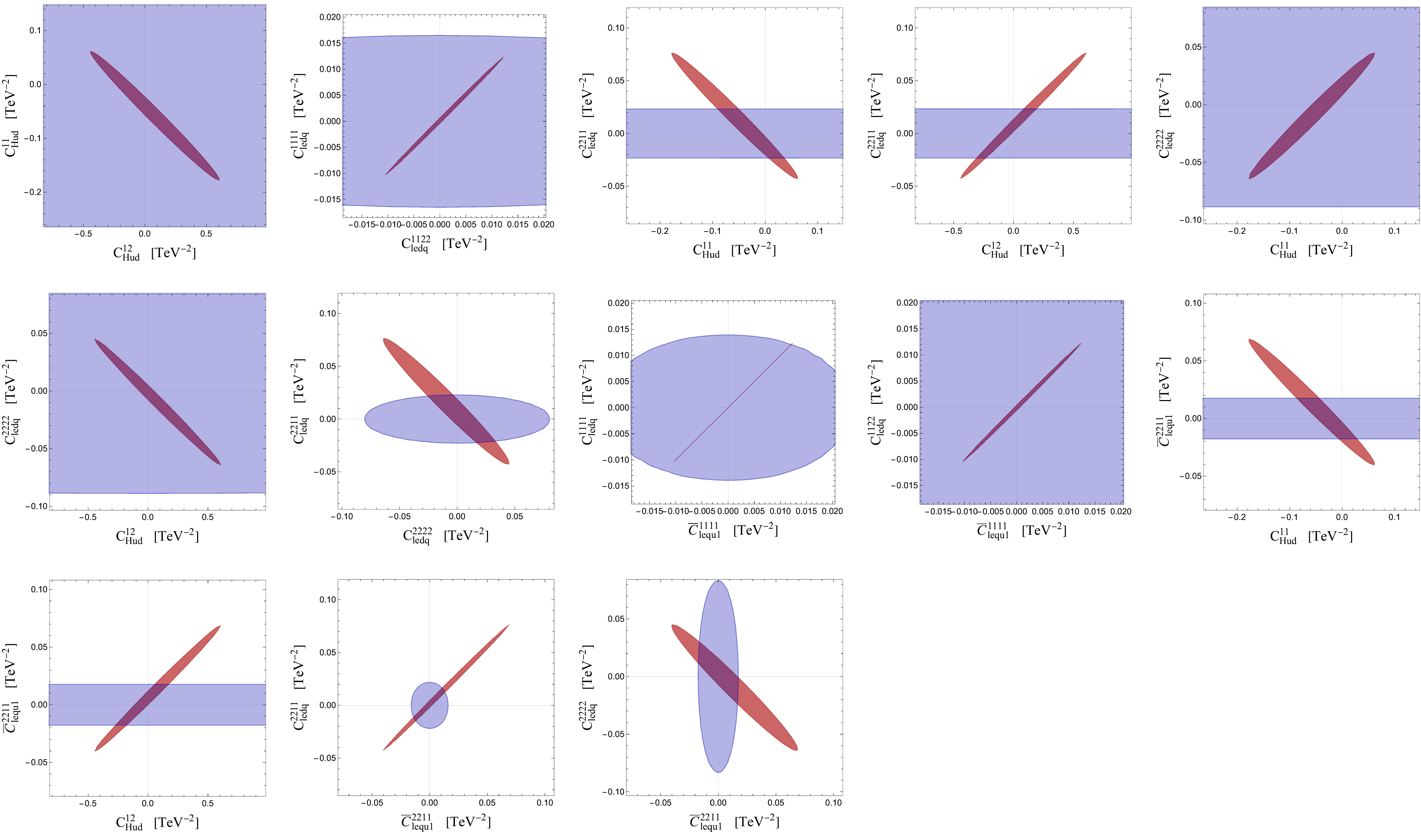}
\caption{The $1\sigma$ allowed region in the L8(RS) scenario, shown for several 2D slices in parameter space. The red and blue regions correspond to those preferred by the low-energy CC and collider observables, respectively.}
\label{fig:L8}
\end{figure}

All best-fit values in L8(RS) are consistent with zero within 2.5$\sigma$. The larger deviations from zero appearing in L2(RH) have been diluted due to the additional operators. However, as shown in Fig.~\ref{fig:L8}, several of the operators in L8(RS) are highly correlated. For example, $\left[C_{Hud}\right]_{11}$ and $\left[C_{Hud}\right]_{12}$ are negatively correlated in L8 (RS), while they have almost no correlation in L2(RH). In fact, if $\left[C_{Hud}\right]_{12} =0$ then $\left[C_{Hud}\right]_{11}<0$ by approximately 3$\sigma$. These correlations demonstrate a strong interplay between the right-handed and scalar/pseudoscalar operators, which affects the options for model building. Although we do not show all the eigenvectors, it is worth noting that L8(RS) features a distinct eigenvector that almost reaches the $4\sigma$ level
\beq
0.77{C}_{\substack{Hud\\11}} +0.1 {C}_{\substack{ledq\\2222}}+0.18  {C}_{\substack{Hud\\12}} +0.35 \left(  {C}_{\substack{ledq\\1111}} +{C}_{\substack{ledq\\1122}} +\bar{C}_{\substack{lequ\\1111}}^{(1)} \right)  = 0.0076\times (3.9\pm1)\, {\rm TeV}^{-2}\,,\nn\\
\eeq
while all other eigenvectors have a significance of less than 2$\sigma$. Interestingly, compared to L2(RH), this eigenvector is dominated by the up-down right-handed current operator ($[Q_{Hud}]_{11}$) instead of the up-strange one ($[Q_{Hud}]_{12}$) that appears in L2(RH).  

All fits conducted with only right-handed and scalar/pseudoscalar operators remain robust when expanding to a broader set of observables. Neither the RH nor the pseudoscalar operators enter the EWPO or the collider data at linear order. As such, CLEW2(RH), CLEW6(SPS), and CLEW8(RS) give the same results as previously reported.

Although the operators do not enter the collider observables at linear order, they do contribute quadratically. To achieve a fully consistent analysis that includes these effects, one must also consider genuine dimension-eight operators, as they formally emerge at the same order. Such an analysis is beyond the scope of this work. However, we would like to nonetheless gauge the sensitivity of the collider observables. To do so, we consider the constraints set by the Drell-Yan measurements on the quadratic contributions assuming that only the L8(RS) Wilson coefficients are turned on. The resulting constraints are shown by the blue regions in Fig.~\ref{fig:L8}, along with the region preferred by the low-energy data in red. These preliminary collider constraints are beginning to explore areas of the parameter space not yet ruled out by low-energy CC measurements. However, the collider processes have not probed any of the nonzero values favored by the low-energy observables.

\subsection{Left-handed operators and vertex corrections}
\begin{figure}[ht!]
\centering
\includegraphics[scale=0.34]{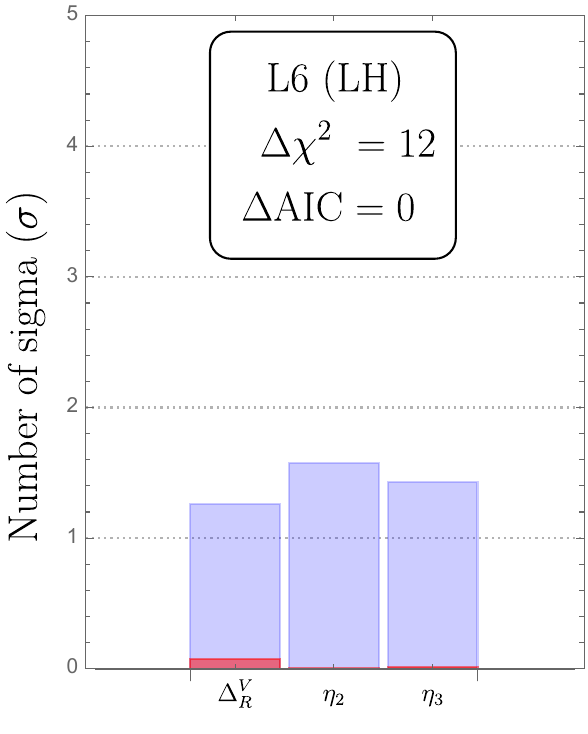}
\includegraphics[scale=0.34]{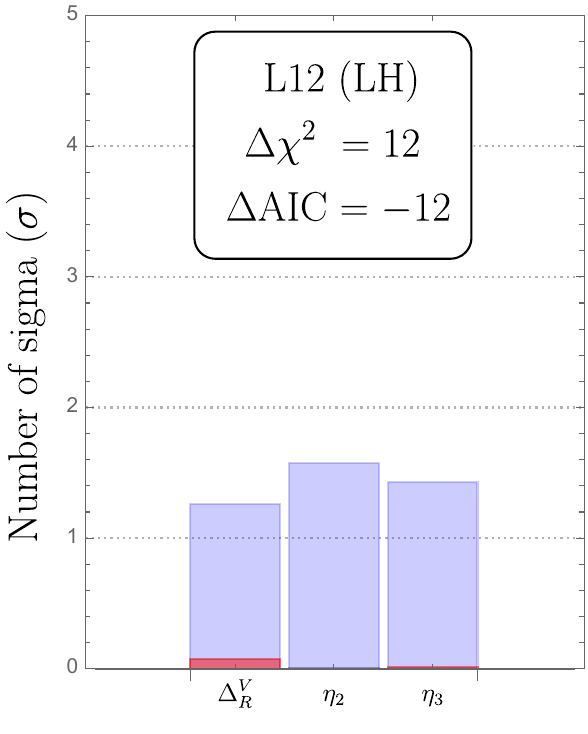}
\includegraphics[scale=0.34]{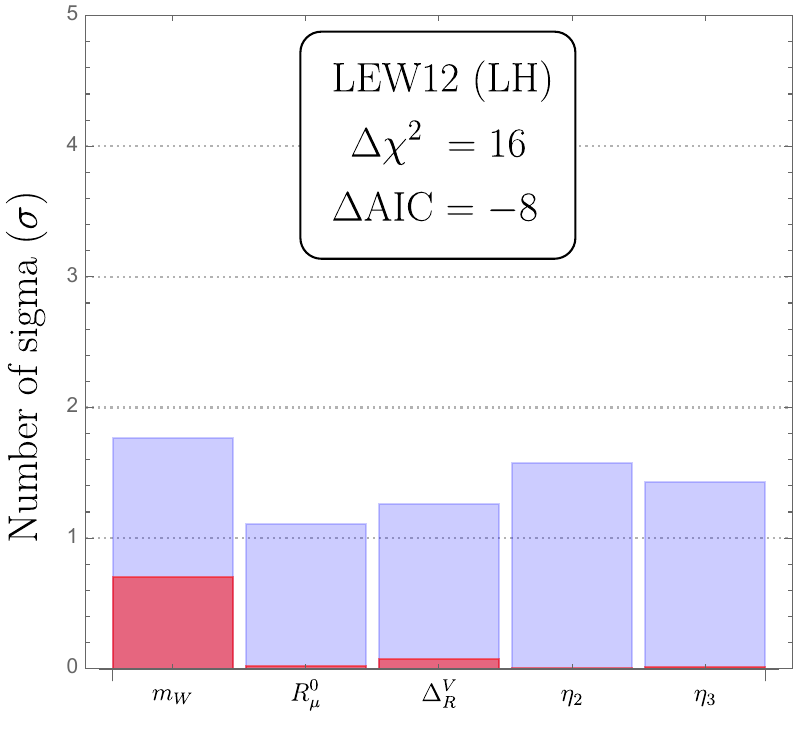}
\includegraphics[scale=0.34]{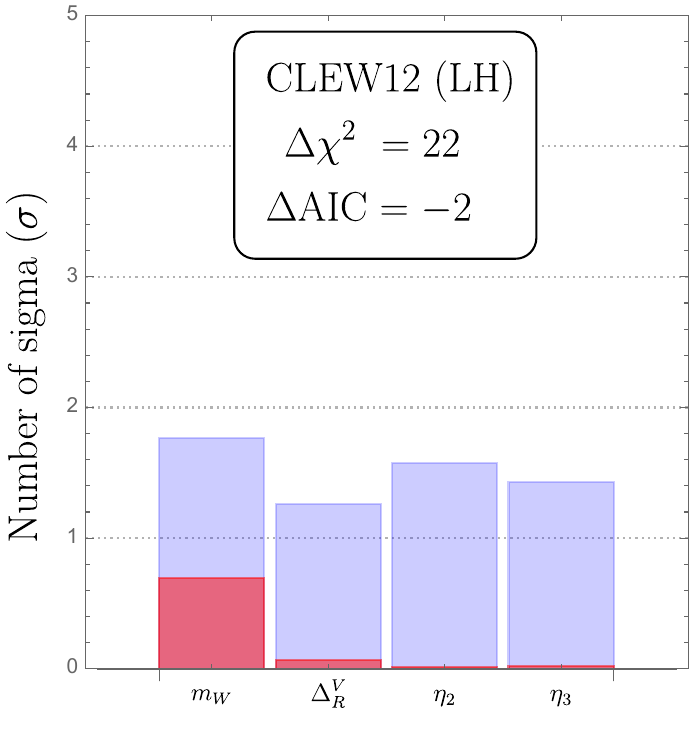}
\caption{Same as Fig.~\ref{fig:ObsL2L6}, for four different SMEFT models. From left to right: L6(LH), L12(LH), LEW12(LH), and CLEW12(LH).}
\label{fig:ObsLH}
\end{figure}

\textbf{Left-handed operators.}
Another way to explain the CKM anomaly could involve modifications of the left-handed charged currents. We investigate this by turning on the six Wilson coefficients associated with left-handed semileptonic four-fermion operators: $C_{lq}^{(u)}$ and $C_{lq}^{(d)}$. The fit, termed L6(LH), yields $\Delta \chi^2=12$, which is on par with L6(SPS) and does not offer an improvement in AIC compared to the SM. The pattern of observables differs between fits, as shown in the left panel of Fig.~\ref{fig:ObsLH}. L6(LH) essentially requires less tension in the $\eta_{2,3}$ parameters related to radiative corrections in superallowed $\beta$ decays, but does not address the tension in kaon observables. The sole eigenvector with a notable nonzero value is given by 
\bea
0.69  {C}^{(d)}_{\substack{lq\\1111}}-0.72{C}^{(u)}_{\substack{lq\\1111}} = 0.0059\times (3.1\pm1)\, {\rm TeV}^{-2}\,.
\eea

In contrast to earlier fits, L6(LH) is not stable under the inclusion of more observables. The four-fermion operators do not modify EWPO but they do affect Drell-Yan processes at the LHC. After taking into account the latter, $\Delta \chi^2$ reduces to 7. This results in a less favorable $\Delta \text{AIC} = -5$ relative to the SM, suggesting that the left-handed four-fermion operators needed to address the CAA conflict with high-energy measurements.

The stability under the inclusion of LHC data might improve if we expand to the full set of SMEFT operators that affect LH currents. We name the fit L12(LH). It encompasses the six operators of L6(LH) as well as the operators associated with vertex corrections $[C^{(3)}_{Hl}]_{ee}$, $[C^{(3)}_{Hl}]_{\mu\mu}$, $[C_{Hq}^{(d)}]_{11}$, $[C_{Hq}^{(d)}]_{22}$, $[C_{Hq}^{(u)}]_{11}$, and the purely leptonic four-fermion operator $[C_{ll}]_{2112}$. As far as the low-energy CC observables are concerned, the additional six operators are redundant. We can see this by comparing the first two panels on the left in Fig.~\ref{fig:ObsLH}, which are identical. However, once we move to LEW12 (3rd panel on the left in Fig.~\ref{fig:ObsLH}), the additional operators can reduce tension in the EWPO, in particular $m_W$ and $R_{0\mu}$, leading to a greater $\Delta\chi^2 = 16$. If we also include the LHC data (the panel on the right of Fig.~\ref{fig:ObsLH}), we see that $\Delta\chi^2$ of CLEW12 grows to 22. This increase is not tied to a singular observable, but results from minor changes across multiple bins. However, even with these improvements, the AIC of CLEW12 falls short compared to the SM, yielding $\Delta \text{AIC}=-2$. This further underscores that purely left-handed SMEFT operators do not provide an effective solution to the CAA.
\\
\\
\textbf{Vertex corrections.} We have seen that LHC measurements can strongly constrain the semileptonic four-fermion operators. In light of this, we will investigate SMEFT operators that provide vertex corrections, which are less stringently probed by DY measurements. Of the 22 operators discussed in this section, only seven belong to this category: $[C_{Hl}^{(3)}]_{ee,\mu\mu}$, $[C_{Hq}^{(d)}]_{11,22}$, $[C_{Hq}^{(u)}]_{11}$, and $[C_{Hud}]_{11,12}$. We will fit them to the low-energy CC observables and the EWPO simultaneously, naming it LEW7(V), with V signifying vertex.

\begin{figure}[ht!]
	\centering
	\includegraphics[scale=0.4]{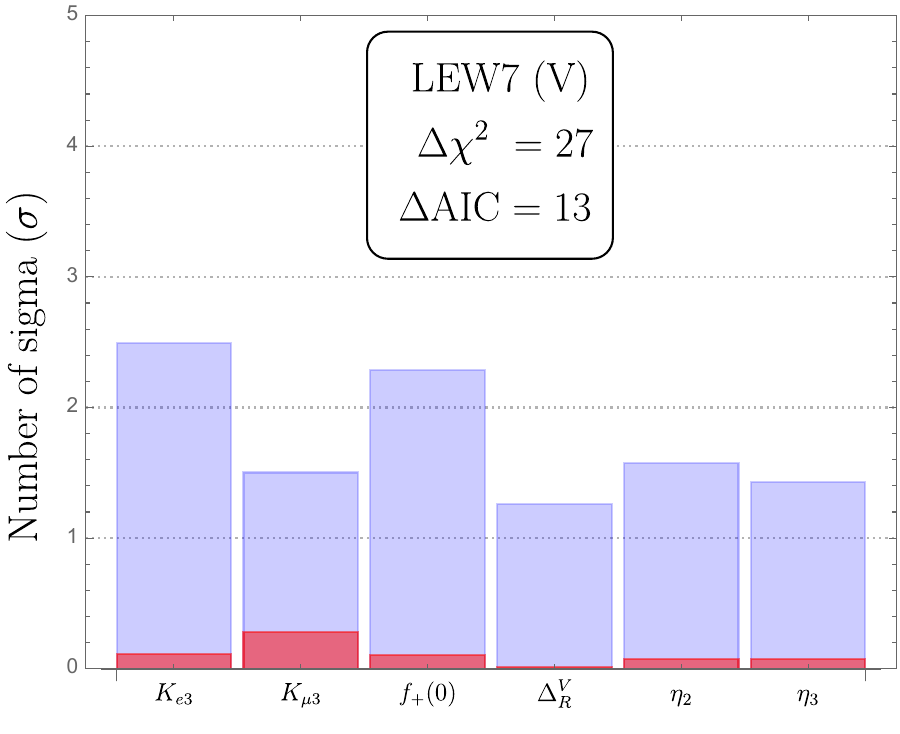}
 	\includegraphics[scale=0.4]{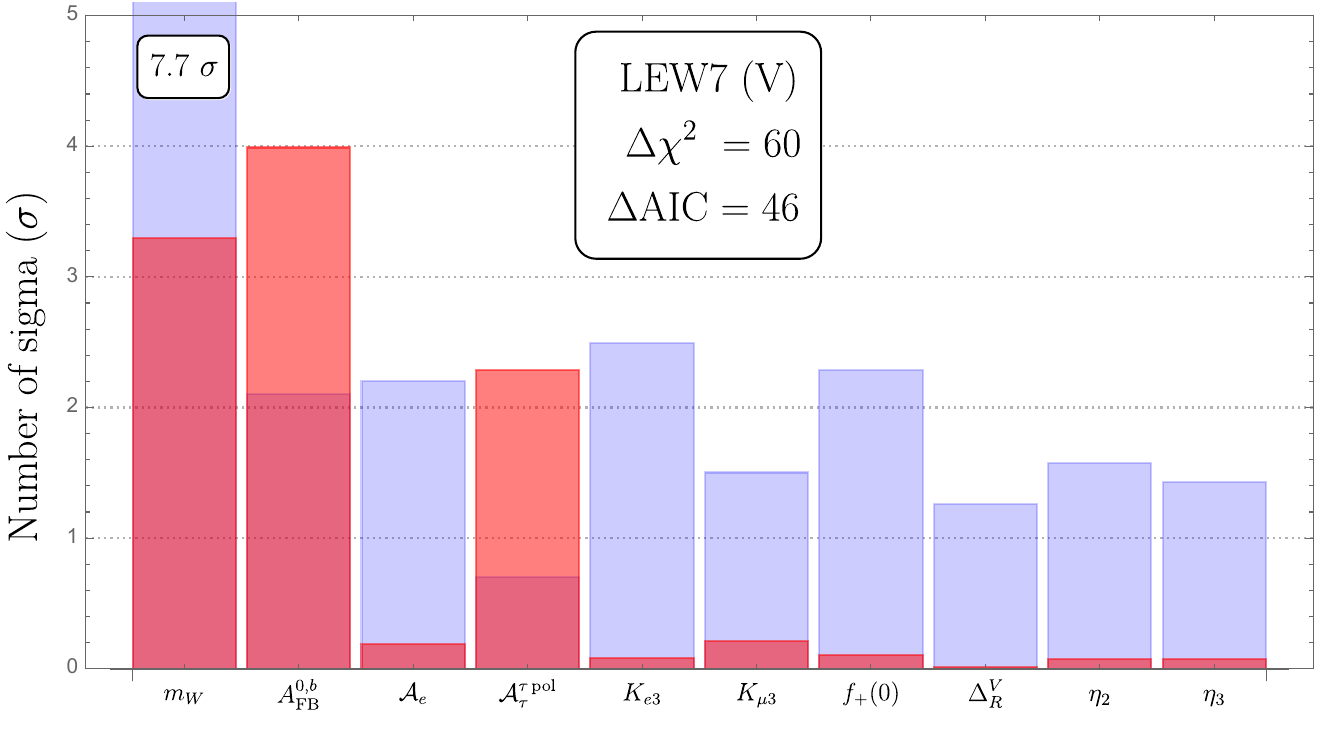}
	\caption{Same as Fig.~\ref{fig:ObsL2L6}, for
 two different fits involving right- and left-handed vertex corrections.  Left: LEW7(V) \textit{without} the CDF $m_W$. Right: LEW7(V) \textit{including} the CDF $m_W$.}
 \label{fig:V7Charts}
\end{figure}

The best-fit values are given in Table~\ref{tab:V7}, while the impact on the observables is shown in Fig.~\ref{fig:V7Charts}. The value of $[C_{Hud}]_{12}$ is close to that obtained in L2(RH), indicating that the RH up-strange interaction accounts for the kaon processes. The role of $[C_{Hud}]_{11}$, however, is diluted by left-handed vertex corrections that can also modify $\beta$-decay processes. In fact, the only eigenvector that is nonzero at more than $2\sigma$ is given by
\bea
0.96 {C}_{\substack{Hud\\12}} +0.12 \hat{C}_{\substack{Hl\\22}}^{(3)}-0.19\hat{C}_{\substack{Hq\\11}}^{(u)} +0.18 \hat{C}_{\substack{Hq\\11}}^{(d)} = -0.011\times (3.9\pm 1)\, {\rm TeV}^{-2}\,,
\eea
which is dominated by the up-strange RH current. 

\begin{table*}[!t]
    \centering
    \begin{tabular}{|c|c|}
\hline
Vertex Corrections & LEW		\\
 \hline
$[C_{Hud}]_{11}$ & $ -0.0047 \pm 0.049 $ \\
$[C_{Hud}]_{12}$ & $ -0.037 \pm 0.013 $ \\
$[\hat{C}_{Hl}^{(3)}]_{11}$ & $ -0.0057 \pm 0.0059 $ \\
$[\hat{C}_{Hl}^{(3)}]_{22}$ & $ -0.0063 \pm 0.0052 $ \\
$[\hat{C}_{Hq}^{(u)}]_{11}$ & $ 0.23 \pm 0.41 $ \\
$[\hat{C}_{Hq}^{(d)}]_{11}$ & $ 0.20 \pm 0.44 $ \\
$[\hat{C}_{Hq}^{(d)}]_{22}$ & $ -0.010 \pm 0.12 $ \\
\hline
    \end{tabular}
    \caption{Best fit for LEW7(V) including right- and left-handed vertex corrections. Wilson coefficients are given in units of TeV$^{-2}$.}
    \label{tab:V7}
\end{table*}

LEW7(V) has a $\Delta\chi^2=27$ compared to the SM, which is slightly better than LEW2(RH) due to mild improvements in $m_W$ (too small to appear in Fig.~\ref{fig:V7Charts}) at the price of five additional operators. This leads to $\Delta$AIC$=13$, compared to 18 for the L2(RH) scenario. One advantage of LEW7(V) is its ability to partially reconcile with the CDF measurement of $m_W$, see the right panel of Fig.~\ref{fig:V7Charts}.

\subsection{The full 22 fit}

\begin{figure}[ht!]
\centering
\includegraphics[scale=0.4]{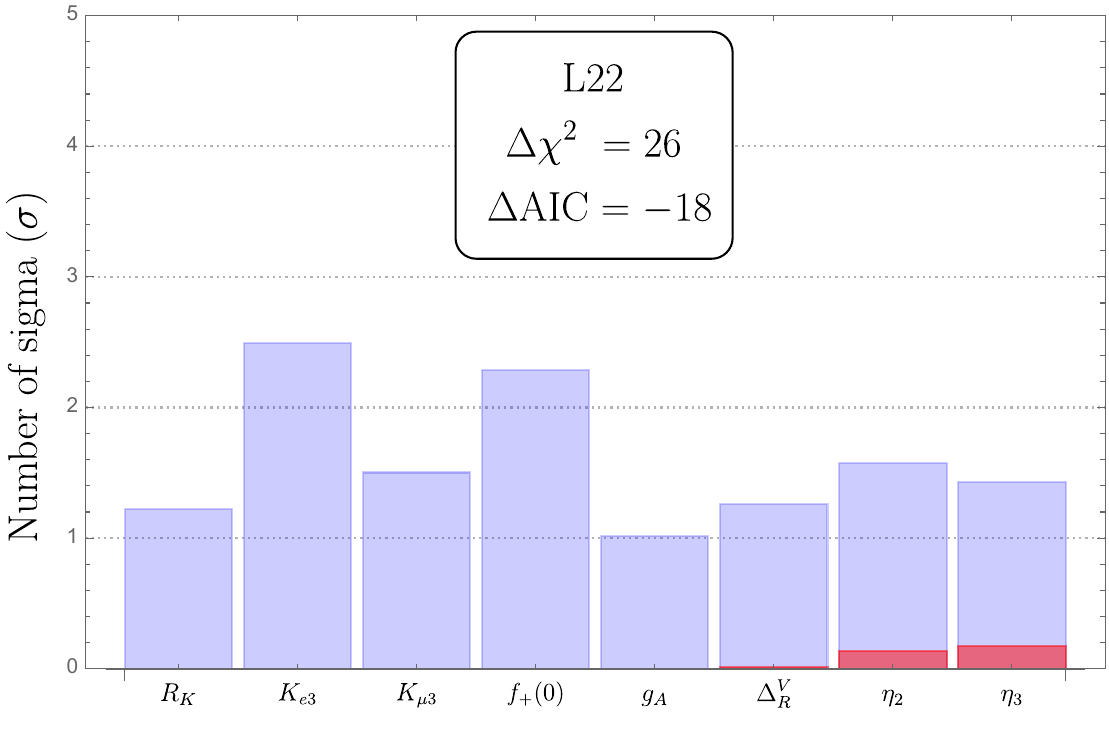}
\includegraphics[scale=0.4]{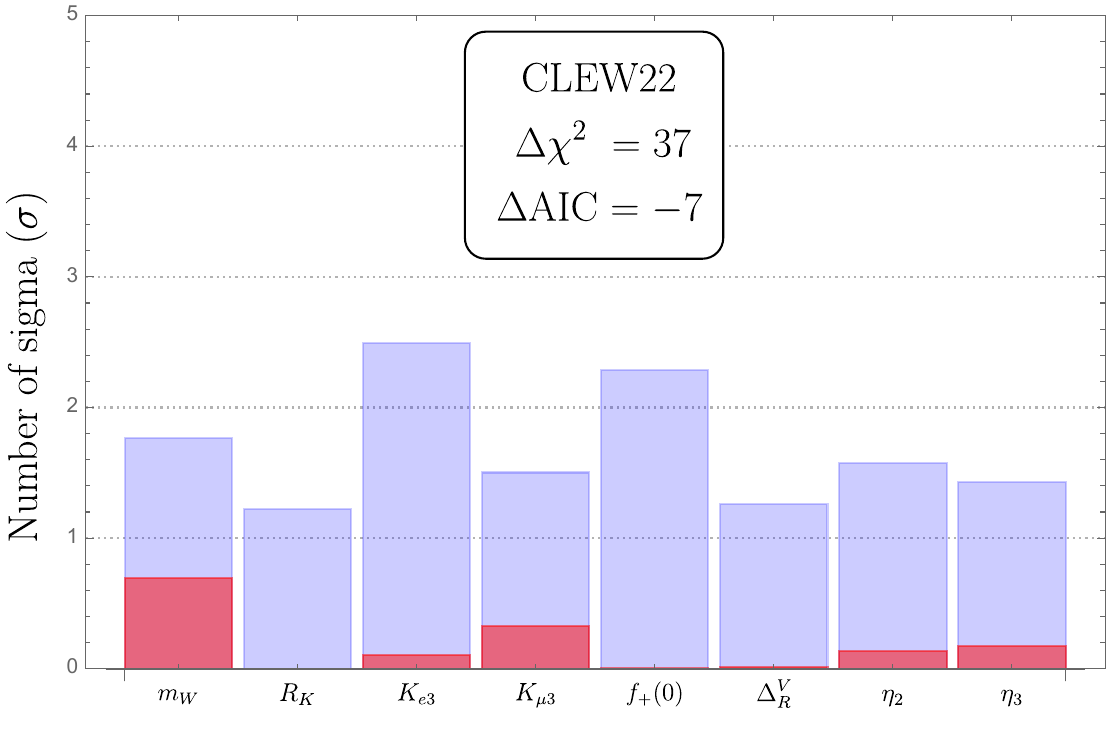}
\includegraphics[scale=0.4]{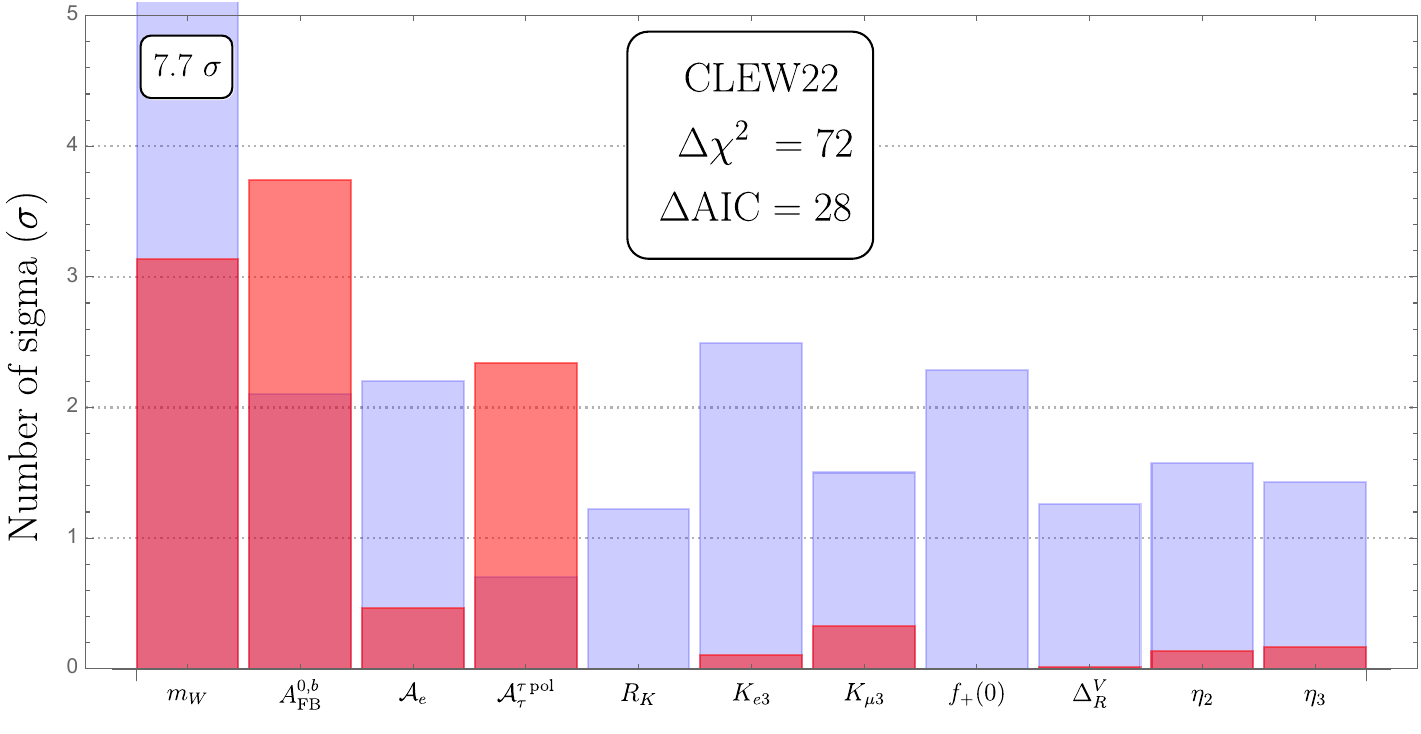}
\caption{Same as Fig.~\ref{fig:ObsL2L6}, for three different SMEFT fits involving the 22 operators in the right panel of Table~\ref{tab:globalWC}.  Left: L22 fit. Right: CLEW22 \textit{without} CDF $m_W$. Bottom: CLEW22 \textit{including} CDF $m_W$.}
\label{fig:Obs22}
\end{figure}

Now we turn the crank and fit all 22 operators simultaneously. For L22, we obtain $\Delta\chi^2 = 26$, only slightly better than L8(RS). This is reflected in the left panels of Figs.~\ref{fig:Obs22} and~\ref{fig:ObsL2L6}. The L22 fit allows for essentially the same improvements over the SM as L8(RS), while the additional small gain in $\Delta\chi^2$ is due to a slightly better description of the neutron lifetime and $R_K$. 
When moving to LEW22, we see that, similar to LEW12(LH), the additional operators can account for some small discrepancies in the EWPO, again mainly for $m_W$ and $R_{0\mu}$. Moving to CLEW22, we find another small improvement of $\Delta \chi^2$, but the pattern of observables essentially stays the same, and therefore we only show the results for CLEW22 in Fig.~\ref{fig:Obs22}.

Moving from L22 to LEW22 and finally to CLEW22 significantly reduces the number of free directions in the fit. The resulting central values, uncertainties, and correlations can be seen explicitly in the Supplementary Material. In particular, in CLEW22 there are no free directions remaining and the magnitudes of each Wilson coefficient are constrained below $1/\text{TeV}^2$. All Wilson coefficients are consistent with zero within $2\sigma$, while three eigenvectors emerge, which deviate more than $2\sigma$ from zero. These linear combinations involve a large number of Wilson coefficients. Truncating the contributions to the normalized eigenvectors at $0.3$, we find\\ 

\begin{table*}[ht!]
	\centering
	\begin{tabular}{|c|cc|cc|cc|}
 \hline
   & L & & LEW & &  CLEW & \\
   \hline
    & $\Delta \chi^2$ & $\Delta$AIC  & $\Delta \chi^2$ & $\Delta$AIC  & $\Delta \chi^2$ & $\Delta$AIC \\
  \hline
  2(RH) & 22 & 18 & 22 & 18 & 22 & 18\\ 
  6(SPS) & 13 &0 & 13 & 0 & 13 & 0 \\ 
8(RS) & 25 & 9  & 25  & 9 & 25  & 9\\
 6(LH) & 12 & 0  & 12  & 0 & 7  & -5 \\
 12(LH) & 12  & -12 & 16  & -8 & 22  & -2 \\
 7(V) & 23  & 9  & 27  & 13 & 27 & 13 \\
  22(All) & 26 & -18  & 31 &-13 & 37 & -7 \\
  \hline
	\end{tabular}
	\caption{$\Delta\chi^2$ and $\Delta$AIC values with respect to the SM for various SMEFT scenarios and sets of observables.}
\label{tab:Deltachi}
\end{table*}

\begin{align}
0.42\,C_{\substack{Hq\\11}}^{(u)}-0.40\,C_{\substack{Hq\\11}}^{(d)}-0.37\left( C_{\substack{Hl\\22}}^{(3)}+C_{\substack{Hud\\11}}\right)-0.59\,\bar C_{\substack{lequ\\1111}}^{(3)} &= 0.0091 \times (2.4\pm1)\,{\rm TeV}^{-2}\,, \nnw 
0.60\,C_{\substack{Hl\\22}}^{(3)}-0.58\,C_{\substack{ll\\2112}} -0.30\,C_{\substack{ledq\\2222}} &= 0.0025 \times (3.1\pm1)\,{\rm TeV}^{-2}\,, \nnw 
-0.96\,C_{\substack{Hl\\11}}^{(3)} &= 0.0051 \times (2.9\pm1)\,{\rm TeV}^{-2}\,.
\end{align}
However, the CLEW22 fit has the worst information criterion with $\Delta \text{AIC}=-7$, indicating that adding more parameters is simply not worth it. 
\\
\\
\textbf{The role of CDF.} In this section, we focused on the CAA and found that right-handed currents have the best performance. The possible advantage of including additional operators is that they can potentially account for the CDF measurement of $m_W$. We already saw this in the fits with left-handed operators which slightly improved the description of the EWPO. While the inclusion of the CDF result is the main topic of the next section, we can already see what happens in the context of the CLEW22 fit. As shown in the bottom panel of Fig.~\ref{fig:Obs22}, it is possible to reduce the tension in $m_W$ to $3\sigma$. However, the 22 operators discussed here can only do so by introducing a large tension with other EWPO. Although $\Delta \mathrm{AIC} = 28$ is still very good, $A_{FB}^{0,b}$ obtained a $4\sigma$ discrepancy, which implies that the fit is not optimal.
In the next section, we transition to a scenario that does not rely on flavor assumptions, encompassing all 37 coefficients in Table~\ref{tab:globalWC}. Our aim is to determine whether we can more accurately represent both the CAA and $m_W$ at the same time.

\subsection{Intermediate conclusions}

We have investigated the CAA by including all SMEFT operators that contribute to low-energy CC observables, without making use of flavor assumptions. Although there are 22 such operators, a detailed study of various fits indicates that many operators do not play a big role in the CAA. We summarize the performance of the fits in Table~\ref{tab:Deltachi}.

The best fit (with the highest $\Delta$AIC) is given by just including the two RH CC operators. The next best fit is obtained by adding LH vertex corrections, which can slightly improve the EWPO but at the price of additional parameters. A similar quality of fit is obtained by combining the RH CC operators with scalar/pseudoscalar four-fermion operators. Other possibilities that do not include RH CC lead to an AIC that is comparable to or worse than that of the SM.

Incorporating the CDF measurement of $m_W$ results in a suboptimal performance for CLEW22. Although it has a positive $\Delta$AIC, it keeps tension in $m_W$ at 3$\sigma$ and severely compromises other observables. To convincingly address both anomalies at the same time, we must include more operators (the left panel of Table~\ref{tab:globalWC}). This will be discussed in the following section.

\section{A flavor-independent global analysis}
\label{sec:global_37}

In the final analysis, we would like to investigate the interplay of the CDF measurement of $m_W$ and the CAA. 
We will expand our set of operators to include the Wilson coefficients listed in the left panel of Table~\ref{tab:globalWC}. The fit incorporates 37 Wilson coefficients, the SM parameter $\la$, as well as a collection of matrix elements and parameters characterizing theoretical uncertainties.

\begin{figure}[t]
	\centering
\includegraphics[width=0.9\textwidth]{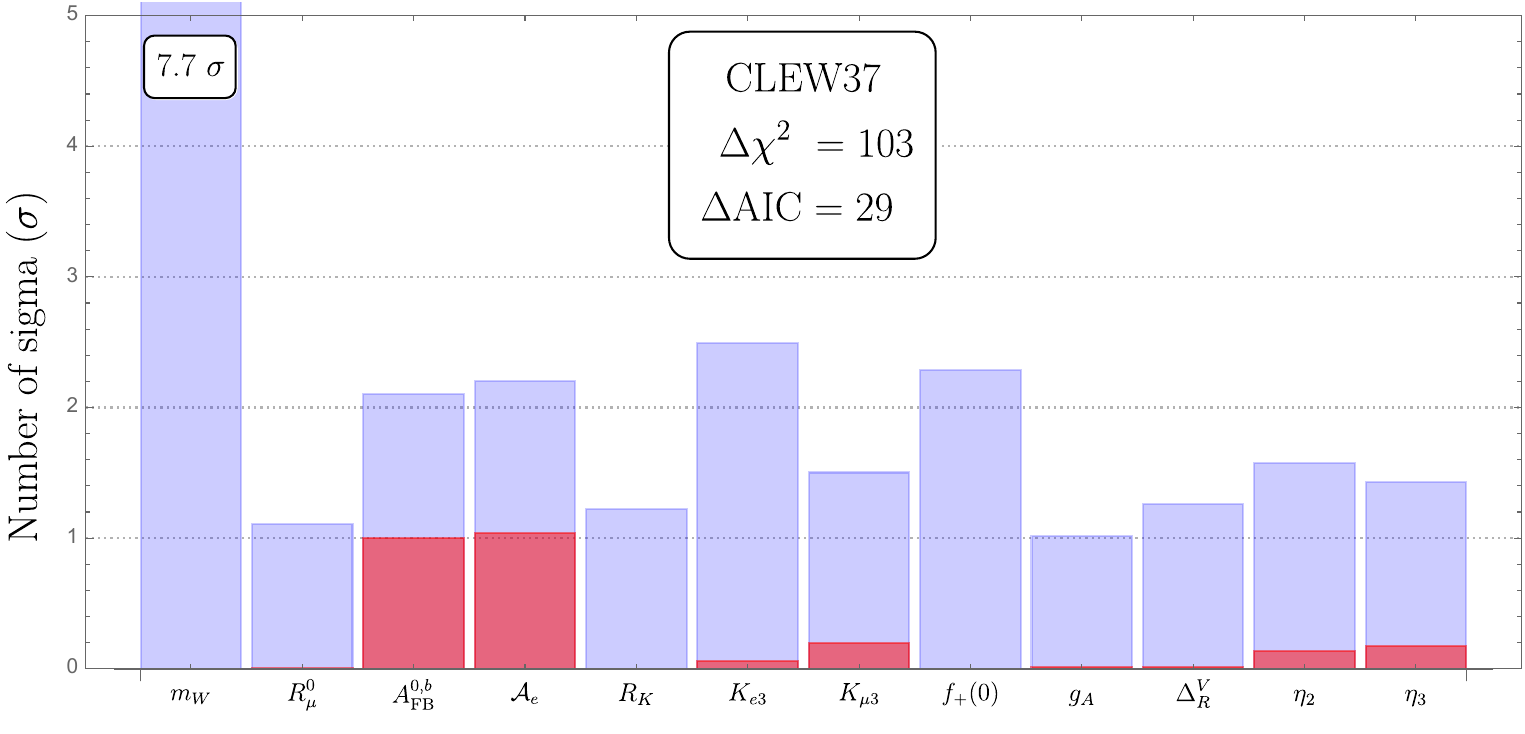}
	\caption{Same as Fig.~\ref{fig:ObsL2L6}, for all of the 37 operators to the CLEW data set.}
\label{fig:37}
\end{figure}

We present the predicted observables at the best-fit point of the global fit CLEW37 in Fig.~\ref{fig:37}. 
CLEW37 successfully removes tension in the CDF W mass and the CAA, as well as several other observables. We obtain $\Delta \chi^2 = 103$ and $\Delta \text{AIC} = 29$, indicating a significant improvement over the SM, comparable to that of the CLEW22 fit. The best-fit results show that all Wilson coefficients are consistent with zero within about 2$\sigma$ (see the Supplemental Material for the central values, uncertainties, and correlations). However, as in the CLEW22 fit, there are several eigenvectors that are nonzero with significance $\gtrsim 3\sigma$. Neglecting the contributions to the normalized eigenvectors that are less than 0.3, we find 

\begin{align}
0.51\, C_{ST}+0.33\,C_{\substack{Hl\\22}}^{(3)}+0.45\,C_{\substack{ledq\\2222}}-0.40\, \bar C_{\substack{lequ\\2211}}^{(1)}-0.34 \,C_{\substack{ll\\2112}} &= -0.0016 \times (3.7\pm1)\,{\rm TeV}^{-2}\,, \nnw 
0.41\,C_{ST}-0.47\,C_{\substack{Hl\\22}}^{(3)}+0.42\,C_{\substack{ll\\2112}} &= -0.0030 \times (6.5\pm1)\,{\rm TeV}^{-2}\,, \nnw 
-0.83\,C_{\substack{He\\11}}-0.34\,C_{\substack{Hl\\11}}^{(1)}-0.31\,C_{\substack{Hl\\11}}^{(3)} &= \phantom{-}0.0093 \times (3\pm1)\,{\rm TeV}^{-2}\,.
\end{align}

The significant improvement in $\Delta$AIC is mainly due to the large tension of the CDF W mass. Fitting CLEW37 with the PDG average of $m_W$ instead, we obtain $\Delta \chi^2 = 47$ and $\Delta \text{AIC} = -27$, a performance worse than the SM. In this case, there are too many operators that do not contribute significantly to $\Delta \chi^2$. 

Even with the CDF $m_W$, the inclusion of all 37 operators is inefficient, resulting in overfitting and a suboptimal $\Delta$AIC. In the next section, we perform a systematic analysis of various scenarios to pinpoint the SMEFT operators that are most important in addressing the CAA and the $m_W$ anomaly.

\subsection{Finding the optimal fit}
\label{Sec:MA}

Section~\ref{22fit} focused on the CAA and we investigated scenarios with various subsets of SMEFT operators. In these cases, we handpicked the operators that were likely to provide the most efficient way to account for the apparent violation of the CKM unitarity. Now that we are also including the W mass anomaly, this dissection by hand is complicated by the large number of possible subsets. Therefore, we implement a more systematic approach to find the optimal fit.

Recall that we define a `model' as the SMEFT Lagrangian with a specific subset of Wilson coefficients turned on. For example, we found that models with $C_{Hud}$ tend to give the highest $\Delta$AIC and thus provide a more likely explanation of the CAA. Although we have explored a fair number of models, they represent only a fraction of the potential combinations of SMEFT operators. However, evaluating every combination of the 37 operators would amount to  $2^{37} = \mathcal O(10^{11})$ fits, an impractical endeavor. 
We therefore group the operators into ten categories that are summarized in Table~\ref{tab:modelave}. 
Our underlying theoretical motivation is that a particular BSM scenario is unlikely to produce just one quark/lepton flavor component in a specific category. We therefore turn on, or off, all Wilson coefficients within a certain category simultaneously. This assumption will be partially relaxed in Section~\ref{sec:Open}. For now, we consider all possible combinations of these ten categories, resulting in $2^{10}=1024$ models.

\begin{table}[t]
\centering
\begin{tabular}{|c|c|c|c|c|c|}
\hline
 Category & Operators & Description & \# of Ops. & $W_\theta^\text{PDG}$ & $W_\theta^\text{CDF}$ \\
 \hline \hline 
 \RNum{1}. & $C_{ST}$ & Oblique corrections & 1 & 0.55 & 1.00 \\
 \RNum{2}. & $C_{Hud}$ & RH charged currents & 2 & 0.99 & 0.96 \\
 \RNum{3}. & $C^{(1)}_{Hl}$~~$C^{(3)}_{Hl}$ & LH lepton vertices & 6 & 0.01 & 0.11 \\
 \RNum{4}. & $C_{He}$ & RH lepton vertices & 3 & 0.09 & 0.42 \\
 \RNum{5}. & $C^{(u)}_{Hq}$~~$C^{(d)}_{Hq}$ & LH quark vertices & 5 & 0.03 & 0.13 \\ 
 \RNum{6}. & $C_{Hu}$~~$C_{Hd}$ & RH quark vertices & 5 & 0.06 & 0.32 \\
 \RNum{7}. & $C_{ll}$ & Lepton 4-fermion & 1 & 0.37 & 0.87 \\
 \RNum{8}. & $C^{(u)}_{lq}$~~$C^{(d)}_{lq}$ & Semileptonic 4-fermion & 6 & 0.03 & 0.03 \\
 \RNum{9}. & $C_{ledq}$~~$C_{lequ}^{(1)}$ & Scalar 4-fermion & 6 & 0.02 & 0.04 \\
 \RNum{10}. & $C^{(3)}_{lequ}$ & Tensor 4-fermion & 2 & 0.13 & 0.13 \\
 \hline
\end{tabular}
\caption{We divide the 37 operators identified in the left panel of Table~\ref{tab:globalWC} into ten categories. In the third column, each category is described by the type of operators within it, which are listed in the second column. The fourth column counts how many operators among the 37 are included in each category. The fifth column gives the total weights of all models that contain the corresponding category, as described in Eq.~\eqref{eq:Wt}. The sixth column repeats this using the CDF $m_W$.}
\label{tab:modelave}
\end{table}

\subsection{PDG value of $m_W$}

\begin{figure}[t]
	\centering
	\includegraphics[scale=0.6]{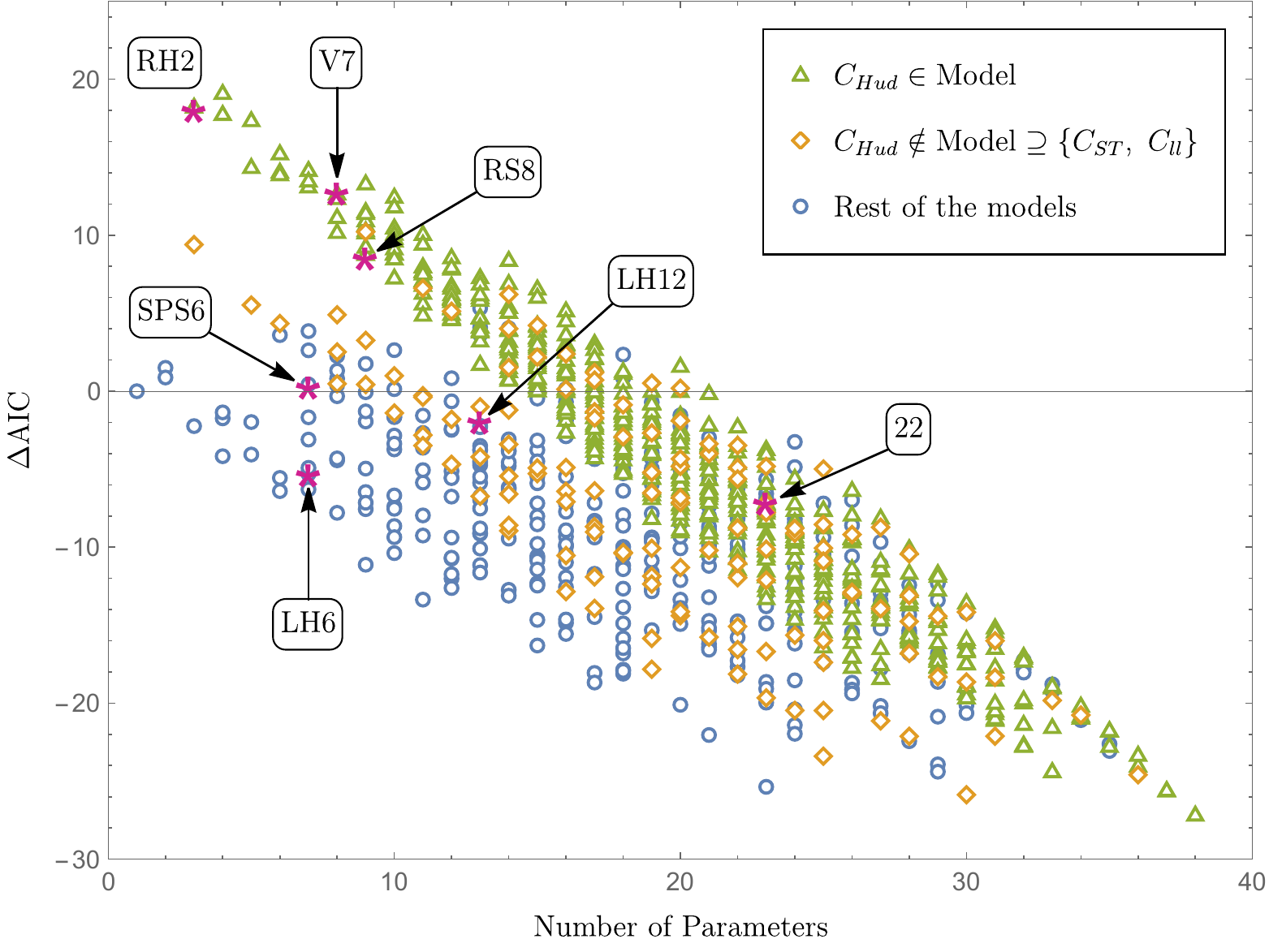}
	\caption{ 
 \textit{Excluding} the CDF $m_W$, we plot $\Delta$AIC for the 1024 models with respect to their number of parameters. Models containing the category $C_{Hud}$ are marked with green triangles. Orange diamonds represent those that contain both $C_{ST}$ and $C_{ll}$, but not $C_{Hud}$. The rest of the models are denoted by blue circles. The purple stars denote the seven models we have analyzed in Section~\ref{22fit} and are labeled by their acronyms.}
 \label{fig:AIC_PDG}
\end{figure}

We start by performing these fits with the PDG value of $m_W$. We show the resulting $\Delta {\rm AIC}$ as a function of the number of parameters for all 1024 models in Fig.~\ref{fig:AIC_PDG}. 
The figure shows that the models can be divided into roughly three `branches': 
\begin{enumerate}
    \item Models that include the right-handed current coefficients $C_{Hud}$ (green triangle)
    \item Models that include both $C_{ST}$ and $C_{ll}$, but not $C_{Hud}$ (orange diamond)
    \item Rest of the models (blue circle)
\end{enumerate}
The best-performing models fall into the first category which includes right-handed charged currents. In fact, the optimal model contains $\la$,
$\left[C_{Hud}\right]_{11}$, $\left[C_{Hud}\right]_{12}$,
and $C_{ST}$ as fit parameters and has $\Delta{\rm AIC} = 19$. The best-fit results are given by
\begin{align}\label{bestfitPDG}
C_{\substack{Hud\\11}} &= (-0.030 \pm 0.008)\,\text{TeV}^{-2} \,, \nnw
C_{\substack{Hud\\12}} &= (-0.040 \pm 0.011)\,\text{TeV}^{-2} \,, \nnw
C_{ST} &= (-0.0038 \pm 0.0022)\,\text{TeV}^{-2} \,.
\end{align}
The values for $C_{Hud}$ are the same as those found in the L2(RH) discussed in the previous section (see Table~\ref{tab:L8}). The nonzero value of $C_{ST}$ accounts for the slight discrepancy in $m_W$ that is present even when the CDF measurement is excluded. In fact, the observables and matrix elements most improved in this model closely resemble those of L2(RH), which are shown in the left panel of Fig.~\ref{fig:ObsL2L6}. In addition, the tension in $m_W$ is reduced from approximately $2\sigma$ to less than $1\sigma$.

The second-best model (with $\Delta {\rm AIC} = 18$) is nothing more than L2(RH), while the third-best model includes $C_{Hud}$ and $C_{ll}$. The two models L7 (V) and L8 (RS) that we studied in Section~\ref{22fit} also fall into this family, with the additional parameters causing a penalty in AIC.

Of the 41 models selected for their performance, where their values of $\Delta$AIC are within 10 units below the best model, only two exclude the right-handed operator, while they include both $C_{ll}$ and $C_{ST}$ (marked by orange diamonds in Fig.~\ref{fig:AIC_PDG}). 
A three-parameter fit with only $\la$, $C_{ll}$ and $C_{ST}$ has a $\Delta \text{AIC}=9$, with both $C_{ll}$ and $C_{ST}$ nonzero at more than 3$\sigma$,
\begin{equation}
    C_{ll}  = (-0.013 \pm 0.004) \, \, {\rm TeV}^{-2}, \qquad C_{ST} = - ( 0.0083 \pm 0.0026 ) \, \, {\rm TeV}^{-2}\,.
\end{equation}
The combination of $C_{ll}$ and $C_{ST}$ performs significantly better than having just one of the two. $C_{ll}$ can improve low-energy observables at the cost of a poorer description of several EWPO. Similarly, $C_{ST}$ can improve $m_W$ a bit but worsens other observables. However, the combination performs better across the chart.

The nine-parameter model with $C_{ST}$, $C_{ll}$, and six scalar/pseudoscalar operators yields $\Delta \text{AIC}=10$. It performs better than the L6(SPS) model, which only contains the scalar/pseudoscalar operators and has a $\Delta \text{AIC}=1$, shown in Fig.~\ref{fig:AIC_PDG} by a purple star right above the SM line ($\Delta \text{AIC}=0$).
The remaining three models studied in Section~\ref{22fit}, also marked by purple stars, all have a worse AIC than the SM and thus are disfavored.  

Among all models that contain neither $C_{Hud}$ nor the pair $\{C_{ST}$,~$C_{ll}\}$ (marked by blue circles), the best performance, $\Delta {\rm AIC} = 5$, is achieved by a model consisting of 13 parameters, including $C_{ST}$, the left-handed quark vertices $C_{Hq}^{(u)}$ and $C_{Hq}^{(d)}$, and the scalar/pseudoscalar four-fermion operators $C_{ledq}$ and $C_{lequ}^{(1)}$.\\ 

\begin{figure}[t]
	\centering
	\includegraphics[scale=0.6]{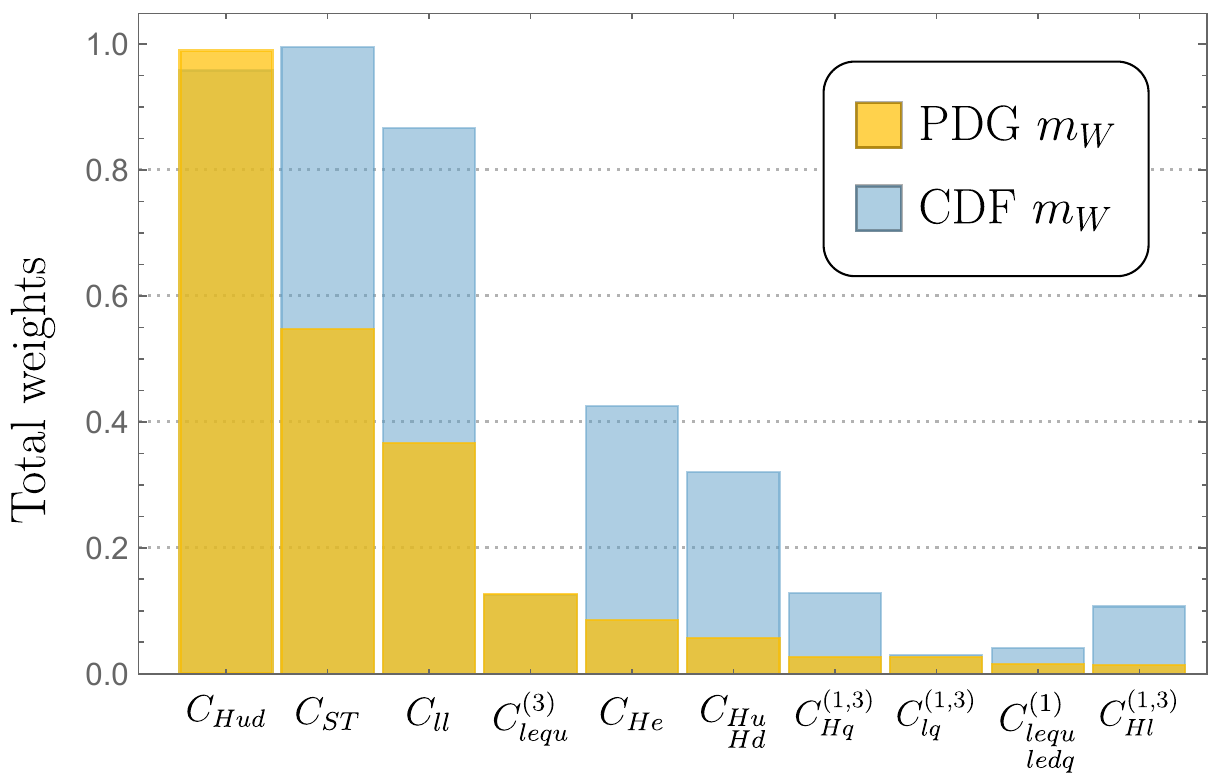}
	\caption{
 Model-averaged weights for each category of operators, as defined in Eq.~\eqref{eq:Wt}. Yellow bars: PDG world average of $m_W$. Blue bars: CDF measurement of $m_W$.}
 \label{fig:weights}
\end{figure}

Using the sum of weights $W_\theta$ to assess the importance of a given operator, the yellow bars in Fig.~\ref{fig:weights} show that $C_{Hud}$ is the most important\footnote{A weight of $0.99$ means that the 512 models that do not contain right-handed currents collectively carry only 1\% of the weight as defined in Eq.~\eqref{eq:Wt}.} with $W=0.99$. $C_{ST}$ and $C_{ll}$ are the second and third most important with $W=0.55$ and $0.33$, respectively, followed by the tensor operator $C_{lequ}^{(3)}$ and the right-handed leptonic vertex corrections $C_{He}$, with $W\sim 0.1$. All other operators have $W < 0.1$. Using Eq.~\eqref{eq:ma}, we can provide the model-averaged best-fit values of the most important Wilson coefficients,
\begin{align}\label{RH_MA}
    C_{\substack{Hud \\ 11}} &= -0.029 \pm 0.016 \; {\rm TeV}^{-2}, &\qquad  C_{\substack{Hud \\ 12}} &= - 0.039 \pm 0.014 \; {\rm TeV}^{-2}, \nn\\
    C_{ST} &= -0.0045 \pm 0.0032 \; {\rm TeV}^{-2}, &\qquad C_{ll} &= -0.001 \pm 0.012 \; {\rm TeV}^{-2}.
\end{align}
Compared to Eq.~\eqref{bestfitPDG}, we see that even after model averaging there is evidence for an up-strange RH current at the $2.7\sigma$ level, with $\sigma$ defined in Eq.~\eqref{eq:ma}. The evidence for an up-down RH current is diluted to roughly $2\sigma$. $C_{ST}$ deviates from zero by slightly more than $1\sigma$, while $C_{ll}$ is compatible with zero. The values in Eq.~\eqref{RH_MA} may provide guidance for model building. 

\subsection{CDF value of $m_W$}
\label{sec:AIC_CDF}

\begin{figure}[t]
	\centering
        \hspace*{-1cm}
	\includegraphics[scale=0.60]{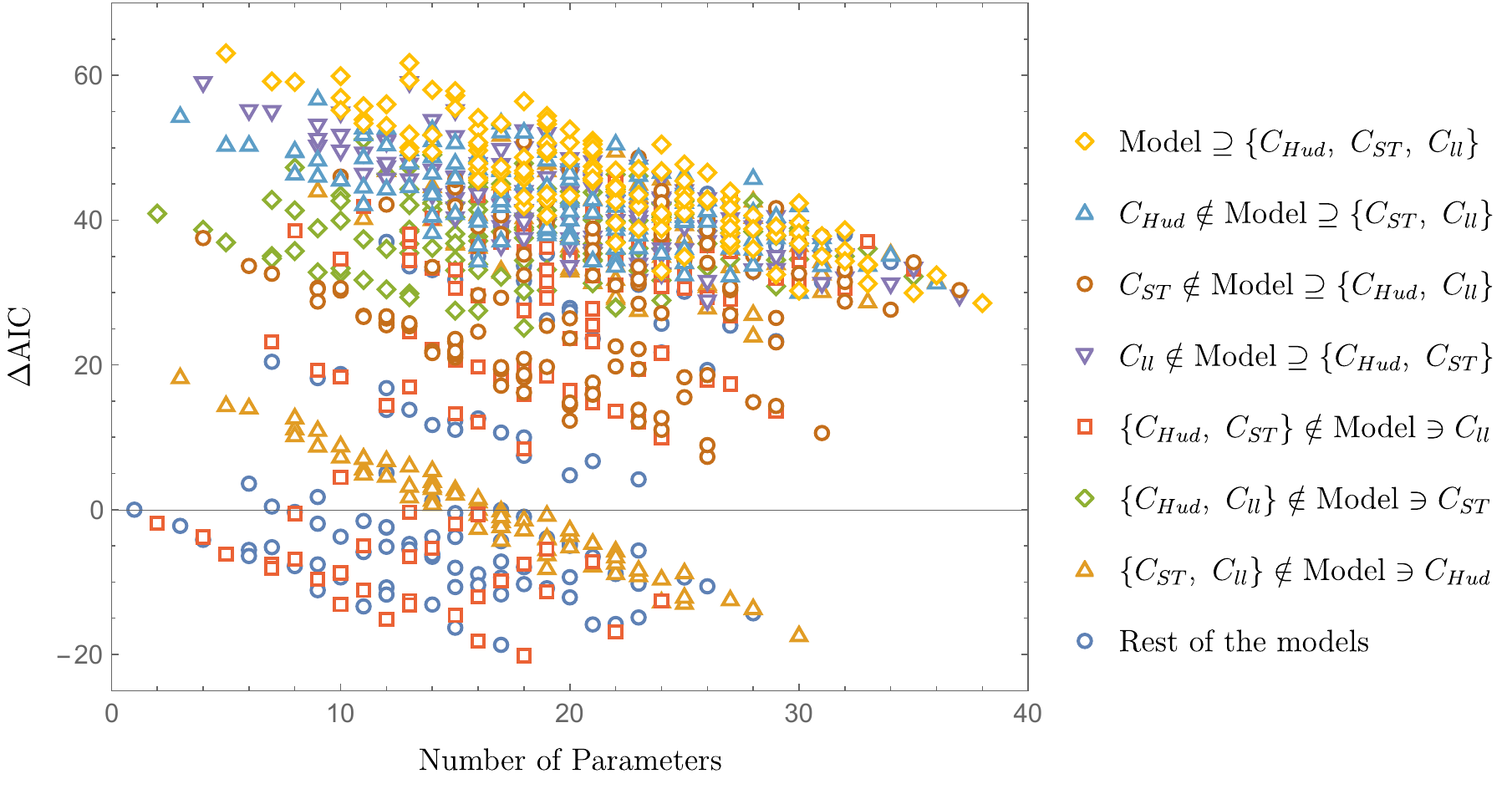}
 \vspace{-0.5cm}
	\caption{
 \textit{Including} the CDF $m_W$, we plot $\Delta$AIC for the 1024 models with respect to their corresponding number of parameters. The legend shows various categories of Wilson coefficients contained in the models, denoted by the corresponding markers.}
 \label{fig:AIC_CDF}
\end{figure}

\begin{figure}[t]
	\centering
    \includegraphics[scale=0.4]{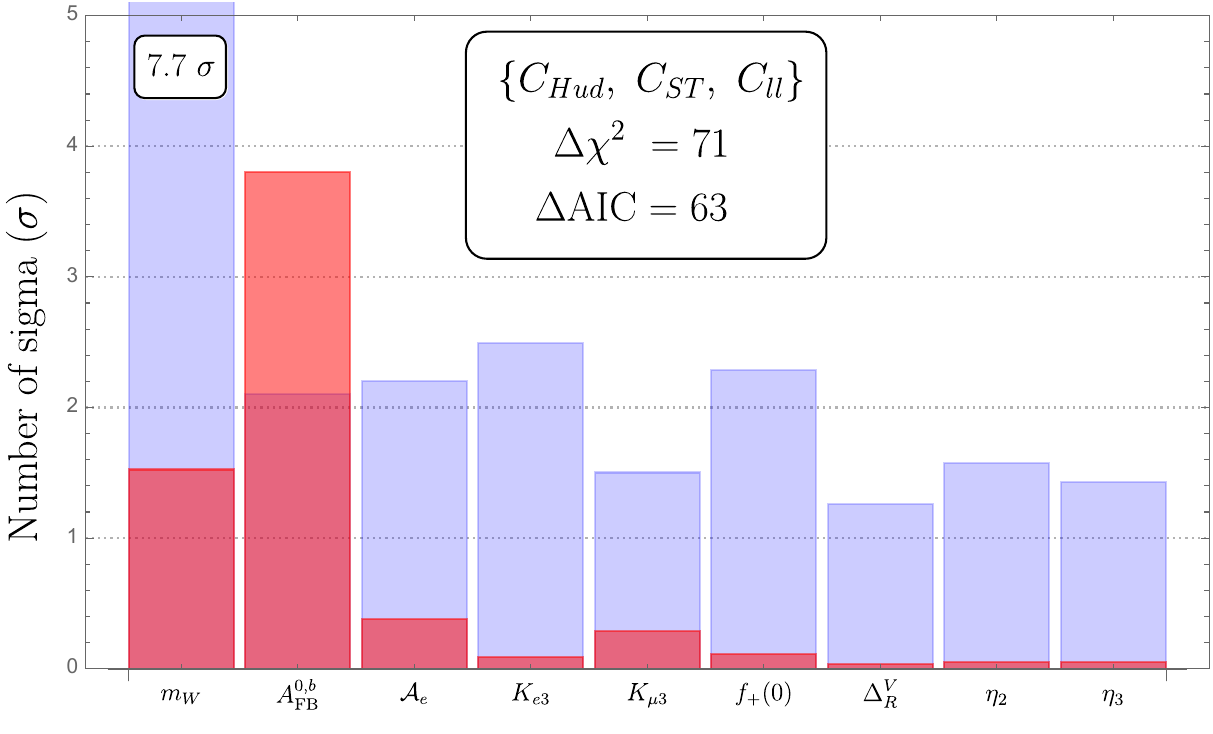}
    \includegraphics[scale=0.4]{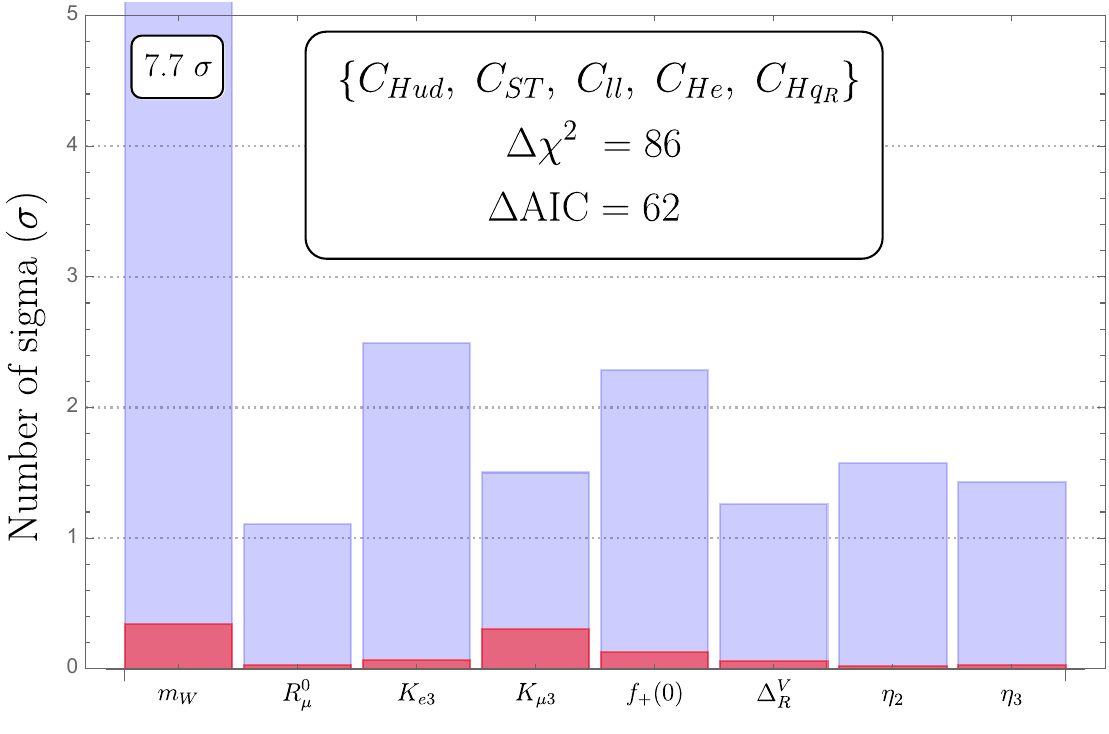}
	\caption{
 Same as Fig.~\ref{fig:ObsL2L6}, for the optimal fits with the CDF $m_W$. Left: Best model. Right: Second-best model.}
\label{fig:win_CDF}
\end{figure}

We now repeat our analysis including the CDF $m_W$. We plot $\Delta{\rm AIC}$ as a function of the number of parameters for all 1024 fits in Fig.~\ref{fig:AIC_CDF}. Although it is harder to identify distinct `branches' compared to the PDG results in Fig.~\ref{fig:AIC_PDG}, the same Wilson coefficients appear in the highly performing models: $C_{Hud}$, $C_{ST}$, and $C_{ll}$ (models that include these operators are denoted by yellow diamonds). In fact, the best model contains only these four operators with their best-fit values given by
\begin{align}
    C_{\substack{Hud \\ 11}} &= \phantom{-} 0.025 \pm 0.024 \; {\rm TeV}^{-2} \,,&\qquad 
    C_{\substack{Hud \\ 12}} &= -0.033 \pm 0.011 \; {\rm TeV}^{-2} \, ,\nnw
     C_{ST} &= -0.022 \pm 0.004 \; {\rm TeV}^{-2}\,,&\qquad 
    C_{\substack{ll \\ 2112}} &= -0.028 \pm 0.011 \; {\rm TeV}^{-2} \,.
\end{align}
This fit has a very high $\Delta \text{AIC}=63$. We show how it improves certain observables and matrix elements in the left panel of Fig.~\ref{fig:win_CDF}. In particular, the four operators can reduce the tension in $m_W$ to $1.5\sigma$ and take care of the problems in kaon and $0^+ \to 0^+$ decays. However, they increase the tension in forward-backward asymmetry of the bottom quark, $A^{0,b}_{FB}$. We also see this in Fig.~\ref{fig:Obs22} associated with the CLEW22 analysis.

Interestingly, the second-best model contains 12 independent Wilson coefficients. In addition to those four in the best model, it includes three RH lepton vertices ($C_{He}$) and five RH quark vertices ($C_{Hu}$ and $C_{Hd}$). Despite the inclusion of eight additional parameters, it achieves a similar $\Delta \text{AIC} = 62$. The best-fit values are given in the third column of Table~\ref{tab:ModelAVCDF}, and the improvements in observables are illustrated on the right of Fig.~\ref{fig:win_CDF}. Compared to those of the best-fit model on the left, we see that the high tension in $A^{0,b}_{FB}$ disappears due to the nonzero values of $[C_{He}]_{11}$ and $[C_{Hd}]_{33}$, which both have a significance level of $2\sigma$. This effect can be understood from Eqs.~\eqref{eq:Zee} and~\eqref{AFB_Chd}. Apart from the RH currents, this model somewhat resembles the $U(3)^5$ scenario we studied in Section~\ref{resultsU3} with $\Delta \text{AIC}=57$. In fact, the best-fit values are not inconsistent with $U(3)^5$, except for the appearance of $C_{Hud}$. Specifically, all $[C_{He}]_{ii}$ have a comparable central value. 

\begin{table}[t]
\centering
\begin{tabular}{|c|c|c|c|}
\hline 
& $\Delta$AIC = 63  & $\Delta$AIC = 62 & Model Average \\
\hline \hline
$[C_{Hud}]_{11}$ &  $\phantom{-}0.025 \pm 0.024$ & $\phantom{-}0.062 \pm 0.027$  & $\phantom{-}0.025 \pm 0.037$ \\
$[C_{Hud}]_{12}$ & $-0.033 \pm 0.011$ & $-0.028 \pm 0.011$  & $-0.032 \pm 0.016$\\
$C_{ST}$         & $-0.022 \pm 0.004$  &$-0.030 \pm 0.005$  & $-0.024 \pm 0.006$\\
$C_{ll}$         & $-0.028 \pm 0.011$ & $-0.046\pm 0.013$  & $-0.036 \pm 0.017$ \\
$[C_{He}]_{11}$  &  --  & $-0.016\pm 0.007$ & $-0.017 \pm 0.009$\\ 
$[C_{He}]_{22}$  &  --  & $-0.021 \pm 0.017$ & $-0.016 \pm 0.025$\\ 
$[C_{He}]_{33}$  &  -- &  $-0.035 \pm 0.018$ & $-0.032 \pm 0.020$ \\
$[C_{Hu}]_{11}$  &  -- & $0.13 \pm 0.77$ & $0.16 \pm 0.79$ \\ 
$[C_{Hu}]_{22}$  &  -- & $0.05 \pm 0.16$ & $0.06 \pm 0.17$\\
$[C_{Hd}]_{11}$  &  -- & $0.5 \pm 2.7$ & $0.7 \pm 2.9$\\ 
$[C_{Hd}]_{22}$  &  -- & $-0.6 \pm 1.3$ & $-0.6 \pm 1.3$\\ 
$[C_{Hd}]_{33}$  &  -- & $-0.29 \pm 0.12$ & $-0.34 \pm 0.19$ \\
\hline
\end{tabular}
\caption{Best-fit and $1\sigma$ ranges for the best model ($\Delta$AIC = 63), the next-to-best model ($\Delta$AIC = 62) and after model averaging, including the CDF $m_W$. Wilson coefficients are given in units of TeV$^{-2}$. }
\label{tab:ModelAVCDF}
\end{table}

The best model that does not contain all three categories of $C_{Hud}$, $C_{ST}$ and $C_{ll}$ has a $\Delta \text{AIC}=59$. It only contains $C_{Hud}$ and $C_{ST}$, corresponding to the leftmost purple triangle in Fig.~\ref{fig:AIC_CDF}. The best model without $C_{Hud}$ has $\Delta \text{AIC}=56$. It contains $C_{ST}$, $C_{ll}$, and six scalar four-fermion operators ($C_{ledq}$ and $C_{lequ}^{(1)}$). 

Models without $C_{ST}$ are notably inferior, lagging more than 10 units in $\Delta$AIC behind the best model, making them strongly disfavored. Models that do not contain any of the operators in the categories $C_{Hud}$, $C_{ST}$, or $C_{ll}$ perform even worse, as marked by the blue circles in Fig.~\ref{fig:AIC_CDF}.

We conclude that the most important operators are given by $C_{Hud}$, $C_{ST}$, and $C_{ll}$. Operators $C_{He}$, $C_{Hu}$, and $C_{Hd}$, while slightly less important, remain relevant. This is reflected in the model-averaged weights that are illustrated by the blue bars in Fig.~\ref{fig:weights}. The model-averaged best-fit values of these coefficients are given in the fourth column of Table~\ref{tab:ModelAVCDF}. The model-averaged results indicate a preference for nonzero values of $C_{ST}$ and, to a lesser extent, $C_{ll}$, $[C_{Hud}]_{12}$, $[C_{He}]_{11}$, $[C_{He}]_{33}$, and $[C_{Hd}]_{33}$. It is essentially a mixture of the best and second-best models.

\subsection{Opening the categories}
\label{sec:Open}

As mentioned in Section~\ref{sec:AIC_CDF}, the second-best model with the CDF $m_W$ has a $\Delta \text{AIC}$ very close to that of the best model. In addition to the four operators that appear in the best model ($[C_{Hud}]_{11,12}$, $C_{ST}$, and $C_{ll}$), eight operators are included in the categories of right-handed lepton and quark vertices, namely $[C_{He}]_{11,22,33}$, $[C_{Hu}]_{11,22}$, and $[C_{Hd}]_{11,22,33}$. This leads to an improvement in $\chi^2$ by 15. It is possible that not all of the eight are necessary. To investigate this, we open up the categories and consider all combinations of them, resulting in $2^{8} = 256$ models, while keeping $[C_{Hud}]_{11,12}$, $C_{ST}$, and $C_{ll}$ in every model by default.

In doing so, we actually find three models that perform better than our previous best model, all with $\Delta \text{AIC} \simeq 68$. None of these three models contains $C_{Hu}$. By including both $C_{Hu}$ and $C_{Hd}$ in category V, as defined in Table~\ref{tab:modelave}, we may have inadvertently introduced unnecessary complexity into some fits. Instead, the best three models all contain $[C_{He}]_{11}$ and $[C_{Hd}]_{33}$, which confirms our observation in Section~\ref{sec:AIC_CDF}, that these operators help reduce the tension in $A^{0,b}_{FB}$. Based on this analysis, the most important SMEFT operators are $[C_{Hud}]_{11,12}$, $C_{ST}$, $C_{ll}$, $[C_{He}]_{11}$, and $[C_{Hd}]_{33}$. 

\section{Falsifying explanations of the Cabibbo anomaly}
\label{sect:falsifying}

In Section~\ref{Sec:MA}, we concluded that the preferred explanation for CAA within SMEFT involves nonzero values of the RH CC coefficients $[C_{Hud}]_{11}$ and $[C_{Hud}]_{12}$. The CAA and CDF $m_W$ can be explained by $[C_{Hud}]_{12}$, $C_{ST}$, and $C_{ll}$ that deviate from zero by more than $2.5\sigma$. In this Section we discuss how these Wilson coefficient could lead to signals in other observables, which are currently less sensitive but may gain in sensitivity thanks to theoretical and experimental developments in the near future.
Some of these observables, such as nucleon axial coupling $g_A$ discussed in Section~\ref{sec:ga}, only receive contributions from the operators appearing in Table~\ref{tab:operators1} at the tree level in SMEFT. In this case, the CLEW setup discussed in this paper could be easily extended to incorporate new results as soon as they appear. 
For other observables, such as $K \rightarrow \pi\pi$, $h \rightarrow \gamma\gamma$, or collider observables that are quadratic in the RH coefficients, the operators that contribute to the CAA and the $m_W$ anomaly are only a subset of all possible SMEFT contributions. 
In these cases, a signal of the size predicted by the nonzero Wilson coefficients in Section~\ref{Sec:MA} would offer a tantalizing confirmation of the explanations identified so far. The absence of a signal would not necessarily rule out the CAA and $m_W$ anomalies, but would imply correlations with additional operators not considered in our analysis, such as $\Delta S = 1$ four-quark operators or Higgs-gauge couplings, thus providing additional guidance to model building.

\subsection{RH CC at low energy: $g_A$ and the $\Delta I =3/2$ $K\rightarrow \pi \pi$ amplitude}
\label{sec:ga}

\subsubsection*{$\boldsymbol g_A$:}
At low energy, the axial nucleon coupling $g_A$ is uniquely sensitive to RH CC \cite{Bhattacharya:2011qm,Alonso:2013hga}.
The value extracted from neutron decay can be related to the axial coupling in pure QCD via
\begin{equation}\label{eq:ga1}
 \frac{g_A}{g_V} = g^{\rm QCD}_A \left( 1 + \delta^{g_A}_{\rm RC} - v^2 \frac{\left[C_{Hud} \right]_{11}}{V_{ud}}\right),
\end{equation}
where $\delta^{g_A}_{\rm RC}$ denotes radiative corrections to the ratio of the axial and vector couplings.
A RH coupling of the $u$ and $d$ quarks to the $W$ boson of the size necessary to explain the CAA (as, for example, in Eq.~\eqref{RH_MA}) would shift $g_A$ by about $0.2\%$. Eq.~\eqref{eq:ga1} can thus provide a sensitive test of RH CC if the experimental value of $g_A/g_V$, $g_A$ in pure QCD and the radiative corrections to $g_A$ are all controlled at the permille level. $g_A/g_V$ is currently measured with $0.1\%$ precision~\cite{ParticleDataGroup:2022pth}
\begin{equation}
    \frac{g_A}{g_V} = 1.2754 \pm 0.0013, 
\end{equation}
where the error is inflated by a scale factor of 2.7, due to a discrepancy between different experimental results. The best measurement from PERKEO-III~\cite{Markisch:2018ndu} has an even smaller relative uncertainty of $0.4 \cdot 10^{-3}$. 
The FLAG average of $g^{\rm QCD}_A$ \cite{FlavourLatticeAveragingGroupFLAG:2021npn} is
\begin{align}
    g^{\rm QCD}_A &= 1.246 \pm 0.028 \,,\qquad N_f = 2 + 1 + 1 \,, \\
    g^{\rm QCD}_A &= 1.248 \pm 0.023 \,,\qquad N_f = 2 + 1 \,, 
\end{align}
with about $2\%$ error. The result of the CalLat collaboration reaches a subpercent uncertainty~\cite{Chang:2018uxx,Walker-Loud:2019cif} 
\begin{align}
g_A^{\rm QCD}  = 1.2711 \pm 0.0124\,,
\end{align}
with prospects of reaching $0.5\%$ uncertainty in the near term and possible $0.2\%$ in the exascale computing era~\cite{Chang:2018uxx,Walker-Loud:2019cif}.
The final theoretical input is the radiative correction $\delta_{\rm RC}^{g_A}$. Dispersive evaluations
of the vector and axial $W\gamma$ box contributions to $\delta^{g_A}_{\rm RC}$ estimated that the correction is well below $10^{-3}$~\cite{Hayen:2020cxh,Gorchtein:2021fce}.
In addition, $g_A$ gets contributions from hard photon exchanges between hadrons (`three-point function'). These have been estimated to be at the level of a few percent using chiral perturbation theory ($\chi$PT) techniques~\cite{Cirigliano:2022hob}, but are affected by large uncertainties. Ref.~\cite{Cirigliano:2022hob} quoted
\begin{equation}
    \delta^{g_A}_{\rm RC}   \in \{ 0.014, 0.026\}, 
\end{equation}
with the range determined by an unknown combination of electromagnetic low-energy constants, which was estimated by varying the renormalization scale in LO $\chi$PT loops, and by the uncertainties in the couplings of the axial current to two nucleons and a pion that affect NLO corrections to $\delta^{g_A}_{\rm RC}$. For $g_A$ to provide a stringent test of the CAA, it is therefore necessary to improve the theoretical calculations of both  $g_A^{\rm QCD}$ and 
$\delta^{g_A}_{\rm RC}$.

\subsubsection*{$\boldsymbol K \rightarrow \pi \pi$:}
In addition to semileptonic processes, $C_{Hud}$ also induces nonleptonic four-fermion operators. At low energy, one finds~\cite{Cirigliano:2016yhc}
\begin{equation}
\label{eq:Leff2}
{\cal L}_{\rm LEFT} = - \sum^{2}_{a=1} \,\left(  C^{ij\, lm}_{a\, LR} \mathcal O^{ij\, lm}_{a\, LR} + C^{{ij\, lm}\,*}_{a\, LR} \big(\mathcal O^{ij\, lm}_{a\, LR}    \big)^\dagger  \right)\ ,      
\end{equation}
with the operators given by
\begin{eqnarray}\label{eq:4q1}
\mathcal O^{ij\, lm}_{1\, LR} = \bar d^m \gamma^\mu P_L u^l \, \bar u^i \gamma_\mu P_R d^j\ , \qquad  \mathcal O^{ij\, lm}_{2\, LR}  =\bar d_\al^m   \gamma^\mu P_L u_\bt^l \, \bar u_\bt^i  \gamma_\mu P_R d_\al^j\ ,
\end{eqnarray}
where $\al,\,\bt$ are color indices.
Taking into account the QCD renormalization group evolution, the matching coefficients at low energy are given by
\begin{equation}
    C^{ij l m}_{1\, LR}(\mu = 3 \, {\rm GeV})   = 0.9 V_{l m}^* \left[C_{Hud}\right]_{i j}, \qquad C^{ij l m}_{2\, LR}(\mu = 3 \, {\rm GeV})  = 0.4 V_{l m}^* \left[C_{Hud}\right]_{i j}.\end{equation}
The imaginary part of the Wilson coefficients in Eq.~\eqref{eq:4q1} contributes to electric dipole moments (EDM) and to CP violation in mesonic decays. As a result, the imaginary part of $\left[C_{Hud}\right]_{11}$ and $\left[C_{Hud}\right]_{12}$ are strongly constrained, as we shall see below. 
The real part of the $\Delta S=1$ operators affects the decay $K \rightarrow \pi \pi$. In particular, the amplitude $A_2$ for the decay into two pions with isospin $I=2$ is suppressed in the SM and is therefore particularly sensitive to BSM corrections. 
The RH CC corrections to $A_2$ can be expressed as \cite{Cirigliano:2016yhc}
\begin{equation}
    A_2 = A^{\rm SM}_2 + \frac{F_0}{2\sqrt{6}}\,\bigg[\big(C_{1LR}^{udus}-C_{1LR}^{usud^*}\big)\mathcal A_{1\, LR}
+\big(C_{2LR}^{udus}-C_{2LR}^{usud^*}\big)\mathcal A_{2\, LR}\bigg]. \label{eq:A2expr}
\end{equation}
The LEC $\mathcal A_{1\, LR}$ and $\mathcal A_{2\, LR}$ can be extracted from Lattice QCD calculations of $\epsilon^\prime/\epsilon$ ~\cite{Blum:2012uk,Cirigliano:2016yhc,RBC:2015gro,Blum:2015ywa,RBC:2020kdj}
\begin{eqnarray}
\mathcal A_{1\, LR}(3\, \textrm{GeV}) &=& \frac{1}{\sqrt{3} F_0}  \, \langle (\pi\pi)_{I=2}|\mathcal  Q_7|K^0\rangle
 + \mathcal O\left(m^2_K\right) \simeq (2.2\pm0.13) \, {\rm GeV}^2\ ,\nn\\
\mathcal A_{2\, LR}(3\, \textrm{GeV}) &=& \frac{1}{\sqrt{3} F_0}  \, \langle (\pi\pi)_{I=2}|\mathcal  Q_8|K^0\rangle
 + \mathcal O\left(m^2_K\right)\simeq (10.1\pm0.6) \, {\rm GeV}^2\ ,
\label{LEC}
\end{eqnarray}
where $\mathcal Q_{7,8}$ denote electroweak penguin operators. 
With these values of the LECs, we find
\begin{equation}\label{LECs}
 \textrm{Re} (A_2) = \textrm{Re}(A^{\rm SM}_{2} ) + \left[ 11 \cdot 10^{-8} \, {\rm GeV} \right] {\rm Re}  \left( V_{us}^*\, \left[C_{Hud}\right]_{11} - V_{ud}^*\, \left[C_{Hud}\right]_{12}   \right) {\rm TeV}^2  \,,\end{equation}
where $\textrm{Re}(A_2) = 1.479(4) \cdot 10^{-8}$ GeV. 
Due to the chiral enhancement of the LR operators, the suppression of the SM amplitude, and a further enhancement by $V_{us}^{-1}$, values of $[C_{Hud}]_{12}$ that can explain the CAA can provide a $\sim 30\%$ correction to $\textrm{Re}(A_2)$. 
As in the case of $g_A$, the sensitivity of $K \rightarrow \pi\pi$ to new physics is limited by the theoretical prediction for the SM value. The RBC/UKQCD collaboration~\cite{Blum:2015ywa,RBC:2015gro} reported 
\begin{equation}
    \textrm{Re}(A^{\rm SM}_{2} ) = 1.50 (4)_{\rm stat} (14)_{\rm syst} \times 10^{-8}\, {\rm GeV}.
\end{equation}

\begin{figure}
\centering
\includegraphics[width=0.45\textwidth]{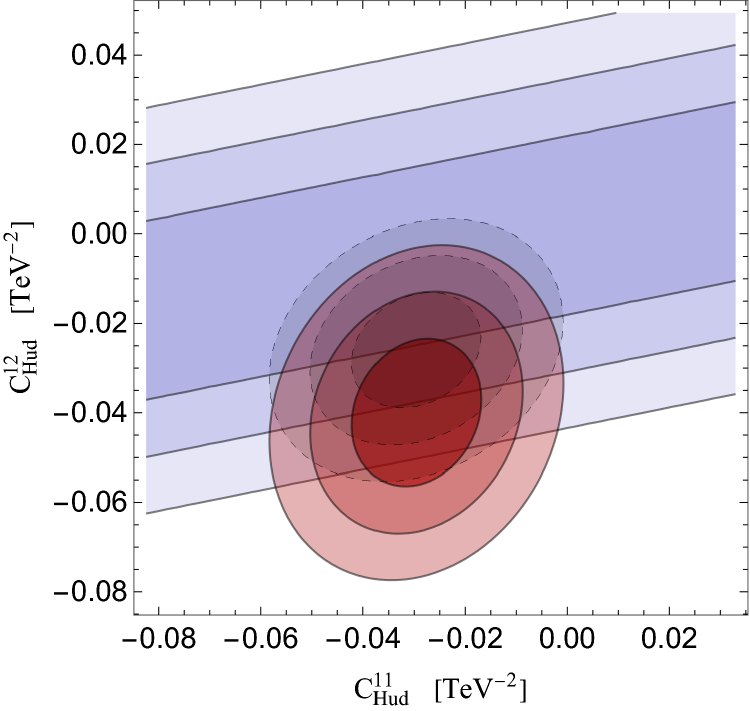}\hfill
\includegraphics[width=0.47\textwidth]{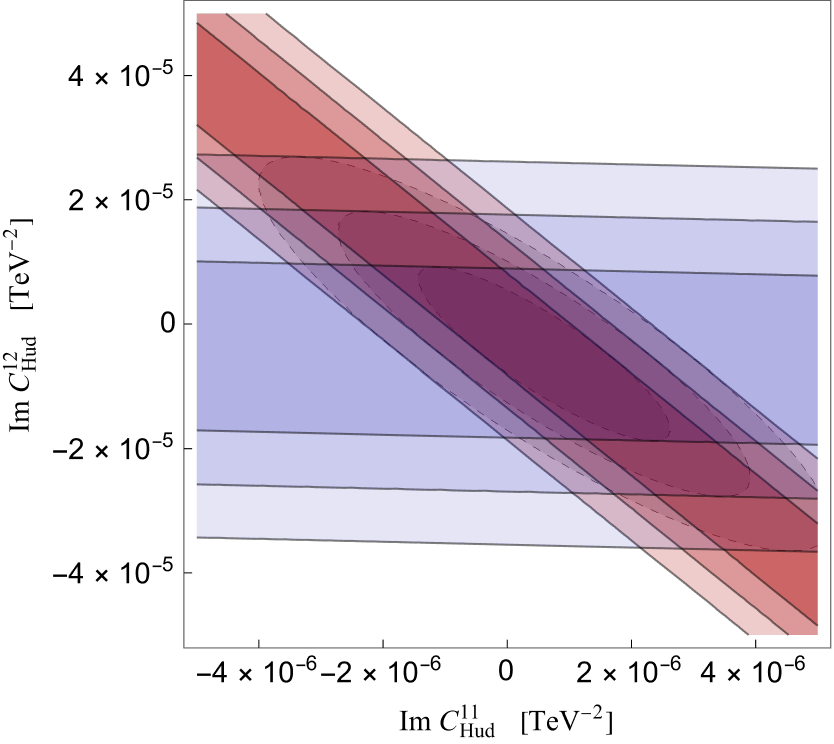}
\caption{
Left panel: the figure shows the L2(RH) scenario with the fit to L observables shown in red, while the constraint from $A_2$ is shown in blue. The combination of the two is depicted by the dashed black lines. Right panel: the constraints from EDMs (red) and $\varepsilon'/\varepsilon_K$ (blue), as well as their combination (black, dashed), on the imaginary parts of the $C_{Hud}$ couplings.
}\label{fig:Kpipi}
\end{figure}

In Fig.~\ref{fig:Kpipi} we repeat the L2(RH) fit, including $K \rightarrow \pi\pi$ with the assumption that $C_{Hud}$ is the only source of $\Delta S=1$ operators. The red ellipses are the results of the fit in Section~\ref{22fit}, the constraint from $A_2$ is shown by the blue bands, and the black ellipses denote the joint fit. 
We can see that the regions preferred by the fits to $\beta$ and kaon decays and the constraints from $A_2$ are compatible at the $1\sigma$ level. The joint fit gives 
\begin{equation}
    [C_{Hud}]_{11} = -0.030\pm 0.0084\,, \qquad  [C_{Hud}]_{12} = - 0.026 \pm 0.0085\,.
\end{equation}
The addition of $A_2$ to the fit currently somewhat shifts the best fit point of the L2(RH) scenario discussed in Section~\ref{22fit} (see Table~\ref{tab:L8}). The preferred value of $\left[C_{Hud}\right]_{12}$ is most affected by $A_2$, although the shift is not greater than $\sim 1\sigma$. Future improvements in the lattice determination of $A_2^{\rm SM}$ could provide a more sensitive probe of the RH couplings.

\paragraph{EDMs and $\epsilon^\prime/\epsilon$:}
The phases of the RH CC coefficients $[C_{Hud}]_{11}$ and $[C_{Hud}]_{12}$ induce tree-level corrections to the neutron EDM, to atomic EDMs, and to direct CP-violation in kaon decays ($\epsilon^\prime/\epsilon$), through the non-leptonic operators $\mathcal O_{1\, LR}$ and $\mathcal O_{2\, LR}$. These contributions were studied in Refs.~\cite{Cirigliano:2016yhc,Alioli:2017ces}, and lead to very strong constraints, shown in the right panel of Fig.~\ref{fig:Kpipi}. The constraints on the imaginary parts of $[C_{Hud}]_{11}$ and $[C_{Hud}]_{12}$ can be naively translated into scales in the 250 to 500 TeV range, much larger than the scales associated with the real part. 

Similarly, the CP-violating partner of $Q_{HWB}$ contributes to the electron EDM at one loop. In a single coupling scenario, its coefficient is restricted to below $3 \times 10^{-6}\,\text{TeV}^{-2}$ (95\% CL)~\cite{Cirigliano:2019vfc,Roussy:2022cmp}. 
While the combination of right-handed charged currents and oblique corrections provides an attractive explanation for the CAA and tensions in EWPO, when matching to concrete UV-complete models, some care must be taken to ensure that their phases are aligned with the SM.

\subsection{Collider signatures}

$C_{HWB}$ and $C_{HD}$ are degenerate in EWPO. The degeneracy is broken in Higgs observables, $WZ$, and $WW$ production data~\cite{Ellis:2020unq,Ethier:2021ydt,Ethier:2021bye,Bagnaschi:2022whn}.
With new and more precise data from the LHC, Higgs properties and diboson production are becoming highly competitive with EWPO~\cite{Ellis:2020unq}. For example, 
a SMEFT fit to $H \rightarrow \gamma \gamma$ data~\cite{ATLAS:2022tnm} yields, in the single coupling assumption, 
\begin{equation}
    C_{HWB}  = 0.0020^{+0.0044}_{-0.0042} \, \, {\rm TeV}^{-2}, \qquad C_{HD}  = -0.21^{+0.42}_{-0.44} \, \, {\rm TeV}^{-2}.
\end{equation}
These can be translated values of $C_{ST}$ 
\begin{equation}\label{eq:SThiggs}
 \left. C_{ST} \right|_{HWB} = 0.0018^{+0.0040}_{-0.0038} \, \, {\rm TeV}^{-2}, \qquad \left. C_{ST}\right|_{HD} = -0.087^{+0.17}_{-0.18} \, \, {\rm TeV}^{-2}.
 \end{equation}
The relatively weak sensitivity to $C_{HD}$ implies that, even under the single coupling assumption, $h \rightarrow \gamma\gamma$ is not sufficient to exclude the $C_{ST}$ explanation for $m_W$. Eq.~\eqref{eq:SThiggs} however illustrates the potential of Higgs measurements to provide constraints that are orthogonal to EWPO. 
In addition to $C_{HWB}$ and $C_{HD}$, the quark bilinear operators in Table~\ref{tab:globalWC} give contributions to Higgs and diboson production that are enhanced with respect to the SM and grow with energy. In particular, $C_{Hq}^{(3)}$ contributes to the production of $HW$ and $WZ$, $C_{Hq}^{(1)}$, $C_{Hq}^{(3)}$,
$C_{Hu}$ and $C_{Hd}$ to $HZ$ and $WW$.
Going forward, it will be important to perform combined fits to EWPO, Higgs data, and diboson production at the LHC~\cite{Ellis:2020unq,Ethier:2021ydt,Ethier:2021bye,Bagnaschi:2022whn}. It will be desirable to minimize the flavor assumptions and to combine these fits with Drell-Yan and low-energy data, in particular $\beta$ decay and parity-violating electron scattering data.  

Higgs and diboson data are also sensitive to the right-handed current operator $C_{Hud}$, although in quadratic order. $C_{Hud}$ is not strongly constrained by the charged-current Drell-Yan processes because it gives rise to corrections with the same energy dependence as the SM background. On the other hand, $C_{Hud}$ gives large corrections to $WH$ and $WZ$ associated production.
In the former case, contact interactions between two quarks, one Higgs, and a $W$ induce corrections to $WH$ that are enhanced by $s/m_W^2$ compared to the SM. In $WZ$ production, the presence of a right-handed current affects a cancellation between the $t$- and $s$-channel diagrams in the SM and also leads to corrections that increase in energy as $s/m_W^2$. 

Corrections to the signal strengths of $WH$ production were discussed in Ref.~\cite{Alioli:2017ces,Alioli:2018ljm} and can be written as
\begin{equation}
    \mu_{WH} = \frac{\sigma_{W^+ H} +\sigma_{W^- H}}{\sigma^{\rm SM}_{W^+ H} +\sigma^{\rm SM}_{W^- H}} = 1 + \sum_{i j} a_{ij} \left| v^2 C_{\substack{Hud\\ ij}}\right|^2.
\end{equation}
At NLO in QCD the coefficients $a_{ij}$ are \cite{Alioli:2017ces}
\begin{align}
    a_{11}(13\,  {\rm TeV}) &= 1.6(1) \cdot 10^2\,,  \qquad  a_{11}(14\,  {\rm TeV}) = 1.7(1) \cdot 10^2\,, \nnw
    a_{12}(13\,  {\rm TeV}) &= 0.9(2) \cdot 10^2\,,  \qquad  a_{12}(14\,  {\rm TeV}) = 1.0(1) \cdot 10^2\,,
\end{align}
where the error comes from PDF and scale uncertainties.
The latest results from the ATLAS and CMS collaborations are
\cite{ATLAS:2022vkf,CMS:2022dwd}
\begin{equation}
    \left. \mu_{WH}(13\, {\rm TeV}) \right|_{\rm ATLAS} = 1.2 \pm 0.2\,,
    \qquad \left. \mu_{WH}(13\, {\rm TeV}) \right|_{\rm CMS} = 1.4 \pm 0.3\,,
\end{equation}
leading to 
\begin{equation}
    \left| [C_{Hud}]_{11} \right| < 0.95 \,\, {\rm TeV}^{-2}\,, \qquad   \left|  [C_{Hud}]_{12} \right| < 1.3 \,\, {\rm TeV}^{-2}\,.
\end{equation}
These limits are about one order of magnitude too weak to constrain the region preferred by the CAA in Eq.~\eqref{RH_MA}. 
As the scaling is quadratic in the SMEFT coefficients, measurements of the signal strength alone will not be sufficient to competitively constrain $C_{Hud}$. However, the enhancement of the SMEFT corrections is more pronounced at high Higgs or $W$ transverse momentum as well as large $HW$ invariant mass, so that dedicated high $p_T$ measurements could further constrain right-handed operators~\cite{Alioli:2017ces,Alioli:2018ljm}. 

For $WZ$, we calculated the cross section by extending the \texttt{POWHEG} implementation of $WZ$ production in the SM~\cite{Melia:2011tj} to right-handed $W$ couplings.
Corrections to the inclusive cross section are a factor of ten smaller compared to $WH$. For example, at 13 TeV 
\begin{equation}
    \mu_{WZ} = \frac{\sigma_{W^+ Z} +\sigma_{W^- Z}}{\sigma^{\rm SM}_{W^+ Z} +\sigma^{\rm SM}_{W^- Z}} = 1 + 19 \left[ v^2 C_{Hud}\right]_{11}^2 + 9 \left[ v^2 C_{Hud}\right]^2_{12}.
\end{equation}
However, the absolute cross section is larger, and there exist precise measurements of the total cross section and differential measurements at high transverse momentum or invariant mass~\cite{CMS:2021icx}, where the contribution of RH CCs is enhanced. 
At the High-Luminosity LHC (HL-LHC), couplings of the size $\left[C_{Hud}\right]_{12} \sim 0.1$ TeV$^{-2}$, which are relevant to the Cabibbo anomaly, will generate hundreds of events with $M(WZ) \gtrsim 1$ TeV, so that at least part of the parameter space identified in Section~\ref{Sec:MA} will be probed. 
As RH CCs affect Higgs observables and diboson production at $\mathcal O(\Lambda^{-4})$, the derivation of consistent bounds requires the inclusion of genuine dimension-eight operators~\cite{Corbett:2023yhk}. Should deviations from the SM be observed, differential observables such as angular distributions could offer valuable insights into the chiral structure of the SMEFT operators~\cite{Alioli:2017ces,Alioli:2018ljm}.

\section{Conclusions}
\label{sect:conclusion}

In this work, we performed a model-independent SMEFT global analysis to investigate potential BSM explanations of the Cabibbo Angle Anomaly. The SMEFT framework offers a systematic approach to analyze experimental data globally, but its practical implementation is challenging due to the vast number of dimension-six operators and the corresponding effective couplings. To manage this, flavor assumptions are often invoked to simplify the analysis. However, such assumptions can reintroduce model dependence and miss certain BSM explanations.

To avoid this,  as discussed in Section~\ref{sect:strategy}, we embrace a `flavor-assumption-independent' analysis, which implements an approximate decoupling of the global fit into flavor-conserving and flavor-changing sectors, corresponding to an approximate factorization of the likelihood function.
Therefore, we expect that the results we obtain in the flavor-conserving sector would not be significantly changed in a truly global analysis that explicitly includes all `cross talk' generated by Wilson coefficients that contribute to multiple classes of observables (for example, the set of observables considered here and FCNCs).

Observables across three key data sets (dubbed CLEW) that are relevant to the CAA were analyzed: low-energy charged-current semileptonic processes (L), electroweak-precision observables (EW), and collider data of the Drell-Yan tails at the LHC (C). 
We performed fits by minimizing the $\chi^2$ and used the Akaike Information Criterion (AIC) to determine the relative quality of the resulting fits.
The AIC penalizes models for excessive complexity and ensures a balance between goodness of fit and simplicity when selecting models. 
Before considering the model-independent analysis outlined above, we first revisited a scenario that does make flavor assumptions, namely, a $U(3)^5$ scenario. We subsequently considered a case that includes all 22 operators contributing to the L observables (see the right panel of Table~\ref{tab:globalWC}), before finally moving to the global analysis with 37 Wilson coefficients (see the left panel of Table~\ref{tab:globalWC}).
The salient features of our results can be summarized as follows:

\begin{itemize}

\item 
As discussed in Section~\ref{resultsU3}, in the $U(3)^5$ scenario it is impossible to separate the EWPO from both low- and high-energy CC observables because of their interdependence on similar operators and their comparable sensitivities to BSM physics. Focusing only on fitting to the EWPO can result in unacceptably large BSM effects in low-energy decay processes, which cannot be offset by semileptonic four-fermion operators without upsetting collider constraints. Therefore, a consistent analysis should include both sets of observables. The traditional set of EWPO as considered in the literature is no longer sufficient, and we advocate always considering the combination with low- and high-energy CC observables.

\item 
For the intermediate flavor-symmetry-independent fit discussed in Section~\ref{22fit}, we explored the CAA without relying on flavor assumptions. 
The corresponding fit involves 22 potential Wilson coefficients, the results of which are provided in the Supplemental Material.
Although initially considering all operators, it became evident that many play an inconsequential role in alleviating the CAA, motivating analyses involving subsets of Wilson coefficients. The optimal fit (based on the highest $\Delta$AIC) involves only two RH CC operators. These SMEFT operators are strongly suppressed when minimal flavor violation is employed, illustrating the risk of flavor assumptions.
Including LH vertex corrections can slightly improve the description of the EWPO, but results in a lower $\Delta$AIC. Combining RH currents with scalar/pseudoscalar four-fermion operators offers a fit of similar quality. Other combinations without RH currents might reduce $\chi_{\rm min}^2$, but often lead to an AIC similar to or worse than that of the SM. The more general scenario involving 22 operators can address low-energy new physics phenomena such as the CAA. However, when including the CDF measurement of the W mass, the performance of CLEW22 is not optimal. Although it improves the AIC compared to the SM, it does retain a 3$\sigma$ tension in $m_W$ and negatively impacts other observables. Addressing anomalies, such as the W mass, requires incorporating additional operators.

\item 
The model-independent CLEW37 scenario is not only able to address the CAA, but can also successfully remove the tension due to the CDF W mass. Details of these fits are also provided in the Supplemental Material.
The resulting AIC shows a clear improvement over the SM when including the CDF $m_W$, as discussed in Section \ref{sec:global_37}. On the other hand, if one uses the PDG value of $m_W$, this scenario leads to an AIC that is worse than the SM. The negative effect of the number of parameters outweighs the improvement in $\chi^2$.

Inspired by this, we conducted an analysis focusing on specific SMEFT `models', each containing only a subset of the 37 Wilson coefficients. These subsets are systematically organized by the ten categories we outline in Table~\ref{tab:modelave}. Overall, we considered 1024 such models and compared their quality with that of the SM. 
As summarized in Fig.~\ref{fig:AIC_PDG}, we find that scenarios involving RH CC operators have large $\Delta$AICs, with the best model containing both $C_{Hud}$ and $C_{ST}$. If one includes the CDF W mass, the best scenario contains one additional operator, namely, the purely leptonic interaction $Q_{ll}$ (see Figs.~\ref{fig:AIC_CDF} and~\ref{fig:win_CDF} for details).

\item 
While the model with $C_{Hud}$, $C_{ST}$, and $C_{ll}$ has the optimal AIC, a more complicated model comes very close. This model contains SMEFT operators for the right-handed neutral currents $C_{He}$, $C_{Hu}$, and $C_{Hd}$. Although this brings eight additional fit parameters, and thus a severe penalty on the AIC, the fit improves several EWPO with respect to the optimal model (see Fig.~\ref{fig:win_CDF}). Furthermore, by opening the individual categories, we determined that the $C_{Hu}$ coefficients are not essential. However, incorporating $C_{He}$ and $C_{Hd}$ does result in an enhancement.

\item 
Finally, in Section~\ref{sect:falsifying} we discuss possible observables that could falsify or verify the BSM explanations of the CAA, assuming the current deviation from CKM unitarity is confirmed. Given the previous discussion, we focus on the most likely explanations in terms of right-handed currents $C_{Hud}$ and the oblique parameter $C_{ST}$. The former can be probed at low energies by the axial charge of the nucleon and the isospin $I=2$ channel in $K\to \pi\pi$. In both cases, the required improvements would have to come from the theory side. At the high-energy frontier, promising probes include $WH$ and $WZ$ production. Future measurements at the HL-LHC are expected to be able to rule out at least part of the parameter space that is relevant to solutions of the CAA.
Similarly, the HL-LHC will provide improved constraints on $C_{ST}$ through measurements of $h\to\ga\ga$ and diboson production. Incorporating these observables into the global analysis would require more operators to participate in the CLEW framework. We leave them for future investigations.

\item 
Our analysis can be extended in several directions. First, the SMEFT models discussed in Section~\ref{sec:global_37} depend on activating or deactivating operators in different categories.
A more detailed analysis would allow each individual operator to be turned on or off by itself. The price to pay is a huge increase in the total number of models. 
Another way to generalize our analysis is by including observables that currently have weaker sensitivity but could improve in the future, such as nonleptonic kaon decays, as well as Higgs and heavy-gauge boson production at high-energy colliders, as discussed in Section~\ref{sect:falsifying}. 
Finally, our analysis can be extended to include low-energy neutral current probes, such as measurements of parity violation in atoms or low-energy electron-proton scattering. 

\end{itemize}

Throughout this paper, we have consciously avoided any reference to UV completions in an effort to stay as close to the EFT philosophy as possible. However, the CAA, regardless of whether we include the CDF $m_W$ measurement or not, prefers nonzero values of $C_{Hud}$ and $C_{ST}$. Interestingly, both of these effective operators are generated in models that feature vector-like quarks, as seen in Ref.~\cite{Belfatto:2023tbv}. For example, TeV-scale vector-like quarks in the same representation as left-handed SM quarks, $(\bold{3},\,\bold{2},\,\frac{1}{6})$, can address both the CAA (by generating $C_{Hud}$ at tree level) and $m_W$ (by generating $C_{ST}$ at one-loop level)~\cite{Belfatto:2023tbv}.\\

In summary, we have performed a detailed analysis of the Cabibbo Angle Anomaly and the W-boson mass anomaly in the framework of the Standard Model Effective Field Theory. We have presented best-fit values for Wilson coefficients that lead to the most optimal solutions of these anomalies. Our results provide clear targets for model building. 
We hope that the analysis method presented in this work will prove useful for scrutinizing other (future) anomalies and their most likely beyond-the-Standard-Model explanations.

\section*{Acknowledgements} 
We thank Adam Falkowski and Mart\'{i}n Gonz\'alez-Alonso for providing us with the $\chi^2$ function
used in their work \cite{Falkowski:2020pma}. 
We express our gratitude to Lukas Allwicher for assistance with the \texttt{HighPT} package.  We acknowledge several stimulating discussions with T. Bhattacharya and J. Kumar. 
JdV acknowledges support from the Dutch Research Council (NWO) in the form of a VIDI grant.
VC and WD acknowledge support from the U.S. DOE under Grant No. DE-FG02-00ER41132.
EM acknowledges support from the U.S. DOE and from Los Alamos National Laboratory’s Laboratory Directed Research and Development program under project 20210190ER. Los Alamos National Laboratory is operated by Triad National Security, LLC, for the National Nuclear Security Administration of U.S. Department of Energy (Contract No. 89233218CNA000001). TT acknowledges support from the University of Siegen under the Young Investigator Research Group grant. TT is grateful to the Mainz Institute for Theoretical Physics (MITP) of the DFG Cluster of Excellence PRISMA+ (Project ID 39083149) for its hospitality and its partial support during the completion of this work. We acknowledge support from the DOE Topical Collaboration ``Nuclear Theory for New Physics", award No. DE-SC0023663.\\

\appendix  

\section{The Standard Model Effective Field Theory}\label{app:SMEFT}

We begin this appendix by recalling the main notation we adopt for the fermionic fields, the Higgs doublet and the covariant derivative.
We denote by $l^T = (\nu_L, e_L)$ and $q^T = (u_L, d_L)$ the left-handed lepton and quark $SU(2)$ doublets, 
while $u = u_R$, $d = d_R$, and $e = e_R$ are the right-handed up-type, down-type, and charged-lepton fields. 
We use $p,r,s,$ and $t$ for generation indices and work in a basis in which the electron and down-quark Yukawa matrices are diagonal. This implies that the fields $d_{L,R}$, $e_{L,R}$ correspond to the mass eigenstates, while for the up-type quarks we have $u_L = V^\dagger u_L^{\rm mass}$, where $V$ is the CKM matrix.

The Higgs doublet in the unitary gauge is given by 
\bea 
H = U(x)\bma 0\\  \frac{h(x)+v}{\sqrt{2}}\ema\,.
\eea
Here $h$ is the Higgs field, $U(x)$ encodes the Goldstone modes, and $v$ is the Higgs vacuum expectation value, which, in the absence of BSM effects, takes the value $v\simeq 246$ GeV. 
We use the following convention for the covariant derivative,
\bea
D_\mu &= \partial_\mu  + i \, g_1 Y B_\mu + i \, g_2 \, W_\mu^a \, t^a+ i g_3 G_\mu^A T^A\,,
\eea
where $B_\mu$, $W_\mu^a$, and $G_\mu^A$ ($g_1$, $g_2$, and $g_3$) are the gauge fields of $U(1)$, $SU(2)$, and $SU(3)_c$ (couplings). Furthermore, $Y$ is the hypercharge, while $t^a$ and $T^A$ represent the generators of $SU(2)$ and $SU(3)_c$ in the representation of the field where $D_\mu$ is acting on.

To obtain a prediction for $m_W$, we will use measurements of the Fermi constant (from muon decay), $G_F$, the $Z$ mass, $m_Z$, and the fine-structure constant, $\alpha_{em}$ as input parameters. Within SMEFT, the usual expressions for these observables obtain corrections from dimension-six operators, which will propagate to the predictions for other observables. In particular, we have~\cite{ALEPH:2005ab}~\footnote{In principle, the SMEFT involves additional operators, $Q_{H,H\Box,HW,HB,HG}$, that affect the gauge couplings, $g_{1,2,3}$ as well as the Higgs kinetic term and its vev. However, these effects can be captured by redefinitions of the gauge couplings and vev, and do not lead to measurable effects in the observables we consider. } 
\bea\label{eq:inputparams}
\Big[ G_F \Big]_{\rm SMEFT} &=& \Big[ G_F \Big]_{\rm SM} + \frac{1}{\sqrt{2}} \left(C^{(3)}_{\substack{Hl\\11}}+C^{(3)}_{\substack{Hl\\22}} -   C_{ll} \right) \,,\nn\\
\left[m_Z^2\right]_{\rm SMEFT} &=& \left[m_Z^2\right]_{\rm SM}+ \frac{1}{8} v^4 (g_1^2 + g_2^2) \, C_{HD} + \frac{1}{2} v^4 g_1 g_2 \, C_{HWB} \,,\nn\\
\left[\al_{em}\right]_{\rm SMEFT} &=& \left[\al_{em}\right]_{\rm SM} - \frac{v^2 g_1^3 g_2^3}{2 \pi (g_1^2 + g_2^2)^2} C_{HWB} \,,
\eea
where, at tree level, $\Big[ G_F \Big]_{\rm SM} = \frac{1}{\sqrt{2}v^2}$, $\left[m_Z^2\right]_{\rm SM} = \frac{g_1^2+g_2^2}{4}v^2$, and $ \left[\al_{em}\right]_{\rm SM}=\frac{1}{4\pi}\frac{g_1^2 g_2^2}{g_1^2+g_2^2}$.
The experimental values for the quantities in Eq.~\eqref{eq:inputparams} expressions are~\cite{ParticleDataGroup:2022pth}
\bea
G_F = 1.1663787(6)\cdot 10^{-5}\, {\rm GeV}^{-2}\,,\quad m_Z = 91.1876(21)\, {\rm GeV}\,,\quad \frac{1}{\al_{em}} = 137.035999180(10)\,.
\eea

\subsection{Translation to EWPO}\label{app:EWPOtransl}
The traditional EWPO observables mainly constrain the flavor-diagonal couplings of the $Z$ boson. In the mass basis, these can be written as 
\bea \label{eq:Z1}
\vL_{Z} =  g_Z\sum_{f=u,d,e,\nu}\bar f \g_\mu \left[ g_L^{(f)} P_L+g_R^{(f)}  P_R\right] f Z^\mu\,,
\eea
where $ g_Z = -2\, 2^{1/4}\sqrt{G_F}m_Z$. The couplings $g_{L,R}^{(f)} = \left[g_{L,R}^{(f)} \right]_{\rm SM}+\dt g_{L,R}^{(f)}$ are matrices in the flavor space. The SM contributions are diagonal in flavor, $\left[g_L^{(f)}\right]_{\rm SM} =\left[T_3-s_w^2 Q\right] \boldsymbol{1}$ and $\left[g_R^{(f)}\right]_{\rm SM} =-s_w^2 Q\, \boldsymbol{1}$, with $T_3$ and $Q$ denoting the third component of weak isospin and the electromagnetic charge of $f$, respectively\footnote{Our normalization is such that e.g.\ $T_3=1/2$ and $Q = 2/3$ for $f=u$.}. 
The shifts due to dimension-six operators are given by \cite{Berthier:2015oma}
\bea\label{eq:Z2}
\dt g_L^{(\ell)} &=& \left[g_L^{(\ell)}\right]_{\rm SM} \dt g_Z - \frac{v^2}{2}\left(C_{Hl}^{(1)}+C_{Hl}^{(3)}\right)-\dt s_w^2\,\boldsymbol{1}\,,\nn\\
\dt g_L^{(\nu)} &=& \left[g_L^{(\nu)}\right]_{\rm SM} \dt g_Z - \frac{v^2}{2}\left(C_{Hl}^{(1)}-C_{Hl}^{(3)}\right)\,,\nn\\
\dt g_L^{(u)} &=& \left[g_L^{(u)}\right]_{\rm SM} \dt g_Z - \frac{v^2}{2}V\left(C_{Hq}^{(1)}-C_{Hq}^{(3)}\right)V^\dagger+\frac{2}{3}\dt s_w^2\,\boldsymbol{1}\,,\nn\\
\dt g_L^{(d)} &=& \left[g_L^{(d)}\right]_{\rm SM} \dt g_Z - \frac{v^2}{2}\left(C_{Hq}^{(1)}+C_{Hq}^{(3)}\right)-\frac{1}{3}\dt s_w^2\,\boldsymbol{1}\,,\nn\\
\dt g_R^{(\ell)} &=& \left[g_R^{(\ell)}\right]_{\rm SM} \dt g_Z - \frac{v^2}{2}C_{He}-\dt s_w^2\,\boldsymbol{1}\,,\nn\\
\dt g_R^{(u)} &=& \left[g_R^{(u)}\right]_{\rm SM} \dt g_Z - \frac{v^2}{2}C_{Hu}+\frac{2}{3}\dt s_w^2\,\boldsymbol{1}\,,\nn\\
\dt g_R^{(d)} &=& \left[g_R^{(d)}\right]_{\rm SM} \dt g_Z - \frac{v^2}{2}C_{Hd}-\frac{1}{3}\dt s_w^2\,\boldsymbol{1}\,,
\eea
where $s_w^2=\sin^2\theta_w$
is expressed in terms of $G_F$, $m_Z$ and $\alpha_{em}$ as
\begin{equation}\label{eq:sw}
    s_w^2
= \frac{1}{2}\left[1-\sqrt{1-\frac{4\pi \al_{em}}{\sqrt{2} G_F m_Z^2}}\right],
\end{equation} and 
\bea
\dt g_Z &=& -\frac{v^2}{2}\left[\frac{1}{2} C_{HD} +C_{\substack{Hl\\11}}^{(3)}+C_{\substack{Hl\\22}}^{(3)}-\frac{1}{2}C_{\substack{ll\\2112}}-\frac{1}{2}C_{\substack{ll\\1221}}\right]\,,\nn\\
\dt s_w^2 &=& v^2\frac{c_w s_w}{s_w^2-c_w^2}\left[ C_{HWB}+ c_w s_w
\left(\frac{1}{2} C_{HD} +C_{\substack{Hl\\11}}^{(3)}+C_{\substack{Hl\\22}}^{(3)}-\frac{1}{2}C_{\substack{ll\\2112}}-\frac{1}{2}C_{\substack{ll\\1221}}\right)
\right]\,.
\eea
with $c_w = \cos\theta_w$.
In addition, several electroweak precision observables are sensitive to the (left-handed) couplings of the W boson, which can be written as
\bea
\vL_W =- \frac{g_W}{\sqrt{2}}\left[\bar u\g^\mu \left( g_L^{Wq} P_L+g_R^{Wq} P_R\right) d+\bar \nu \g^\mu \left( g_L^{W\ell} P_L\right) e\right] W^+_\mu +{\rm h.c.}
\eea
Here $g_W = \frac{\sqrt{4\pi \al_{em}}}{s_w}$ and
\bea
g_L^{Wq} &=&V\left[\left(1+ \frac{\dt s_w^2}{2 s_w^2}+\frac{v^2}{2}\frac{c_w}{s_w}C_{HWB}\right)\boldsymbol{1}+v^2 C_{Hq}^{(3)}\right]\,,\nn\\
g_R^{Wq} &=&\frac{v^2}{2} C_{Hud}\,,\nn\\
g_L^{W\ell} &=&\left(1+ \frac{\dt s_w^2}{2 s_w^2}+\frac{v^2}{2}\frac{c_w}{s_w}C_{HWB}\right)\boldsymbol{1}+v^2 C_{Hl}^{(3)}\,.
\eea
Finally, the expression for $m_W$ in SMEFT is given by
\begin{align}
	\frac{\delta m_W^2}{m_W^2} &= 
	v^2 \ \frac{s_w c_w}{s_w^2 - c_w^2} 
	\left[ 2 \, C_{H WB} + \frac{c_w}{2 s_w} \, C_{H D} + 
	\frac{s_w}{c_w} \left( 2 \, C_{H l}^{(3)} - \hat C_{ll} \right) \right].
\label{eq:mW2}
\end{align}

\subsection{Translation to low-energy basis}
\label{app:epstransl}
For the decays of kaon and pion, and $\beta$ decays, the often used low-energy effective Lagrangian is given by~\cite{Cirigliano:2009wk,Cirigliano:2012ab,Gonzalez-Alonso:2016etj,Falkowski:2020pma}
\bea
& & {\cal L}_{\beta} 
=
- \frac{G_F}{\sqrt{2} } \,  V_{uD} 
\, \Bigg[ \ 
\   \bar{\ell}  \gamma_\mu  (1 - \gamma_5)   \nu_{\ell}  
\Bigg( \Big( 1   +   \epsilon^{\ell D}_L - \epsilon_L^\mu\Big) \  
 \, \bar{u}   \gamma^\mu  (1 - \gamma_5)  D 
+ \epsilon^{\ell D}_R  
\, \bar{u}   \gamma^\mu  (1 + \gamma_5)  D \Bigg)
\nonumber\\
&& +        \bar{\ell} (1 - \gamma_5) \nu_{\ell}  \,  \left( \epsilon^{\ell D}_S \, \bar{u} D
+ \epsilon^{\ell D}_P \,  \bar{u} \gamma_5 D \right) 
+ \epsilon^{\ell D}_T    \  \   \bar{\ell}   \sigma_{\mu \nu} (1 - \gamma_5) \nu_{\ell}    \,  \bar{u}   \sigma^{\mu \nu} (1 - \gamma_5) D
\ \Bigg]+{\rm h.c.} \,,
\eea
where $D=\{d,s\}$.
The tree-level expressions for the $\epsilon$ couplings in terms of SMEFT coefficients are given by
\begin{subequations}
	\bea
	V_{uD} \,\epsilon_L^{\ell D} &=& v^2 \left[ V \, C_{ H q }^{(3)} \right]_{1D}  - v^2   \left[  V C_{  l q }^{(3)} \right]_{\ell \ell 1 D}  + v^2   V_{u D} \left[  C_{ H l }^{(3)} \right]_{\ell \ell}\,, \nn\\
&=& \frac{v^2}{2}\left[V C_{Hq}^{(d)}-C_{Hq}^{(u)}V\right]_{uD}-\frac{v^2}{2}\left[V C_{lq}^{(d)}-C_{lq}^{(u)}V\right]_{\ell\ell uD}  + v^2   V_{uD} \left[  C_{ H l }^{(3)} \right]_{\ell \ell}\nn\\
	V_{uD} \, \epsilon_R^{\ell D} &=&  \frac{v^2}{2} \left[ C_{H ud } \right]_{1D}
 \,,\nn\\
	V_{uD} \,\epsilon_S^{\ell D} &=&  - \frac{v^2}{2}   \left[  V C^{\dagger}_{  l e d q } +  C^{(1)\,\dagger}_{  l e q u } \right]_{\ell \ell 1 D} 
 = - \frac{v^2}{2}  \left[  V C^{\dagger}_{  l e d q } +  \bar{C}^{(1)\,\dagger}_{  l e q u } V \right]_{\ell \ell 1 D} 
 \,, \nn\\
	V_{uD} \,\epsilon_P^{\ell D} &=&  + \frac{v^2}{2}   \left[  V C^{\dagger}_{  l e d q } -    C^{(1)\,\dagger}_{  l e q u } \right]_{\ell \ell 1 D}  = + \frac{v^2}{2}  \left[  V C^{\dagger}_{  l e d q } -    \bar{C}^{(1)\,\dagger}_{  l e q u } V\right]_{\ell \ell 1 D}   \,,\nn\\
	V_{uD} \,\epsilon_T^{\ell D} &=&   - \frac{v^2}{2}   \left[   C^{(3)\,\dagger}_{  l e q u } \right]_{\ell \ell 1 D}  =  - \frac{v^2}{2}   \left[  \bar{C}^{(3)\,\dagger}_{  l e q u } V \right]_{\ell \ell 1 D}  \,.
	\label{eq:match}\eea
\end{subequations}
Finally, $ \epsilon_L^\mu$ arises from the SMEFT correction to $G_F$ as extracted from muon decay, see Eq.~\eqref{eq:inputparams}, and is given by
\bea
\epsilon_L^\mu   &=& v^2 \left[C_{\substack{H l\\ 11}}^{(3)}  + C_{\substack{H l\\ 22}}^{(3)}  - C_{\substack{l l\\ 2112}} \right]\,.
\eea

\section{Experimental and theoretical input to observables}\label{app:obs}

\subsection{Electroweak precision observables}\label{app:EWPO}

\begin{table}[t]
	\footnotesize	\centering
	\scalebox{0.9}{
		\begin{tabular}{|c|lc|lc||c|lc|lc|}
			\hline
			\hline
			Obs.\ &  \multicolumn{2}{c|}{Expt.\ Value} & \multicolumn{2}{c||}{SM Prediction} &Obs.\ &  \multicolumn{2}{c}{Expt.\ Value} &   \multicolumn{2}{|c|}{SM Prediction}   \\
			\hline
			\hline
			\ccg$\Gamma_Z\, ({\rm GeV})$ \ccg &\ccg 2.4955(23) &\ccg \cite{ALEPH:2005ab,Janot:2019oyi} &\ccg 2.49414(56) &\ccg \cite{deBlas:2021wap}&
			$m_W$(GeV)   & 80.4335(94)&\cite{CDF:2022hxs}  &  80.3545(42)&\cite{deBlas:2021wap} \\
			\ccg $\sigma_{\rm had}^0$(nb)&\ccg 41.480(33)	&\ccg \cite{ALEPH:2005ab,Janot:2019oyi}&\ccg  41.4929(53)&\ccg \cite{deBlas:2021wap}&
			$\Gamma_W\, ({\rm GeV})$ & 2.085(42)&\cite{ParticleDataGroup:2022pth}&2.08782(52)&\cite{deBlas:2021wap}\\
			\ccg $R^0_e$&\ccg  20.804(50)&\ccg \cite{ALEPH:2005ab,Janot:2019oyi} &\ccg 20.7464(63)&\ccg \cite{deBlas:2021wap} &
			$R_{Wc}$ &0.49(4)& \cite{ParticleDataGroup:2022pth}&0.50 &\\
			\ccg $R^0_\mu$&\ccg 20.784(34)&\ccg \cite{ALEPH:2005ab,Janot:2019oyi} &\ccg  &\ccg & 
			$R_{\sigma}$ &0.998(41)&\cite{CMS:2014mgj}&1&\\\cline{6-10}
			\ccg $R^0_\tau$&\ccg 20.764(45)&\ccg \cite{ALEPH:2005ab,Janot:2019oyi} &\ccg &\ccg & 
			\ccg Br$(W\to e\nu)$&\ccg 0.1071(16) & \ccg \cite{ParticleDataGroup:2022pth}&\ccg 0.108386(24)&\ccg \cite{deBlas:2021wap}\\
			\ccg $A^{0,e}_{\rm FB}$&\ccg 0.0145(25)&\ccg \cite{ALEPH:2005ab,Janot:2019oyi} &\ccg 0.016191(70)&\ccg \cite{deBlas:2021wap}& 
			\ccg Br$(W\to \mu\nu)$& \ccg 0.1063(15) &\ccg  \cite{ParticleDataGroup:2022pth}&\ccg 0.108386(24)&\ccg \cite{deBlas:2021wap}\\
			\ccg $A^{0,\mu}_{\rm FB}$&\ccg 0.0169(13)&\ccg \cite{ALEPH:2005ab,Janot:2019oyi} &\ccg &\ccg & 
			\ccg Br$(W\to \tau\nu)$&\ccg  0.1138(21)&\ccg  \cite{ParticleDataGroup:2022pth}&\ccg 0.108386(24)&\ccg \cite{deBlas:2021wap}\\\cline{6-10}
			\ccg $A^{0,\tau}_{\rm FB}$&\ccg 0.0188(17)&\ccg \cite{ALEPH:2005ab,Janot:2019oyi} &\ccg &\ccg & 
			$\frac{\Gamma(W\to \mu\nu)}{\Gamma(W\to e \nu)}$ &0.982(24)&  \cite{ParticleDataGroup:2022pth}& 1 &\\\cline{1-5}
			\ccgg $R_b^0$&\ccgg 0.21629(66)&\ccgg \cite{ALEPH:2005ab} &\ccgg 0.215880(19)&\ccgg \cite{deBlas:2021wap}& 
			$\frac{\Gamma(W\to \mu\nu)}{\Gamma(W\to e \nu)}$ & 1.020(19)& \cite{ParticleDataGroup:2022pth}&&\\
			\ccgg $R_c^0$&\ccgg 0.1721(30)&\ccgg \cite{ALEPH:2005ab} &\ccgg 0.172198(20)&\ccgg \cite{deBlas:2021wap}& 
			$\frac{\Gamma(W\to \mu\nu)}{\Gamma(W\to e \nu)}$ &  1.003(10)&\cite{ParticleDataGroup:2022pth}&&\\
			\ccgg $A^{0,b}_{\rm FB}$&\ccgg 0.0996(16)&\ccgg \cite{ALEPH:2005ab} &\ccgg 0.10300(23)&\ccgg \cite{deBlas:2021wap}& 
			$\frac{\Gamma(W\to \tau\nu)}{\Gamma(W\to e \nu)}$ &0.961(61)& \cite{ParticleDataGroup:2022pth}&&\\
			\ccgg $A^{0,c}_{\rm FB}$&\ccgg 0.0707(35)&\ccgg \cite{ALEPH:2005ab} &\ccgg  0.07358(18)&\ccgg \cite{deBlas:2021wap}& 
			$\frac{\Gamma(W\to \tau\nu)}{\Gamma(W\to \mu \nu)}$ & 0.992(13)& \cite{ParticleDataGroup:2022pth}&&\\
			\ccgg ${\cal A}_{c}$&\ccgg 0.67(3)&\ccgg \cite{ALEPH:2005ab} &\ccgg 0.66775(14)&\ccgg \cite{deBlas:2021wap}& 
			$A_4(0-0.8)$ & 0.0195(15)&\cite{ATLAS:2018gqq}& 0.0144(7)&\cite{Breso-Pla2021Mar}\\
			\ccgg ${\cal A}_{b}$&\ccgg 0.923(20)&\ccgg \cite{ALEPH:2005ab} &\ccgg 0.934727(25)&\ccgg \cite{deBlas:2021wap}&  
			$A_4(0.8-1.6)$ & 0.0448(16)&\cite{ATLAS:2018gqq}& 0.0471(17)&\cite{Breso-Pla2021Mar}\\
			\ccgg ${\cal A}_{e}$&\ccgg 0.1516(21)&\ccgg \cite{ALEPH:2005ab} &\ccgg 0.14692(32)&\ccgg \cite{deBlas:2021wap}& 
			$A_4(1.6-2.5)$ & 0.0923(26)&\cite{ATLAS:2018gqq}& 0.0928(21)&\cite{Breso-Pla2021Mar}\\
			\ccgg ${\cal A}_{\mu}$&\ccgg 0.142(15)&\ccgg \cite{ALEPH:2005ab} &\ccgg &\ccgg &  
			$A_4(2.5-3.6)$ & 0.1445(46)&\cite{ATLAS:2018gqq}& 0.1464(21)&\cite{Breso-Pla2021Mar}\\\cline{6-10}
			\ccgg ${\cal A}_{\tau}$&\ccgg 0.136(15)&\ccgg \cite{ALEPH:2005ab} &\ccgg &\ccgg  &
			\ccg $g_V^{(u)}$ &\ccg  0.201(112)&\ccg \cite{D0:2011baz} &\ccg  0.192&\ccg \cite{Baur2001Aug}\\\cline{1-5}
			\ccg ${\cal A}_{e}^{\tau {\rm \,pol}}$&\ccg 0.1498(49)&\ccg \cite{ALEPH:2005ab} &\ccg &\ccg & 
			\ccg $g_V^{(d)}$ &\ccg  -0.351(251)&\ccg \cite{D0:2011baz} &\ccg -0.347&\ccg \cite{Baur2001Aug} \\
			\ccg ${\cal A}_{\tau}^{\tau {\rm \,pol}}$&\ccg 0.1439(43)&\ccg \cite{ALEPH:2005ab} &\ccg &\ccg &
			\ccg $g_A^{(u)}$\ccg  &\ccg 0.50(11)&\ccg  \cite{D0:2011baz} &\ccg 0.501&\ccg \cite{Baur2001Aug}\\\cline{1-5}
			\ccgg ${\cal A}_s$ &\ccgg  0.895(91)&\ccgg \cite{SLD:2000jop}&\ccgg 0.935637(26)&\ccgg \cite{deBlas:2021wap}& 
			\ccg  $g_A^{(d)}$ &\ccg  -0.497(165)&\ccg \cite{D0:2011baz} &\ccg -0.502&\ccg \cite{Baur2001Aug}\\\cline{6-10}
			\ccgg $R_{uc}$ & \ccgg 0.166(9)&\ccgg \cite{ParticleDataGroup:2022pth}&\ccgg 0.172220(20)&\ccgg \cite{deBlas:2021wap}&
			&&&&\\
			\hline
			\hline
	\end{tabular}}%
	\caption{Input parameters and EWPOs used in the analysis. Each shaded block indicates a set of correlated observables. Entries without an explicit SM prediction share their SM value with the observable above.}
\label{tab:EWPO}%
\end{table}%

The observables used are listed in Table \ref{tab:EWPO}. The Table includes `traditional' EWPOs measured at LEP and SLC in the first column, while the second column involves several observables measured at hadron colliders. These additional constraints are needed when going beyond flavor-universal SMEFT scenarios. Observables that may be less familiar are defined by
\bea
{R}_{uc}  = \frac{\Gamma(Z\to \bar uu)+\Gamma(Z\to \bar cc)}{\sum_q \Gamma(Z\to \bar qq)}\,,\quad R_{Wc} = \frac{\Gamma(W\to cs)}{\Gamma(W\to ud)+\Gamma(W\to cs)}\,,\quad 
R_\sigma^2=\frac{\sigma_t}{\sigma_t^{\rm SM}}\,,
\eea
where $\sigma_t$ is the single-top production cross section in the $t$ channel. Furthermore, $g_{V,A}^{(q)} = g_L^{(q)}\pm g_R^{(q)}$ and $A_4(a-b)$ correspond to the forward-backward asymmetry in $pp\to \bar \ell\ell$ in the rapidity range $a\leq Y\leq b$ \cite{Breso-Pla2021Mar}.
For each observable, we include the leading tree-level SMEFT corrections up to $\Or(1/\Lambda^2)$. We follow Ref.~\cite{Berthier:2015oma} for the traditional EWPOs, while we use the results of \cite{Breso-Pla2021Mar} for the measurements of $A_4$. SMEFT corrections for the remaining observables depend on the modified couplings of $W$ and $Z$ and can be straightforwardly derived from their definitions.

\subsection{Low-energy charged-current observables}
\label{app:input}

\subsubsection{Neutron and nuclear $\beta$ decays}\label{app:beta}
We include measurements of $0^+\to 0^+$ transitions, neutron decay, as well as decay correlation measurements, following Ref.~\cite{Falkowski:2020pma}. For measurements of neutron decays, the observables include the lifetime, as well as several correlation coefficients, $\tilde A_n$, $\tilde B_n$, $ a_n$, $\tilde a_n$, and $\lambda_{AB}$. 
The included ${\cal F} t$ values from superallowed decays are listed
in Table~\ref{tab:superallowed}, while the correlation measurements used in the nuclei are collected in Table~\ref{tab:NuclearData}. For definitions of the observables, see Ref.~\cite{Falkowski:2020pma}. In addition to these observables, the fit includes data from measurements of mirror nuclei, as described in~\cite{Falkowski:2020pma}. However, the impact of these observables is small in the scenario we consider, while they are more important in models involving right-handed neutrinos. Finally, we include the measurement of the ratio of the axial and vector form factors in $\Lambda\to p e\nu_e$~\cite{ParticleDataGroup:2022pth}.

As mentioned in Section~\ref{sec:betaObs} the theoretical expressions do not just involve the SM and SMEFT parameters, but also depend on several hadronic and nuclear matrix elements. The relevant hadronic input for the SM consists of the vector and axial nucleon charges, $g_V$ and $g_A$, and radiative corrections, $\Delta_R^{V,A}$, while the contributions due to BSM interactions require the scalar and tensor charges, $g_S$, $g_T$. Pseudoscalar interactions are suppressed by $\Or(m_e \Lambda_\chi/m_\pi^2)$, and we neglect their contributions to $\bt$ decays.
We summarize these hadronic parameters in the first column of Table~\ref{tab:hadronic}. 

In addition, the fit includes several ratios of Gamow-Teller (GT) and Fermi (F) matrix elements, $\rho = \frac{g_A}{g_V}\frac{M_{GT}}{M_F}$, which appear in the mirror transitions mentioned above. Finally, Ref.~\cite{Falkowski:2020pma} included the parameters $\eta_1$ and $\eta_{2,3}$ to capture theoretical uncertainties due to radiative corrections $\dt_R'$, and contributions dependent on nuclear structure, $\dt_{NS}$, respectively. We treat $\eta_{1,2,3}$ as Gaussian with a mean of $\mu(\eta_{1,2,3})=0$ and $\sigma(\eta_{1,2,3})=1$, while we do not include any theoretical input of $\rho_i$ which is determined by the fit. 
\begin{table}[h]
\footnotesize
\begin{center}
\begin{tabular}{ ccccccccccccccccc}
\hline\hline
$^{10}$C   & 
$^{14}$O   & 
$^{22}$Mg  &
$^{26m}$Al &
$^{26}$Si & 
$^{34}$Cl  &
$^{34}$Ar  & 
$^{38m}$K  & 
$^{38}$Ca  & 
$^{42}$Sc  & 
$^{46}$V   & 
$^{50}$Mn  & 
$^{54}$Co  & 
$^{62}$Ga  & 
$^{74}$Rb  \\
\hline\hline
\end{tabular}
\end{center}
\caption{
The parent nuclei participating in superallowed decays for which we use their measured ${\cal F}t$ values in the fits, following Refs.~\cite{Falkowski:2020pma,Hardy2020Oct}.}
\label{tab:superallowed}
\end{table}

\begin{table}[h]
\begin{center}
\begin{tabular}{
l@{\hspace{5mm}} c @{\hspace{3mm}}  c@{\hspace{3mm}}  c@{\hspace{3mm}}  c@{\hspace{3mm}}
 r@{\hspace{5mm}}  
c@{\hspace{3mm}}  c 
}
\hline\hline
Parent		& $J_i$	& $J_f$	& Type	& Observable 
& Ref.
\\ \hline
$^6$He		& 0		& 1		& GT/$\beta^-$		& $\tilde{a}$	& \cite{Johnson:1963zza}
\\
$^{32}$Ar		& 0		& 0		&  F/$\beta^+$		& $\tilde{a}$	& \cite{Adelberger:1999ud} 
\\
$^{38m}$K	& 0		& 0		&  F/$\beta^+$		& $\tilde{a}$	& \cite{Gorelov:2004hv}
\\
$^{60}$Co		& 5		& 4		& GT/$\beta^-$		& $\tilde{A}$		& \cite{Wauters:2010gh} \\
$^{67}$Cu		& 3/2		& 5/2		& GT/$\beta^-$		& $\tilde{A}$	& \cite{Soti:2014xua} 
\\
$^{114}$In	& 1		& 0		& GT/$\beta^-$		& $\tilde{A}$	& \cite{Wauters:2009jw}
\\
$^{14}$O/$^{10}$C &	&		& F-GT/$\beta^+$	& $P_F/P_{GT}$ 	&  \cite{Carnoy:1991jd}
\\
$^{26}$Al/$^{30}$P &	&		& F-GT/$\beta^+$	& $P_F/P_{GT}$    & \cite{Wichers:1986es}
\\ \hline\hline
\end{tabular}
\end{center}
\caption{
Data from correlation measurements in pure Fermi and pure Gamow-Teller decays used in the fits. Table adapted from~\cite{Falkowski:2020pma}.}
\label{tab:NuclearData}
\end{table}

\begin{table}[h]
\footnotesize
\begin{center}
\begin{tabular}{ |llc|llc|}
\hline\hline
\multicolumn{3}{|c|}{Baryon decays} & \multicolumn{3}{c|}{$K$ and $\pi$ decays}\\\hline
$g_A $ &  $1.246(28)$ & \cite{FlavourLatticeAveragingGroupFLAG:2021npn} & $f_+(0)$ & $0.9698(17)$&\cite{FlavourLatticeAveragingGroupFLAG:2021npn} \\
$g_S$& $1.02(10)$ & \cite{FlavourLatticeAveragingGroupFLAG:2021npn} & $\frac{f_{K^\pm}}{f_{\pi^\pm}}$ & 1.1932(21)& \cite{FlavourLatticeAveragingGroupFLAG:2021npn}\\ 
$g_T $& $0.989(34)$ & \cite{FlavourLatticeAveragingGroupFLAG:2021npn}& $f_{K^\pm}$ &$155.7(3)\, {\rm MeV}$ & \cite{FlavourLatticeAveragingGroupFLAG:2021npn}\\
$\Delta_R^V$ & $0.02467(22)$& \cite{Seng:2018yzq}& RC$_\pi$ &$0.0332(3)$& \cite{Feng:2020zdc}\\
$\frac{\Delta_R^A-\Delta_R^V}{2}$ & $0.00013(13)$& \cite{Gorchtein:2021fce}& $c_1^\pi$ & $-2.4(5)$ & \cite{DescotesGenon:2005pw} \\
$g_1/f_1\Big|_{\Lambda\to p}$& $0.72(7)$& \cite{Gonzalez-Alonso:2016etj,QCDSFUKQCD:2010cha}&$\Delta_{\rm CT}$& $-0.0035(80)$ & \cite{FlaviaNetWorkingGrouponKaonDecays:2010lot}\\
&&&$\frac{B_T(0)}{f_+(0)}$&$0.68(3)$ & \cite{Baum:2011rm}\\
\hline\hline
\end{tabular}
\end{center}
\caption{The hadronic input that enters $\bt$ decays. We approximate $g_V=1$, which holds with high precision. Whenever appropriate, the value on a renormalization scale $\mu=2$ GeV is given.}
\label{tab:hadronic}
\end{table}

\subsubsection{Kaon and pion decays}\label{app:meson}

\begin{table}[h]
\footnotesize
\begin{center}
\begin{tabular}{ |lll|llc|}
\hline\hline 
\multicolumn{3}{|c}{$P_{l2}$} & \multicolumn{3}{|c|}{$K_{l3}$} \\ \hline
Observable & value & Ref. & Observable & value & Ref. \\ \hline
$ R_\pi$ &$ 1.2344(30)\cdot 10^{-4}$ &\cite{PiENu:2015seu} &$|\tilde V_{us}^{e}f_+^K(0)| $ & $ 0.21626(40)$ & \cite{Seng:2021nar}\\
$ R_K $ &$ 2.488(9)\cdot 10^{-5}$ & \cite{ParticleDataGroup:2022pth}&$|\tilde V_{us}^{\mu}f_+^K(0)| $ & $ 0.21667(52)$ & \cite{Seng:2021nar}\\
$ \Gamma(K_{\mu2}) $ &$ 5.134(10)\cdot 10^{-7}s^{-1}$ &\cite{ParticleDataGroup:2022pth}& $\log C $& $0.1985(70)$ & \cite{Moulson:2017ive}\\ 
$ {\rm Br}(\pi_{\mu2}) $ &$ 0.9998770(4)$ &\cite{ParticleDataGroup:2022pth}&$2\epsilon_T^{s\mu}\frac{B_T(0)}{f_+(0)}$& $0.0007(71)$ & \cite{Yushchenko:2003xz}\\
$ \tau_{\pi^+} $ &$ 2.6033(5)\cdot 10^{-8}s$ &\cite{ParticleDataGroup:2022pth}&&&\\
\hline\hline
\end{tabular}
\end{center}
\caption{
The experimental input used for kaon and $\pi$ decays. Note that the determinations of $|\tilde V_{us}^{\ell}f_+^K(0)|$ involve a correlation not shown in the table.}
\label{tab:mesons}
\end{table}

The experimental data used for the decays of pion and kaon are collected in Table~\ref{tab:mesons}. Apart from the two-body decay rates (the first column), we include determinations of $V_{us}^{\ell}f_+(0)$ from the $K\to \pi e\nu$ and $K\to \pi \mu\nu$ rates and measurements of the differential distributions (second column). The latter include a determination of the tensor interaction, $\epsilon_T^{s\mu}$, as well as a determination of $\log C$, which is sensitive to scalar interactions.
As $V_{us}^{\ell}f_+(0)$ has been determined assuming the SM form of the decay distribution, we follow~\cite{Gonzalez-Alonso:2016etj} to obtain the correlation between these rates and the tensor interaction.
Using the input from Ref.~\cite{Seng:2021nar}, shown in Table~\ref{tab:mesons}, we then obtain the following correlation matrix
\bea
\bma
|V_{us}^{e}f_+(0)|\\
|V_{us}^{\mu}f_+(0)|\\
-2\epsilon_T^{s\mu}\frac{B_T(0)}{f_+(0)}
\ema
=
\bma
0.21634(40)\\
0.21652(52)\\
0.0007(71)
\ema\,,\qquad 
{\rm Corr} = 
\bma 
1 & 0.57 & 0\\
0.57 & 1 & 0.44\\
0 & 0.44 & 1
\ema\,.
\eea

We follow~\cite{Gonzalez-Alonso:2016etj} for the theoretical expressions of the observables and summarize the relevant hadronic input in the second column of Table~\ref{tab:hadronic}. The two-body decays depend on the decay constants $f_{\pi,K}$, a low-energy constant, $c_1^\pi$, and $S_{ew} = 1.0232(3)$, which captures the SM short-distance corrections.
Instead, the three-body decays depend on the form factor $f_+(0)$, as well as the parameters $\Delta_{\rm CT}$ and $B_T$. The latter two parameters enter into the expression for the shape of the differential distributions that are sensitive to scalar and tensor interactions.
Finally, pion $\bt$ decay depends on radiative corrections, captured by the factor RC$_\pi$.

\subsection{Collider probes}\label{app:LHC}

\begin{table}[H]
	\centering
	{
		\renewcommand{\arraystretch}{1.5}
		\small
		\begin{tabular}{l | c c c c c}
			Process & Experiment & Lumi. & $x_\mathrm{obs}$ & $x$ & Ref.
			\\ \hline\hline
			$pp \to \tau^+\tau^-$ & ATLAS & $139\,\mathrm{fb}^{-1}$ & $m_T^{\rm tot}(\tau_h^1,\tau_h^2,\slashed{E}_T)$ & $m_{\tau\tau}$ & \cite{ATLAS:2020zms}
			\\ 
			$pp \to \mu^+\mu^-$ & CMS & $140\,\mathrm{fb}^{-1}$ & $m_{\mu\mu}$ & $m_{\mu\mu}$ & \cite{CMS:2021ctt}
			\\ 
			$pp \to e^+ e^-$ & CMS & $137\,\mathrm{fb}^{-1}$ & $m_{ee}$ & $m_{ee}$ & \cite{CMS:2021ctt}
			\\ \hline
			$pp \to \tau^\pm\nu$ & ATLAS & $139\,\mathrm{fb}^{-1}$ & $m_T(\tau_h,\slashed{E}_T)$ & $p_T(\tau)$ & \cite{ATLAS:2021bjk}
			\\ 
			$pp \to \mu^\pm\nu$ & ATLAS & $139\,\mathrm{fb}^{-1}$ & $m_T(\mu,\slashed{E}_T)$ & $p_T(\mu)$ & \cite{ATLAS:2019lsy}
			\\ 
			$pp \to e^\pm\nu$ & ATLAS & $139\,\mathrm{fb}^{-1}$ & $m_T(e,\slashed{E}_T)$ & $p_T(e)$ & \cite{ATLAS:2019lsy}
			\\ \hline
			$pp \to \tau^\pm \mu^\mp$ & CMS & $138\,\mathrm{fb}^{-1}$ & $m_{\tau_h\mu}^{\rm col}$ & $m_{\tau\mu}$ & \cite{CMS:2022fsw}
			\\ 
			$pp \to \tau^\pm e^\mp$ & CMS & $138\,\mathrm{fb}^{-1}$ & $m_{\tau_he}^{\rm col}$ & $m_{\tau e}$ & \cite{CMS:2022fsw}
			\\ 
			$pp \to \mu^\pm  e^\mp$ & CMS & $138\,\mathrm{fb}^{-1}$ & $m_{\mu e}$ & $m_{\mu e}$ & \cite{CMS:2022fsw}
		\end{tabular}
	}
	\caption{Experimental searches by the ATLAS and CMS collaborations that are available in \texttt{HighPT} \cite{Allwicher:2022gkm,Allwicher:2022mcg}.}
\label{tab:searches}
\end{table}

\section{Flavor constraints}\label{app:flavor}

In this appendix, we further motivate the flavor structure of the SMEFT operators adopted in Section~\ref{sec:flavors-global} by considering limits from leptonic and semileptonic flavor-changing neutral-current decays of $B$, $D$, and $K$ mesons. This Appendix does not aim to be exhaustive, but to provide a justification of neglecting off-diagonal components.  

FCNC processes are more easily analyzed in LEFT, for which we adopt the operator basis defined in Ref.~\cite{Jenkins:2017jig}.  
Using the notation of Section~\ref{sec:flavors-global}, the tree-level matching coefficients onto neutral current LEFT vector operators with charged leptons $\ell \in \{ e, \mu, \tau \}$ are given by
\begin{align}
 \Big[L^{eu}_{\rm VLL}\Big]_{\ell \ell i j} =& \Big[C^{(u)}_{lq}\Big]_{\ell \ell i j}  - 4\sqrt{2} G_F \left[g_L^{(e)} \right]_{\ell \ell} \left[g_L^{(u)}\right]_{i j}\, ,  \nn \\
 \Big[L^{ed}_{\rm VLL}\Big]_{\ell \ell i j} =& \Big[C^{(d)}_{lq}\Big]_{\ell \ell i j}  - 4\sqrt{2} G_F \left[g_L^{(e)} \right]_{\ell \ell} \left[g_L^{(d)}\right]_{i j}\,, \nn \\
 \Big[L^{eu}_{\rm VRR}\Big]_{\ell \ell i j} =& \Big[C_{eu}\Big]_{\ell \ell i j}  -  4\sqrt{2} G_F \left[g_R^{(e)} \right]_{\ell \ell} \left[g_R^{(u)}\right]_{i j}\,, \nn \\
 \Big[L^{ed}_{\rm VRR}\Big]_{\ell \ell i j} =& \Big[C_{ed}\Big]_{\ell \ell i j}  - 4\sqrt{2} G_F \left[g_R^{(e)} \right]_{\ell \ell} \left[g_R^{(d)}\right]_{i j}\,,\nn \\
 \Big[L^{eu}_{\rm VLR}\Big]_{\ell \ell i j} =& \Big[C_{lu}\Big]_{\ell \ell i j}  - 4\sqrt{2} G_F \left[g_L^{(e)} \right]_{\ell \ell} \left[g_R^{(u)}\right]_{i j}\,, \nn \\
 \Big[L^{ed}_{\rm VLR}\Big]_{\ell \ell i j} =& \Big[C_{ld}\Big]_{\ell \ell i j}  - 4\sqrt{2} G_F \left[g_L^{(e)} \right]_{\ell \ell} \left[g_R^{(d)}\right]_{i j}\,,\nn \\
 \Big[L^{ue}_{\rm VLR}\Big]_{\ell \ell i j} =& \Big[V_{}C_{qe} V^{\dagger}_{}\Big]_{\ell \ell i j}  - 4\sqrt{2} G_F \left[g_R^{(e)} \right]_{\ell \ell} \left[g_L^{(u)}\right]_{i j}\,, \nn \\
 \Big[L^{de}_{\rm VLR}\Big]_{\ell \ell i j} =& \Big[C_{qe}\Big]_{\ell \ell i j}  - 4\sqrt{2} G_F \left[g_R^{(e)} \right]_{\ell \ell} \left[g_L^{(d)}\right]_{i j}\,, \label{eq:LEFTmatch}
\end{align}
with the $Z$ couplings in the SMEFT given in Eqs.~\eqref{eq:Z1}
and~\eqref{eq:Z2}. In Eq.~\eqref{eq:LEFTmatch} 
we listed the contributions from the four-fermion operators $C_{lq}^{(u)}$ and $C_{lq}^{(d)}$, which also induce low-energy charged-current processes. For completeness, we also included the remaining semileptonic operators $C_{eu}$, $C_{ed}$, $C_{lu}$, $C_{ld}$, and $C_{qe}$, which do not contribute to the CC decays. In the Standard Model, off-diagonal couplings arise at one loop and, because of the GIM mechanism, vanish for degenerate quark masses.

The neutral-current LEFT operators involving neutrinos are
\begin{align}
 \Big[L^{\nu u}_{\rm VLL}\Big]_{\nu_\ell \nu_\ell i j} =& \Big[V C^{(d)}_{lq}V^{\dagger}\Big]_{\nu_\ell \nu_\ell i j}  - 4 \sqrt{2} G_F  \left[g_L^{(\nu)}\right]_{\nu_\ell \nu_{\ell}} \left[g_L^{(u)}\right]_{i j}, \nn\\
 \Big[L^{\nu d}_{\rm VLL}\Big]_{\nu_\ell \nu_\ell i j} =& \Big[V^{\dagger} C^{(u)}_{lq} V \Big]_{\nu_\ell \nu_\ell i j}  - 4 \sqrt{2} G_F  \left[g_L^{(\nu)}\right]_{\nu_\ell \nu_{\ell}} \left[g_L^{(d)}\right]_{i j},\nn \\
 \Big[L^{\nu u}_{\rm VLR}\Big]_{\nu_\ell \nu_\ell i j} =& \Big[C_{lu}\Big]_{\ell \ell i j}  - 4 \sqrt{2} G_F  \left[g_L^{(\nu)}\right]_{\nu_\ell \nu_{\ell}} \left[g_R^{(u)}\right]_{i j}, \nn\\
 \Big[L^{\nu d}_{\rm VLR}\Big]_{\nu_\ell \nu_\ell i j} =& \Big[C_{ld}\Big]_{\ell \ell i j}  - 4 \sqrt{2} G_F  \left[g_L^{(\nu)}\right]_{\nu_\ell \nu_{\ell}} \left[g_R^{(d)}\right]_{i j}.
\label{neutrinos}\end{align}
For scalar and tensor operators, neglecting the quark and lepton SM Yukawa couplings, we get
\begin{align}
\Big[ L^{eu}_{\rm SRR} \Big]_{\ell \ell j i} &= - \Big[\bar{C}^{(1)}_{lequ}\Big]_{\ell \ell j i}, 
\quad \Big[ L^{ed}_{\rm SRL} \Big]_{\ell \ell j i} &=  \Big[C_{ledq}\Big]_{\ell \ell j i},  \quad 
\Big[ L^{eu}_{\rm TRR} \Big]_{\ell \ell j i} &= - \Big[ \bar{C}^{(3)}_{lequ}\Big]_{\ell \ell j i}, \label{scalar}
\end{align}
with $L^{ed}_{\rm SRR} = L^{eu}_{\rm SRL} = L^{ed}_{\rm TRR} = 0$.
We neglect contributions from dimension-6 corrections to the Yukawa couplings, which do not play a role in our analysis. 
Using the matching equations in Eqs.~\eqref{eq:LEFTmatch}, \eqref{neutrinos},
and~\eqref{scalar} we can derive constraints on the
off-diagonal components of the SMEFT coefficients from FCNC decays of the kaon, $B$, and $D$ meson.

\subsection{Kaon decays}
The coefficients $\left[C^{(d)}_{Hq} \right]_{12,\, 21}$,
$\left[C^{(d)}_{l q}\right]_{\ell \ell 1 2,\, 21}$,
and $\left[C_{ledq}\right]_{\ell \ell 1 2,\, 21}$
contribute to the decays $K_L \rightarrow \ell^+ \ell^-$ and $K \rightarrow \pi \ell^+ \ell^-$ at tree level. To check the constraints on these couplings, we used packages \texttt{flavio}~\cite{Straub:2018kue}
and \texttt{wilson}~\cite{Aebischer:2018bkb},
which include the branching ratios
$K_L \rightarrow e^+ e^-$ and $K_L \rightarrow \mu^+ \mu^-$. 
Assuming only one SMEFT operator at a time, we find very strong limits on the four-fermion operators
\begin{align} 
& \left| C^{(d)}_{\substack{lq\\ 1112}} \right| <  2.4 \cdot 10^{-3} \, {\rm TeV}^{-2}\,, 
& &  \left| C^{(d)}_{\substack{lq\\ 2212}} \right| < 2.2 \cdot 10^{-4} \, {\rm TeV}^{-2}   \,,\nonumber \\
 & \left| C^{}_{\substack{ledq\\ 1112}} \right| < 8.2 \cdot 10^{-7} \, {\rm TeV}^{-2}  \,,
& & \left| C^{}_{\substack{ledq\\ 2212}} \right| < 1.6 \cdot 10^{-6} \, {\rm TeV}^{-2} \,,
\label{kaonFCNC}\end{align}
at 99.7\% confidence level. Similar constraints apply to the coefficients with $1\leftrightarrow 2$ on the quark flavor indices.
As discussed in Section \ref{sec:flavors-global},
while $[C^{(d)}_{lq}]_{\ell \ell 12}$
and $[C_{ledq}]_{\ell \ell 12}$
could in principle affect nuclear, kaon, and pion charged current decays, the flavor constraints are so strong that it is safe to neglect their contributions.
The bounds in Eq.~\eqref{kaonFCNC} can be weakened by turning on several coefficients at the same time. However, performing a global analysis would also require the inclusion of other observables, such as $K_S \rightarrow \ell^+ \ell^-$ or $K \rightarrow \pi \ell^+ \ell^-$, which would provide additional constraining power.

Vertex corrections $C_{Hq}^{(d)}$ and $C_{Hd}$ contribute to semileptonic processes with charged leptons or neutrinos in the final state. Including the branching ratios $K^+ \rightarrow \pi^+ \bar\nu \nu$ and $K_L \rightarrow \pi^0 \bar\nu \nu$, we find 
\begin{equation}
    \left| C^{(d)}_{\substack{Hq\\12}} \right| < 1.0 \cdot 10^{-4} \, {\rm TeV}^{-2}, \qquad \left| C_{\substack{Hd\\12}} \right| < 1.2 \cdot 10^{-4} \, {\rm TeV}^{-2},
\end{equation}
which once again justifies neglecting $s-d$ off-diagonal components. 

Finally, as shown in Eqs.~\eqref{eq:clqnu} and~\eqref{neutrinos}, $C_{lq}^{(u)}$
induces couplings of $d$-quarks to neutrinos. For example, at tree level, 
\begin{equation}
\Big[L^{\nu d}_{\rm VLL}\Big]_{\nu_\ell \nu_\ell 1 2} = C^{(u)}_{\substack{lq\\ \ell \ell 1 2}} + \lambda  \left(C^{(u)}_{\substack{lq\\ \ell \ell 1 1}} - C^{(u)}_{\substack{lq\\ \ell \ell 2 2}} \right) + \ldots, 
 \end{equation}
with $K \rightarrow \pi \nu \nu$ enforcing
\begin{equation}
   \left| \Big[L^{\nu d}_{\rm VLL}\Big]_{\nu_\ell \nu_\ell 1 2}  \right| < 9.7 \cdot 10^{-5} \, {\rm TeV}^{-2} .
\end{equation}
This constraint on the diagonal couplings can be avoided by setting the 11 and 22 components of $C^{(u)}_{l q}$ to be equal.

\subsection{$B$ decays}
Nuclear $\beta$ decays as well as the charged-current decays of pions and kaons are minimally affected by couplings to $b$ quarks.
In the notation of Eqs.~\eqref{eq:CHqud} and~\eqref{eq:Clq},  we get 
\begin{equation}
    \epsilon^{d\ell}_L = \frac{v^2}{2} \frac{V_{ub}}{V_{ud}} \left(  C^{(d)}_{\substack{Hq\\ 31}}
    -C^{(d)}_{\substack{lq\\ \ell \ell 31}}\right), \qquad 
    \epsilon^{s\ell}_L  = \frac{v^2}{2} \frac{V_{ub}}{V_{us}} \left(  C^{(d)}_{\substack{Hq\\ 32}}
    -C^{(d)}_{\substack{lq\\ \ell \ell 32}}\right).
 \end{equation}
The $V_{ub}$ suppression is such that, for coefficients of $\mathcal O(1 \, {\rm TeV}^{-2})$, $\epsilon^{d \ell}_L$ and $\epsilon^{s \ell}_L$, are about $1 \cdot 10^{-4}$ and $5 \cdot 10^{-4}$, respectively, already too small to explain the Cabibbo anomaly.
The strong constraints from $B\rightarrow \mu \mu$,
$B \rightarrow K \mu \mu$, and $B \rightarrow \pi \mu \mu$ ensure that the contributions of the $bs$ and $bd$ components of $C^{(d)}_{Hq}$ and $C_{lq}^{(d)}$ are irrelevant to the CLEW observables.

\subsection{$D$ and top decays}

In the case of $D$ decays, the experimental limits on the purely leptonic branching ratios are \cite{ParticleDataGroup:2022pth}
\begin{equation}
 \textrm{BR}(D^0 \rightarrow e^+ e^-) < 7.9 \cdot 10^{-8}\,, \qquad  \textrm{BR}(D^0 \rightarrow \mu^+ \mu^-) < 6.2 \cdot 10^{-9}\,, 
 \end{equation}
while the 90\% CL limits of the semileptonic branching ratios are
\begin{align}
    \textrm{BR}(D^+ \rightarrow \pi^+ e^+ e^-) < 1.1 \cdot 10^{-6}\,, \qquad  \textrm{BR}(D^+ \rightarrow \pi^+ \mu^+ \mu^-) < 6.7 \cdot 10^{-8}\,,  
\\
      \textrm{BR}(D_s \rightarrow K^+ e^+ e^-) < 3.7 \cdot 10^{-6}\,, \qquad  \textrm{BR}(D_s \rightarrow K^+ \mu^+ \mu^-) < 1.4 \cdot 10^{-7}\,.
\end{align}

The short-distance contribution to the branching ratio $D \rightarrow \ell^+ \ell^-$ is given by
\begin{align}
 \textrm{BR}(D^0 \rightarrow \ell^+ \ell^-)&= \frac{\tau_{D^0}}{32 \pi} \sqrt{1 - \frac{4  m_\ell^2}{ m_{D^0}^2}} m_{D^0}  f_D^2 m^2_\ell
   \nonumber \\ 
 &\times\Bigg[
 \left|  B + \frac{m_{D^0}^2}{2 m_\ell (m_c + m_u)  } D 
 \right|^2&  + \left(1 - \frac{4 m_\ell^2}{m_{D^0}^2}\right)
 \left|\frac{m_{D^0}^2}{2 m_\ell (m_c + m_u)  } C \right|^2
 \Bigg],
 \end{align}
where $f_D$ is the $D$ meson decay constant, for which we take the FLAG average $f_D = 212(1)$ MeV \cite{Aoki:2019cca}. $B$, $C$ and $D$ are combinations of matching coefficients given by 
\begin{align}
B &= \left( L^{e u}_{\rm VLL}
 - L^{eu}_{\rm VLR} + L^{e u}_{\rm VRR} - L^{u e}_{\rm VLR}\right)_{\ell \ell 1 2}  \,,\\[1ex]
 C&=\left(L^{eu}_{\rm SRR} \right)_{\ell \ell  1 2 }-\left(L^{eu}_{\rm SRL} \right)_{\ell \ell 1 2}+\left(L^{eu*}_{\rm SRL} \right)_{\ell \ell 2 1}-\left(L^{eu*}_{\rm SRR} \right)_{\ell \ell 2 1}\,,\\[1ex]
D&=\left(L^{eu}_{\rm SRR} \right)_{\ell \ell 1 2}-\left(L^{eu}_{\rm SRL} \right)_{\ell \ell 1 2}-\left(L^{eu*}_{\rm SRL} \right)_{\ell \ell 21 }+\left(L^{eu*}_{\rm SRR} \right)_{\ell \ell 21}\,.
\end{align}
The expression for semileptonic decays $D \rightarrow P \ell^+ \ell^-$ is more complicated and depends on vector, scalar, and tensor form factors. See, for example, Ref.~\cite{Duraisamy:2016gsd} for a general parameterization.

Considering the bounds on the leptonic branching ratios, assuming the SMEFT contribution to saturate the experimental bound, and turning on one operator at a time we obtain
\begin{align}
    &\left|C^{(u)}_{\substack{lq\\ 1112} } \right| < 41.5 \,\textrm{TeV}^{-2}\,,& & \left| \bar{C}^{(1)}_{\substack{lequ\\1112}} \right| <  0.01 \, \textrm{TeV}^{-2}\,, \nonumber \\
& \left|C^{(u)}_{\substack{lq\\ 2212} } \right| < 0.06 \, \textrm{TeV}^{-2}\,, & & \left| \bar{C}^{(1)}_{\substack{lequ\\2212}} \right| <  0.006 \, \textrm{TeV}^{-2}\,, \nonumber \\
&\left|C^{(u)}_{\substack{Hq\\ 12} } \right| < 0.06 \, \textrm{TeV}^{-2}\,, & & \left|C_{\substack{Hu\\ 12} } \right| < 0.06 \, \textrm{TeV}^{-2}\,.  \label{eq:DmesonBound}
\end{align}
The limit on the electron coupling $C^{(u)}_{lq}$
is particularly weak because of the chiral suppression in $D^0 \rightarrow e^+ e^-$. Adding $D \rightarrow \pi e^+ e^-$ would constrain $[C^{(u)}_{lq}]_{1112}$ at a level similar to the other vector couplings in Eq.~\eqref{eq:DmesonBound}.
Comparing the bounds in Eq.~\eqref{eq:DmesonBound} with the results of Section~\ref{22fit}, we see that
the constraints on $u$-$c$ FCNC are not strong enough to exclude contributions from $u$-$c$ operators to the neutron and nuclear $\beta$ decays, pion, and kaon CC decays at a level comparable with current uncertainties. 
Therefore, one could imagine performing a joint analysis of $u$-$c$ FCNC processes, $c$-$d$ and $c$-$s$ CC transitions, neutron $\beta$ decays, and pion and kaon decays. Here, for simplicity, we choose to focus only on processes without charm and to set $u$-$c$ FCNC coefficients to zero.    

As shown in Eq.~\eqref{eq:lowEflavor},
top flavor-changing couplings can enter low-energy charged-current observables at $\mathcal O(\lambda)$.
FCNC top decays in SMEFT were studied in Refs.~\cite{Zhang:2014rja,Durieux:2014xla}. Focusing on the $Z$ couplings induced by $C_{H q}^{(u)}$ and $C_{Hu}$, the branching ratio at LO in QCD becomes
\begin{align}
    {\textrm{Br}(t \rightarrow Z q_i)} &=
    \frac{2}{c_w^2}  f\left(\frac{m_Z^2}{m_t^2}\right) \left[f\left(\frac{m_W^2}{m_t^2}\right) \right]^{-1}
 \left( \left[\delta  g_R^{(u)}\right]_{i 3}^{2} + \left[\delta g_L^{(u)}\right]^2_{i 3} \right) \,,\nonumber \\
    f(x)  &= \frac{1}{x} \left(1 - x\right)^2 \left(1 + 2 x\right)\,.
\end{align}
The most recent limits on the branching ratios from the ATLAS collaboration are 
\cite{ATLAS:2023qzr}
\begin{equation}
    {\textrm{Br}(t \rightarrow Z u)} < 13 \cdot 10^{-5}\,, \qquad 
    {\textrm{Br}(t \rightarrow Z c)} < 6.2 \cdot 10^{-5}\,,
\end{equation}
implying 
\begin{equation}
    \frac{v^2}{2}\left[ C^{(u)}_{Hq} \right]_{13} < 8.4 \cdot 10^{-3},\, \qquad 
    \frac{v^2}{2}\left[ C^{(u)}_{Hq}\right]_{23} < 5.8 \cdot 10^{-3}\,.
\end{equation} 
When multiplied by $V_{ts}/V_{us}$, the contribution to $\epsilon^{s\ell}_L$ is of order $10^{-3}$ and negligible.

\bibliographystyle{JHEP} 
\bibliography{bibliography}

\providecommand{\href}[2]{#2}\begingroup\raggedright\begin{thebibliography}{100}

\bibitem{Cabibbo:1963yz}
N.~Cabibbo, \emph{{Unitary Symmetry and Leptonic Decays}},
  \href{http://dx.doi.org/10.1103/PhysRevLett.10.531}{\emph{Phys. Rev. Lett.}
  {\bf 10} (1963) 531--533}.

\bibitem{Kobayashi:1973fv}
M.~Kobayashi and T.~Maskawa, \emph{{CP Violation in the Renormalizable Theory
  of Weak Interaction}},
  \href{http://dx.doi.org/10.1143/PTP.49.652}{\emph{Prog. Theor. Phys.} {\bf
  49} (1973) 652--657}.

\bibitem{ParticleDataGroup:2022pth}
{\scshape Particle Data Group} collaboration, R.~L. Workman et~al.,
  \emph{{Review of Particle Physics}},
  \href{http://dx.doi.org/10.1093/ptep/ptac097}{\emph{PTEP} {\bf 2022} (2022)
  083C01}.

\bibitem{Seng:2018yzq}
C.-Y. Seng, M.~Gorchtein, H.~H. Patel and M.~J. Ramsey-Musolf, \emph{{Reduced
  Hadronic Uncertainty in the Determination of $V_{ud}$}},
  \href{http://dx.doi.org/10.1103/PhysRevLett.121.241804}{\emph{Phys. Rev.
  Lett.} {\bf 121} (2018) 241804},
  [\href{https://arxiv.org/abs/1807.10197}{{\tt 1807.10197}}].

\bibitem{Seng:2018qru}
C.~Y. Seng, M.~Gorchtein and M.~J. Ramsey-Musolf, \emph{{Dispersive evaluation
  of the inner radiative correction in neutron and nuclear $\beta$ decay}},
  \href{http://dx.doi.org/10.1103/PhysRevD.100.013001}{\emph{Phys. Rev. D} {\bf
  100} (2019) 013001}, [\href{https://arxiv.org/abs/1812.03352}{{\tt
  1812.03352}}].

\bibitem{Czarnecki:2019mwq}
A.~Czarnecki, W.~J. Marciano and A.~Sirlin, \emph{{Radiative Corrections to
  Neutron and Nuclear Beta Decays Revisited}},
  \href{http://dx.doi.org/10.1103/PhysRevD.100.073008}{\emph{Phys. Rev. D} {\bf
  100} (2019) 073008}, [\href{https://arxiv.org/abs/1907.06737}{{\tt
  1907.06737}}].

\bibitem{Shiells:2020fqp}
K.~Shiells, P.~G. Blunden and W.~Melnitchouk, \emph{{Electroweak axial
  structure functions and improved extraction of the Vud CKM matrix element}},
  \href{http://dx.doi.org/10.1103/PhysRevD.104.033003}{\emph{Phys. Rev. D} {\bf
  104} (2021) 033003}, [\href{https://arxiv.org/abs/2012.01580}{{\tt
  2012.01580}}].

\bibitem{Hayen:2020cxh}
L.~Hayen, \emph{{Standard model $\mathcal{O}(\alpha)$ renormalization of $g_A$
  and its impact on new physics searches}},
  \href{http://dx.doi.org/10.1103/PhysRevD.103.113001}{\emph{Phys. Rev. D} {\bf
  103} (2021) 113001}, [\href{https://arxiv.org/abs/2010.07262}{{\tt
  2010.07262}}].

\bibitem{Hardy2020Oct}
J.~C. Hardy and I.~S. Towner, \emph{{Superallowed
  ${0}^{+}\ensuremath{\rightarrow}{0}^{+}$ nuclear $\ensuremath{\beta}$ decays:
  2020 critical survey, with implications for ${V}_{\mathit{ud}}$ and CKM
  unitarity}}, \href{http://dx.doi.org/10.1103/PhysRevC.102.045501}{\emph{Phys.
  Rev. C} {\bf 102} (Oct., 2020) 045501}.

\bibitem{Carrasco:2016kpy}
N.~Carrasco, P.~Lami, V.~Lubicz, L.~Riggio, S.~Simula and C.~Tarantino,
  \emph{{$K \to \pi$ semileptonic form factors with $N_f=2+1+1$ twisted mass
  fermions}}, \href{http://dx.doi.org/10.1103/PhysRevD.93.114512}{\emph{Phys.
  Rev. D} {\bf 93} (2016) 114512},
  [\href{https://arxiv.org/abs/1602.04113}{{\tt 1602.04113}}].

\bibitem{FermilabLattice:2018zqv}
{\scshape Fermilab Lattice, MILC} collaboration, A.~Bazavov et~al.,
  \emph{{$|V_{us}|$ from $K_{\ell 3}$ decay and four-flavor lattice QCD}},
  \href{http://dx.doi.org/10.1103/PhysRevD.99.114509}{\emph{Phys. Rev. D} {\bf
  99} (2019) 114509}, [\href{https://arxiv.org/abs/1809.02827}{{\tt
  1809.02827}}].

\bibitem{FlavourLatticeAveragingGroupFLAG:2021npn}
{\scshape Flavour Lattice Averaging Group (FLAG)} collaboration, Y.~Aoki
  et~al., \emph{{FLAG Review 2021}},
  \href{http://dx.doi.org/10.1140/epjc/s10052-022-10536-1}{\emph{Eur. Phys. J.
  C} {\bf 82} (2022) 869}, [\href{https://arxiv.org/abs/2111.09849}{{\tt
  2111.09849}}].

\bibitem{Cirigliano:2022yyo}
V.~Cirigliano, A.~Crivellin, M.~Hoferichter and M.~Moulson, \emph{{Scrutinizing
  CKM unitarity with a new measurement of the
  K\ensuremath{\mu}3/K\ensuremath{\mu}2 branching fraction}},
  \href{http://dx.doi.org/10.1016/j.physletb.2023.137748}{\emph{Phys. Lett. B}
  {\bf 838} (2023) 137748}, [\href{https://arxiv.org/abs/2208.11707}{{\tt
  2208.11707}}].

\bibitem{Feng:2020zdc}
X.~Feng, M.~Gorchtein, L.-C. Jin, P.-X. Ma and C.-Y. Seng,
  \emph{{First-principles calculation of electroweak box diagrams from lattice
  QCD}}, \href{http://dx.doi.org/10.1103/PhysRevLett.124.192002}{\emph{Phys.
  Rev. Lett.} {\bf 124} (2020) 192002},
  [\href{https://arxiv.org/abs/2003.09798}{{\tt 2003.09798}}].

\bibitem{Seng:2020wjq}
C.-Y. Seng, X.~Feng, M.~Gorchtein and L.-C. Jin, \emph{{Joint lattice
  QCD\textendash{}dispersion theory analysis confirms the quark-mixing top-row
  unitarity deficit}},
  \href{http://dx.doi.org/10.1103/PhysRevD.101.111301}{\emph{Phys. Rev. D} {\bf
  101} (2020) 111301}, [\href{https://arxiv.org/abs/2003.11264}{{\tt
  2003.11264}}].

\bibitem{Ma:2021azh}
P.-X. Ma, X.~Feng, M.~Gorchtein, L.-C. Jin and C.-Y. Seng, \emph{{Lattice QCD
  calculation of the electroweak box diagrams for the kaon semileptonic
  decays}},  \href{https://arxiv.org/abs/2102.12048}{{\tt 2102.12048}}.

\bibitem{Yoo:2023gln}
J.-S. Yoo, T.~Bhattacharya, R.~Gupta, S.~Mondal and B.~Yoon, \emph{{Electroweak
  box diagram contribution for pion and kaon decay from lattice QCD}},
  \href{http://dx.doi.org/10.1103/PhysRevD.108.034508}{\emph{Phys. Rev. D} {\bf
  108} (2023) 034508}, [\href{https://arxiv.org/abs/2305.03198}{{\tt
  2305.03198}}].

\bibitem{Cirigliano:2023fnz}
V.~Cirigliano, W.~Dekens, E.~Mereghetti and O.~Tomalak, \emph{{Effective field
  theory for radiative corrections to charged-current processes: Vector
  coupling}}, \href{http://dx.doi.org/10.1103/PhysRevD.108.053003}{\emph{Phys.
  Rev. D} {\bf 108} (2023) 053003},
  [\href{https://arxiv.org/abs/2306.03138}{{\tt 2306.03138}}].

\bibitem{Seng:2022cnq}
C.-Y. Seng and M.~Gorchtein, \emph{{Dispersive formalism for the nuclear
  structure correction \ensuremath{\delta}NS to the \ensuremath{\beta} decay
  rate}}, \href{http://dx.doi.org/10.1103/PhysRevC.107.035503}{\emph{Phys. Rev.
  C} {\bf 107} (2023) 035503}, [\href{https://arxiv.org/abs/2211.10214}{{\tt
  2211.10214}}].

\bibitem{Seng:2023cvt}
C.-Y. Seng and M.~Gorchtein, \emph{{Towards $\it{ab}$-$\it{initio}$ nuclear
  theory calculations of $\delta_\mathrm{C}$}},
  \href{https://arxiv.org/abs/2304.03800}{{\tt 2304.03800}}.

\bibitem{1100705}
H.~Akaike, \emph{A new look at the statistical model identification},
  \href{http://dx.doi.org/10.1109/TAC.1974.1100705}{\emph{IEEE Transactions on
  Automatic Control} {\bf 19} (1974) 716--723}.

\bibitem{Seng:2022epj}
C.-Y. Seng and M.~Gorchtein, \emph{{Electroweak nuclear radii constrain the
  isospin breaking correction to Vud}},
  \href{http://dx.doi.org/10.1016/j.physletb.2022.137654}{\emph{Phys. Lett. B}
  {\bf 838} (2023) 137654}, [\href{https://arxiv.org/abs/2208.03037}{{\tt
  2208.03037}}].

\bibitem{Seng:2022inj}
C.-Y. Seng, \emph{{Model-Independent Determination of Nuclear Weak Form Factors
  and Implications for Standard Model Precision Tests}},
  \href{http://dx.doi.org/10.1103/PhysRevLett.130.152501}{\emph{Phys. Rev.
  Lett.} {\bf 130} (2023) 152501},
  [\href{https://arxiv.org/abs/2212.02681}{{\tt 2212.02681}}].

\bibitem{Ma:2023kfr}
P.-X. Ma, X.~Feng, M.~Gorchtein, L.-C. Jin, K.-F. Liu, C.-Y. Seng et~al.,
  \emph{{Lattice QCD Calculation of Electroweak Box Contributions to
  Superallowed Nuclear and Neutron Beta Decays}},
  \href{https://arxiv.org/abs/2308.16755}{{\tt 2308.16755}}.

\bibitem{Seng:2023cgl}
C.-Y. Seng and M.~Gorchtein, \emph{{Data-driven re-evaluation of $ft$-values in
  superallowed beta decays}},  \href{https://arxiv.org/abs/2309.16893}{{\tt
  2309.16893}}.

\bibitem{Belfatto:2019swo}
B.~Belfatto, R.~Beradze and Z.~Berezhiani, \emph{{The CKM unitarity problem: A
  trace of new physics at the TeV scale?}},
  \href{http://dx.doi.org/10.1140/epjc/s10052-020-7691-6}{\emph{Eur. Phys. J.
  C} {\bf 80} (2020) 149}, [\href{https://arxiv.org/abs/1906.02714}{{\tt
  1906.02714}}].

\bibitem{Grossman:2019bzp}
Y.~Grossman, E.~Passemar and S.~Schacht, \emph{{On the Statistical Treatment of
  the Cabibbo Angle Anomaly}},
  \href{http://dx.doi.org/10.1007/JHEP07(2020)068}{\emph{JHEP} {\bf 07} (2020)
  068}, [\href{https://arxiv.org/abs/1911.07821}{{\tt 1911.07821}}].

\bibitem{Crivellin:2020lzu}
A.~Crivellin and M.~Hoferichter, \emph{{\ensuremath{\beta} Decays as Sensitive
  Probes of Lepton Flavor Universality}},
  \href{http://dx.doi.org/10.1103/PhysRevLett.125.111801}{\emph{Phys. Rev.
  Lett.} {\bf 125} (2020) 111801},
  [\href{https://arxiv.org/abs/2002.07184}{{\tt 2002.07184}}].

\bibitem{Kirk:2020wdk}
M.~Kirk, \emph{{Cabibbo anomaly versus electroweak precision tests: An
  exploration of extensions of the Standard Model}},
  \href{http://dx.doi.org/10.1103/PhysRevD.103.035004}{\emph{Phys. Rev. D} {\bf
  103} (2021) 035004}, [\href{https://arxiv.org/abs/2008.03261}{{\tt
  2008.03261}}].

\bibitem{Crivellin:2020ebi}
A.~Crivellin, F.~Kirk, C.~A. Manzari and M.~Montull, \emph{{Global Electroweak
  Fit and Vector-Like Leptons in Light of the Cabibbo Angle Anomaly}},
  \href{http://dx.doi.org/10.1007/JHEP12(2020)166}{\emph{JHEP} {\bf 12} (2020)
  166}, [\href{https://arxiv.org/abs/2008.01113}{{\tt 2008.01113}}].

\bibitem{Alok:2021ydy}
A.~K. Alok, A.~Dighe, S.~Gangal and J.~Kumar, \emph{{A natural resolution for
  the Cabibbo angle anomaly and $R_{K^{(*)}}$}},
  \href{https://arxiv.org/abs/2108.05614}{{\tt 2108.05614}}.

\bibitem{Crivellin:2021bkd}
A.~Crivellin, M.~Hoferichter, M.~Kirk, C.~A. Manzari and L.~Schnell,
  \emph{{First-generation new physics in simplified models: from low-energy
  parity violation to the LHC}},
  \href{http://dx.doi.org/10.1007/JHEP10(2021)221}{\emph{JHEP} {\bf 10} (2021)
  221}, [\href{https://arxiv.org/abs/2107.13569}{{\tt 2107.13569}}].

\bibitem{Crivellin:2022rhw}
A.~Crivellin, M.~Kirk, T.~Kitahara and F.~Mescia, \emph{{Global fit of modified
  quark couplings to EW gauge bosons and vector-like quarks in light of the
  Cabibbo angle anomaly}},
  \href{http://dx.doi.org/10.1007/JHEP03(2023)234}{\emph{JHEP} {\bf 03} (2023)
  234}, [\href{https://arxiv.org/abs/2212.06862}{{\tt 2212.06862}}].

\bibitem{Belfatto:2021jhf}
B.~Belfatto and Z.~Berezhiani, \emph{{Are the CKM anomalies induced by
  vector-like quarks? Limits from flavor changing and Standard Model precision
  tests}}, \href{http://dx.doi.org/10.1007/JHEP10(2021)079}{\emph{JHEP} {\bf
  10} (2021) 079}, [\href{https://arxiv.org/abs/2103.05549}{{\tt 2103.05549}}].

\bibitem{Belfatto:2023tbv}
B.~Belfatto and S.~Trifinopoulos, \emph{{Cabibbo angle anomalies and oblique
  corrections: The remarkable role of the vectorlike quark doublet}},
  \href{http://dx.doi.org/10.1103/PhysRevD.108.035022}{\emph{Phys. Rev. D}
  (2023) 035022}, [\href{https://arxiv.org/abs/2302.14097}{{\tt 2302.14097}}].

\bibitem{Gonzalez-Alonso:2016etj}
M.~Gonz\'alez-Alonso and J.~Martin~Camalich, \emph{{Global
  Effective-Field-Theory analysis of New-Physics effects in (semi)leptonic kaon
  decays}}, \href{http://dx.doi.org/10.1007/JHEP12(2016)052}{\emph{JHEP} {\bf
  12} (2016) 052}, [\href{https://arxiv.org/abs/1605.07114}{{\tt 1605.07114}}].

\bibitem{Falkowski:2017pss}
A.~Falkowski, M.~Gonz\'alez-Alonso and K.~Mimouni, \emph{{Compilation of
  low-energy constraints on 4-fermion operators in the SMEFT}},
  \href{http://dx.doi.org/10.1007/JHEP08(2017)123}{\emph{JHEP} {\bf 08} (2017)
  123}, [\href{https://arxiv.org/abs/1706.03783}{{\tt 1706.03783}}].

\bibitem{Cirigliano:2021yto}
V.~Cirigliano, D.~D\'\i{}az-Calder\'on, A.~Falkowski, M.~Gonz\'alez-Alonso and
  A.~Rodr\'\i{}guez-S\'anchez, \emph{{Semileptonic tau decays beyond the
  Standard Model}},
  \href{http://dx.doi.org/10.1007/JHEP04(2022)152}{\emph{JHEP} {\bf 04} (2022)
  152}, [\href{https://arxiv.org/abs/2112.02087}{{\tt 2112.02087}}].

\bibitem{CDF:2022hxs}
{\scshape CDF} collaboration, T.~Aaltonen et~al., \emph{{High-precision
  measurement of the $W$ boson mass with the CDF II detector}},
  \href{http://dx.doi.org/10.1126/science.abk1781}{\emph{Science} {\bf 376}
  (2022) 170--176}.

\bibitem{Cirigliano:2022qdm}
V.~Cirigliano, W.~Dekens, J.~de~Vries, E.~Mereghetti and T.~Tong,
  \emph{{Beta-decay implications for the W-boson mass anomaly}},
  \href{http://dx.doi.org/10.1103/PhysRevD.106.075001}{\emph{Phys. Rev. D} {\bf
  106} (2022) 075001}, [\href{https://arxiv.org/abs/2204.08440}{{\tt
  2204.08440}}].

\bibitem{Bagnaschi:2022whn}
E.~Bagnaschi, J.~Ellis, M.~Madigan, K.~Mimasu, V.~Sanz and T.~You, \emph{{SMEFT
  analysis of m$_{W}$}},
  \href{http://dx.doi.org/10.1007/JHEP08(2022)308}{\emph{JHEP} {\bf 08} (2022)
  308}, [\href{https://arxiv.org/abs/2204.05260}{{\tt 2204.05260}}].

\bibitem{Buchmuller:1985jz}
W.~Buchmuller and D.~Wyler, \emph{{Effective Lagrangian Analysis of New
  Interactions and Flavor Conservation}},
  \href{http://dx.doi.org/10.1016/0550-3213(86)90262-2}{\emph{Nucl. Phys. B}
  {\bf 268} (1986) 621--653}.

\bibitem{Grzadkowski:2010es}
B.~Grzadkowski, M.~Iskrzynski, M.~Misiak and J.~Rosiek, \emph{{Dimension-Six
  Terms in the Standard Model Lagrangian}},
  \href{http://dx.doi.org/10.1007/JHEP10(2010)085}{\emph{JHEP} {\bf 10} (2010)
  085}, [\href{https://arxiv.org/abs/1008.4884}{{\tt 1008.4884}}].

\bibitem{Weinberg:1979sa}
S.~Weinberg, \emph{{Baryon and Lepton Nonconserving Processes}},
  \href{http://dx.doi.org/10.1103/PhysRevLett.43.1566}{\emph{Phys. Rev. Lett.}
  {\bf 43} (1979) 1566--1570}.

\bibitem{Alonso:2013hga}
R.~Alonso, E.~E. Jenkins, A.~V. Manohar and M.~Trott, \emph{{Renormalization
  Group Evolution of the Standard Model Dimension Six Operators III: Gauge
  Coupling Dependence and Phenomenology}},
  \href{http://dx.doi.org/10.1007/JHEP04(2014)159}{\emph{JHEP} {\bf 04} (2014)
  159}, [\href{https://arxiv.org/abs/1312.2014}{{\tt 1312.2014}}].

\bibitem{Jenkins:2017jig}
E.~E. Jenkins, A.~V. Manohar and P.~Stoffer, \emph{{Low-Energy Effective Field
  Theory below the Electroweak Scale: Operators and Matching}},
  \href{http://dx.doi.org/10.1007/JHEP03(2018)016}{\emph{JHEP} {\bf 03} (2018)
  016}, [\href{https://arxiv.org/abs/1709.04486}{{\tt 1709.04486}}].

\bibitem{Falkowski:2020pma}
A.~Falkowski, M.~Gonz\'alez-Alonso and O.~Naviliat-Cuncic, \emph{{Comprehensive
  analysis of beta decays within and beyond the Standard Model}},
  \href{http://dx.doi.org/10.1007/JHEP04(2021)126}{\emph{JHEP} {\bf 04} (2021)
  126}, [\href{https://arxiv.org/abs/2010.13797}{{\tt 2010.13797}}].

\bibitem{Towner:2007np}
I.~S. Towner and J.~C. Hardy, \emph{{An Improved calculation of the
  isospin-symmetry-breaking corrections to superallowed Fermi beta decay}},
  \href{http://dx.doi.org/10.1103/PhysRevC.77.025501}{\emph{Phys. Rev. C} {\bf
  77} (2008) 025501}, [\href{https://arxiv.org/abs/0710.3181}{{\tt
  0710.3181}}].

\bibitem{Hardy:2014qxa}
J.~C. Hardy and I.~S. Towner, \emph{{Superallowed $0^+\to 0^+$ nuclear
  \ensuremath{\beta} decays: 2014 critical survey, with precise results for
  $V_{ud}$ and CKM unitarity}},
  \href{http://dx.doi.org/10.1103/PhysRevC.91.025501}{\emph{Phys. Rev. C} {\bf
  91} (2015) 025501}, [\href{https://arxiv.org/abs/1411.5987}{{\tt
  1411.5987}}].

\bibitem{Gonzalez-Alonso:2018omy}
M.~Gonz\'alez-Alonso, O.~Naviliat-Cuncic and N.~Severijns, \emph{{New physics
  searches in nuclear and neutron $\beta$ decay}},
  \href{http://dx.doi.org/10.1016/j.ppnp.2018.08.002}{\emph{Prog. Part. Nucl.
  Phys.} {\bf 104} (2019) 165--223},
  [\href{https://arxiv.org/abs/1803.08732}{{\tt 1803.08732}}].

\bibitem{Hardy:2020qwl}
J.~C. Hardy and I.~S. Towner, \emph{{Superallowed $0^+ \to 0^+$ nuclear $\beta$
  decays: 2020 critical survey, with implications for V$_{ud}$ and CKM
  unitarity}}, \href{http://dx.doi.org/10.1103/PhysRevC.102.045501}{\emph{Phys.
  Rev. C} {\bf 102} (2020) 045501}.

\bibitem{Gorchtein:2018fxl}
M.~Gorchtein, \emph{{\ensuremath{\gamma}W Box Inside Out: Nuclear
  Polarizabilities Distort the Beta Decay Spectrum}},
  \href{http://dx.doi.org/10.1103/PhysRevLett.123.042503}{\emph{Phys. Rev.
  Lett.} {\bf 123} (2019) 042503},
  [\href{https://arxiv.org/abs/1812.04229}{{\tt 1812.04229}}].

\bibitem{ALEPH:2005ab}
{\scshape ALEPH, DELPHI, L3, OPAL, SLD, LEP Electroweak Working Group, SLD
  Electroweak Group, SLD Heavy Flavour Group} collaboration, S.~Schael et~al.,
  \emph{{Precision electroweak measurements on the $Z$ resonance}},
  \href{http://dx.doi.org/10.1016/j.physrep.2005.12.006}{\emph{Phys. Rept.}
  {\bf 427} (2006) 257--454}, [\href{https://arxiv.org/abs/hep-ex/0509008}{{\tt
  hep-ex/0509008}}].

\bibitem{Balkin:2022glu}
R.~Balkin, E.~Madge, T.~Menzo, G.~Perez, Y.~Soreq and J.~Zupan, \emph{{On the
  implications of positive W mass shift}},
  \href{http://dx.doi.org/10.1007/JHEP05(2022)133}{\emph{JHEP} {\bf 05} (2022)
  133}, [\href{https://arxiv.org/abs/2204.05992}{{\tt 2204.05992}}].

\bibitem{Efrati:2015eaa}
A.~Efrati, A.~Falkowski and Y.~Soreq, \emph{{Electroweak constraints on
  flavorful effective theories}},
  \href{http://dx.doi.org/10.1007/JHEP07(2015)018}{\emph{JHEP} {\bf 07} (2015)
  018}, [\href{https://arxiv.org/abs/1503.07872}{{\tt 1503.07872}}].

\bibitem{Han:2004az}
Z.~Han and W.~Skiba, \emph{{Effective theory analysis of precision electroweak
  data}}, \href{http://dx.doi.org/10.1103/PhysRevD.71.075009}{\emph{Phys. Rev.
  D} {\bf 71} (2005) 075009}, [\href{https://arxiv.org/abs/hep-ph/0412166}{{\tt
  hep-ph/0412166}}].

\bibitem{Ciuchini:2013pca}
M.~Ciuchini, E.~Franco, S.~Mishima and L.~Silvestrini, \emph{{Electroweak
  Precision Observables, New Physics and the Nature of a 126 GeV Higgs Boson}},
  \href{http://dx.doi.org/10.1007/JHEP08(2013)106}{\emph{JHEP} {\bf 08} (2013)
  106}, [\href{https://arxiv.org/abs/1306.4644}{{\tt 1306.4644}}].

\bibitem{Falkowski:2014tna}
A.~Falkowski and F.~Riva, \emph{{Model-independent precision constraints on
  dimension-6 operators}},
  \href{http://dx.doi.org/10.1007/JHEP02(2015)039}{\emph{JHEP} {\bf 02} (2015)
  039}, [\href{https://arxiv.org/abs/1411.0669}{{\tt 1411.0669}}].

\bibitem{Berthier:2015oma}
L.~Berthier and M.~Trott, \emph{{Towards consistent Electroweak Precision Data
  constraints in the SMEFT}},
  \href{http://dx.doi.org/10.1007/JHEP05(2015)024}{\emph{JHEP} {\bf 05} (2015)
  024}, [\href{https://arxiv.org/abs/1502.02570}{{\tt 1502.02570}}].

\bibitem{deBlas:2021wap}
J.~de~Blas, M.~Ciuchini, E.~Franco, A.~Goncalves, S.~Mishima, M.~Pierini
  et~al., \emph{{Global analysis of electroweak data in the Standard Model}},
  \href{http://dx.doi.org/10.1103/PhysRevD.106.033003}{\emph{Phys. Rev. D} {\bf
  106} (2022) 033003}, [\href{https://arxiv.org/abs/2112.07274}{{\tt
  2112.07274}}].

\bibitem{deBlas:2022hdk}
J.~de~Blas, M.~Pierini, L.~Reina and L.~Silvestrini, \emph{{Impact of the
  Recent Measurements of the Top-Quark and W-Boson Masses on Electroweak
  Precision Fits}},
  \href{http://dx.doi.org/10.1103/PhysRevLett.129.271801}{\emph{Phys. Rev.
  Lett.} {\bf 129} (2022) 271801},
  [\href{https://arxiv.org/abs/2204.04204}{{\tt 2204.04204}}].

\bibitem{Breso-Pla:2021qoe}
V.~Bres\'o-Pla, A.~Falkowski and M.~Gonz\'alez-Alonso, \emph{{A$_{FB}$ in the
  SMEFT: precision Z physics at the LHC}},
  \href{http://dx.doi.org/10.1007/JHEP08(2021)021}{\emph{JHEP} {\bf 08} (2021)
  021}, [\href{https://arxiv.org/abs/2103.12074}{{\tt 2103.12074}}].

\bibitem{Bruggisser:2022rhb}
S.~Bruggisser, D.~van Dyk and S.~Westhoff, \emph{{Resolving the flavor
  structure in the MFV-SMEFT}},
  \href{http://dx.doi.org/10.1007/JHEP02(2023)225}{\emph{JHEP} {\bf 02} (2023)
  225}, [\href{https://arxiv.org/abs/2212.02532}{{\tt 2212.02532}}].

\bibitem{Ellis:2018gqa}
J.~Ellis, C.~W. Murphy, V.~Sanz and T.~You, \emph{{Updated Global SMEFT Fit to
  Higgs, Diboson and Electroweak Data}},
  \href{http://dx.doi.org/10.1007/JHEP06(2018)146}{\emph{JHEP} {\bf 06} (2018)
  146}, [\href{https://arxiv.org/abs/1803.03252}{{\tt 1803.03252}}].

\bibitem{Ellis:2020unq}
J.~Ellis, M.~Madigan, K.~Mimasu, V.~Sanz and T.~You, \emph{{Top, Higgs, Diboson
  and Electroweak Fit to the Standard Model Effective Field Theory}},
  \href{http://dx.doi.org/10.1007/JHEP04(2021)279}{\emph{JHEP} {\bf 04} (2021)
  279}, [\href{https://arxiv.org/abs/2012.02779}{{\tt 2012.02779}}].

\bibitem{Almeida:2021asy}
E.~d.~S. Almeida, A.~Alves, O.~J.~P. \'Eboli and M.~C. Gonzalez-Garcia,
  \emph{{Electroweak legacy of the LHC run II}},
  \href{http://dx.doi.org/10.1103/PhysRevD.105.013006}{\emph{Phys. Rev. D} {\bf
  105} (2022) 013006}, [\href{https://arxiv.org/abs/2108.04828}{{\tt
  2108.04828}}].

\bibitem{Fan:2022yly}
J.~Fan, L.~Li, T.~Liu and K.-F. Lyu, \emph{{W-boson mass, electroweak precision
  tests, and SMEFT}},
  \href{http://dx.doi.org/10.1103/PhysRevD.106.073010}{\emph{Phys. Rev. D} {\bf
  106} (2022) 073010}, [\href{https://arxiv.org/abs/2204.04805}{{\tt
  2204.04805}}].

\bibitem{Corbett:2021eux}
T.~Corbett, A.~Helset, A.~Martin and M.~Trott, \emph{{EWPD in the SMEFT to
  dimension eight}},
  \href{http://dx.doi.org/10.1007/JHEP06(2021)076}{\emph{JHEP} {\bf 06} (2021)
  076}, [\href{https://arxiv.org/abs/2102.02819}{{\tt 2102.02819}}].

\bibitem{Corbett:2023qtg}
T.~Corbett, J.~Desai, O.~J.~P. \'Eboli, M.~C. Gonzalez-Garcia, M.~Martines and
  P.~Reimitz, \emph{{Impact of dimension-eight SMEFT operators in the
  electroweak precision observables and triple gauge couplings analysis in
  universal SMEFT}},
  \href{http://dx.doi.org/10.1103/PhysRevD.107.115013}{\emph{Phys. Rev. D} {\bf
  107} (2023) 115013}, [\href{https://arxiv.org/abs/2304.03305}{{\tt
  2304.03305}}].

\bibitem{Bellafronte:2023amz}
L.~Bellafronte, S.~Dawson and P.~P. Giardino, \emph{{The Importance of Flavor
  in SMEFT Electroweak Precision Fits}},
  \href{https://arxiv.org/abs/2304.00029}{{\tt 2304.00029}}.

\bibitem{Cirigliano:2012ab}
V.~Cirigliano, M.~Gonzalez-Alonso and M.~L. Graesser, \emph{{Non-standard
  Charged Current Interactions: beta decays versus the LHC}},
  \href{http://dx.doi.org/10.1007/JHEP02(2013)046}{\emph{JHEP} {\bf 02} (2013)
  046}, [\href{https://arxiv.org/abs/1210.4553}{{\tt 1210.4553}}].

\bibitem{Greljo:2017vvb}
A.~Greljo and D.~Marzocca, \emph{{High-$p_T$ dilepton tails and flavor
  physics}}, \href{http://dx.doi.org/10.1140/epjc/s10052-017-5119-8}{\emph{Eur.
  Phys. J. C} {\bf 77} (2017) 548},
  [\href{https://arxiv.org/abs/1704.09015}{{\tt 1704.09015}}].

\bibitem{Alioli:2018ljm}
S.~Alioli, W.~Dekens, M.~Girard and E.~Mereghetti, \emph{{NLO QCD corrections
  to SM-EFT dilepton and electroweak Higgs boson production, matched to parton
  shower in POWHEG}},
  \href{http://dx.doi.org/10.1007/JHEP08(2018)205}{\emph{JHEP} {\bf 08} (2018)
  205}, [\href{https://arxiv.org/abs/1804.07407}{{\tt 1804.07407}}].

\bibitem{Torre:2020aiz}
R.~Torre, L.~Ricci and A.~Wulzer, \emph{{On the W\&Y interpretation of
  high-energy Drell-Yan measurements}},
  \href{http://dx.doi.org/10.1007/JHEP02(2021)144}{\emph{JHEP} {\bf 02} (2021)
  144}, [\href{https://arxiv.org/abs/2008.12978}{{\tt 2008.12978}}].

\bibitem{Allwicher:2022mcg}
L.~Allwicher, D.~A. Faroughy, F.~Jaffredo, O.~Sumensari and F.~Wilsch,
  \emph{{HighPT: A Tool for high-$p_T$ Drell-Yan Tails Beyond the Standard
  Model}},  \href{https://arxiv.org/abs/2207.10756}{{\tt 2207.10756}}.

\bibitem{Allwicher:2022gkm}
L.~Allwicher, D.~A. Faroughy, F.~Jaffredo, O.~Sumensari and F.~Wilsch,
  \emph{{Drell-Yan Tails Beyond the Standard Model}},
  \href{https://arxiv.org/abs/2207.10714}{{\tt 2207.10714}}.

\bibitem{Boughezal:2021tih}
R.~Boughezal, E.~Mereghetti and F.~Petriello, \emph{{Dilepton production in the
  SMEFT at O(1/\ensuremath{\Lambda}4)}},
  \href{http://dx.doi.org/10.1103/PhysRevD.104.095022}{\emph{Phys. Rev. D} {\bf
  104} (2021) 095022}, [\href{https://arxiv.org/abs/2106.05337}{{\tt
  2106.05337}}].

\bibitem{Kim:2022amu}
T.~Kim and A.~Martin, \emph{{Monolepton production in SMEFT to $ \mathcal{O}
  $(1/\ensuremath{\Lambda}$^{4}$) and beyond}},
  \href{http://dx.doi.org/10.1007/JHEP09(2022)124}{\emph{JHEP} {\bf 09} (2022)
  124}, [\href{https://arxiv.org/abs/2203.11976}{{\tt 2203.11976}}].

\bibitem{Boughezal:2022nof}
R.~Boughezal, Y.~Huang and F.~Petriello, \emph{{Exploring the SMEFT at
  dimension eight with Drell-Yan transverse momentum measurements}},
  \href{http://dx.doi.org/10.1103/PhysRevD.106.036020}{\emph{Phys. Rev. D} {\bf
  106} (2022) 036020}, [\href{https://arxiv.org/abs/2207.01703}{{\tt
  2207.01703}}].

\bibitem{Bernard:2007cf}
V.~Bernard, M.~Oertel, E.~Passemar and J.~Stern, \emph{{Tests of non-standard
  electroweak couplings of right-handed quarks}},
  \href{http://dx.doi.org/10.1088/1126-6708/2008/01/015}{\emph{JHEP} {\bf 01}
  (2008) 015}, [\href{https://arxiv.org/abs/0707.4194}{{\tt 0707.4194}}].

\bibitem{Cirigliano2022Aug}
V.~Cirigliano, A.~Crivellin, M.~Hoferichter and M.~Moulson, \emph{{Scrutinizing
  CKM unitarity with a new measurement of the $K_{\mu 3}/K_{\mu 2}$ branching
  fraction}}, \href{http://dx.doi.org/10.48550/arXiv.2208.11707}{\emph{arXiv}
  (Aug., 2022) }, [\href{https://arxiv.org/abs/2208.11707}{{\tt 2208.11707}}].

\bibitem{DAmbrosio:2002vsn}
G.~D'Ambrosio, G.~F. Giudice, G.~Isidori and A.~Strumia, \emph{{Minimal flavor
  violation: An Effective field theory approach}},
  \href{http://dx.doi.org/10.1016/S0550-3213(02)00836-2}{\emph{Nucl. Phys. B}
  {\bf 645} (2002) 155--187}, [\href{https://arxiv.org/abs/hep-ph/0207036}{{\tt
  hep-ph/0207036}}].

\bibitem{Dekens:2021bro}
W.~Dekens, L.~Andreoli, J.~de~Vries, E.~Mereghetti and F.~Oosterhof, \emph{{A
  low-energy perspective on the minimal left-right symmetric model}},
  \href{http://dx.doi.org/10.1007/JHEP11(2021)127}{\emph{JHEP} {\bf 11} (2021)
  127}, [\href{https://arxiv.org/abs/2107.10852}{{\tt 2107.10852}}].

\bibitem{Peskin:1991sw}
M.~E. Peskin and T.~Takeuchi, \emph{{Estimation of oblique electroweak
  corrections}}, \href{http://dx.doi.org/10.1103/PhysRevD.46.381}{\emph{Phys.
  Rev. D} {\bf 46} (1992) 381--409}.

\bibitem{burnham2003model}
K.~Burnham and D.~Anderson, \emph{Model Selection and Multimodel Inference: A
  Practical Information-Theoretic Approach}.
\newblock Springer New York, 2003.

\bibitem{Bjorn:2016zlr}
M.~Bj\o{}rn and M.~Trott, \emph{{Interpreting $W$ mass measurements in the
  SMEFT}}, \href{http://dx.doi.org/10.1016/j.physletb.2016.10.003}{\emph{Phys.
  Lett. B} {\bf 762} (2016) 426--431},
  [\href{https://arxiv.org/abs/1606.06502}{{\tt 1606.06502}}].

\bibitem{Kribs:2020jgn}
G.~D. Kribs, X.~Lu, A.~Martin and T.~Tong, \emph{{Custodial symmetry violation
  in the SMEFT}},
  \href{http://dx.doi.org/10.1103/PhysRevD.104.056006}{\emph{Phys. Rev. D} {\bf
  104} (2021) 056006}, [\href{https://arxiv.org/abs/2009.10725}{{\tt
  2009.10725}}].

\bibitem{Blennow:2022yfm}
M.~Blennow, P.~Coloma, E.~Fern\'andez-Mart\'\i{}nez and M.~Gonz\'alez-L\'opez,
  \emph{{Right-handed neutrinos and the CDF II anomaly}},
  \href{http://dx.doi.org/10.1103/PhysRevD.106.073005}{\emph{Phys. Rev. D} {\bf
  106} (2022) 073005}, [\href{https://arxiv.org/abs/2204.04559}{{\tt
  2204.04559}}].

\bibitem{ThomasArun:2023wbd}
M.~Thomas~Arun, K.~Deka and T.~Srivastava, \emph{{Constraining SMEFT BSM
  scenarios with EWPO and $\Delta_{CKM}$}},
  \href{https://arxiv.org/abs/2301.09273}{{\tt 2301.09273}}.

\bibitem{Bhattacharya:2011qm}
T.~Bhattacharya, V.~Cirigliano, S.~D. Cohen, A.~Filipuzzi, M.~Gonzalez-Alonso,
  M.~L. Graesser et~al., \emph{{Probing Novel Scalar and Tensor Interactions
  from (Ultra)Cold Neutrons to the LHC}},
  \href{http://dx.doi.org/10.1103/PhysRevD.85.054512}{\emph{Phys. Rev. D} {\bf
  85} (2012) 054512}, [\href{https://arxiv.org/abs/1110.6448}{{\tt
  1110.6448}}].

\bibitem{Markisch:2018ndu}
B.~M\"arkisch et~al., \emph{{Measurement of the Weak Axial-Vector Coupling
  Constant in the Decay of Free Neutrons Using a Pulsed Cold Neutron Beam}},
  \href{http://dx.doi.org/10.1103/PhysRevLett.122.242501}{\emph{Phys. Rev.
  Lett.} {\bf 122} (2019) 242501},
  [\href{https://arxiv.org/abs/1812.04666}{{\tt 1812.04666}}].

\bibitem{Chang:2018uxx}
C.~C. Chang et~al., \emph{{A per-cent-level determination of the nucleon axial
  coupling from quantum chromodynamics}},
  \href{http://dx.doi.org/10.1038/s41586-018-0161-8}{\emph{Nature} {\bf 558}
  (2018) 91--94}, [\href{https://arxiv.org/abs/1805.12130}{{\tt 1805.12130}}].

\bibitem{Walker-Loud:2019cif}
A.~Walker-Loud et~al., \emph{{Lattice QCD Determination of $g_A$}},
  \href{http://dx.doi.org/10.22323/1.317.0020}{\emph{PoS} {\bf CD2018} (2020)
  020}, [\href{https://arxiv.org/abs/1912.08321}{{\tt 1912.08321}}].

\bibitem{Gorchtein:2021fce}
M.~Gorchtein and C.-Y. Seng, \emph{{Dispersion relation analysis of the
  radiative corrections to g$_{A}$ in the neutron \ensuremath{\beta}-decay}},
  \href{http://dx.doi.org/10.1007/JHEP10(2021)053}{\emph{JHEP} {\bf 10} (2021)
  053}, [\href{https://arxiv.org/abs/2106.09185}{{\tt 2106.09185}}].

\bibitem{Cirigliano:2022hob}
V.~Cirigliano, J.~de~Vries, L.~Hayen, E.~Mereghetti and A.~Walker-Loud,
  \emph{{Pion-Induced Radiative Corrections to Neutron \ensuremath{\beta}
  Decay}}, \href{http://dx.doi.org/10.1103/PhysRevLett.129.121801}{\emph{Phys.
  Rev. Lett.} {\bf 129} (2022) 121801},
  [\href{https://arxiv.org/abs/2202.10439}{{\tt 2202.10439}}].

\bibitem{Cirigliano:2016yhc}
V.~Cirigliano, W.~Dekens, J.~de~Vries and E.~Mereghetti, \emph{{An $\epsilon'$
  improvement from right-handed currents}},
  \href{http://dx.doi.org/10.1016/j.physletb.2017.01.037}{\emph{Phys. Lett. B}
  {\bf 767} (2017) 1--9}, [\href{https://arxiv.org/abs/1612.03914}{{\tt
  1612.03914}}].

\bibitem{Blum:2012uk}
T.~Blum et~al., \emph{{Lattice determination of the $K \to (\pi\pi)_{I=2}$
  Decay Amplitude $A_2$}},
  \href{http://dx.doi.org/10.1103/PhysRevD.86.074513}{\emph{Phys. Rev. D} {\bf
  86} (2012) 074513}, [\href{https://arxiv.org/abs/1206.5142}{{\tt
  1206.5142}}].

\bibitem{RBC:2015gro}
{\scshape RBC, UKQCD} collaboration, Z.~Bai et~al., \emph{{Standard Model
  Prediction for Direct CP Violation in
  K\textrightarrow{}\ensuremath{\pi}\ensuremath{\pi} Decay}},
  \href{http://dx.doi.org/10.1103/PhysRevLett.115.212001}{\emph{Phys. Rev.
  Lett.} {\bf 115} (2015) 212001},
  [\href{https://arxiv.org/abs/1505.07863}{{\tt 1505.07863}}].

\bibitem{Blum:2015ywa}
T.~Blum et~al., \emph{{$K \rightarrow \pi\pi$ $\Delta I=3/2$ decay amplitude in
  the continuum limit}},
  \href{http://dx.doi.org/10.1103/PhysRevD.91.074502}{\emph{Phys. Rev. D} {\bf
  91} (2015) 074502}, [\href{https://arxiv.org/abs/1502.00263}{{\tt
  1502.00263}}].

\bibitem{RBC:2020kdj}
{\scshape RBC, UKQCD} collaboration, R.~Abbott et~al., \emph{{Direct CP
  violation and the $\Delta I=1/2$ rule in $K\to\pi\pi$ decay from the standard
  model}}, \href{http://dx.doi.org/10.1103/PhysRevD.102.054509}{\emph{Phys.
  Rev. D} {\bf 102} (2020) 054509},
  [\href{https://arxiv.org/abs/2004.09440}{{\tt 2004.09440}}].

\bibitem{Alioli:2017ces}
S.~Alioli, V.~Cirigliano, W.~Dekens, J.~de~Vries and E.~Mereghetti,
  \emph{{Right-handed charged currents in the era of the Large Hadron
  Collider}}, \href{http://dx.doi.org/10.1007/JHEP05(2017)086}{\emph{JHEP} {\bf
  05} (2017) 086}, [\href{https://arxiv.org/abs/1703.04751}{{\tt 1703.04751}}].

\bibitem{Cirigliano:2019vfc}
V.~Cirigliano, A.~Crivellin, W.~Dekens, J.~de~Vries, M.~Hoferichter and
  E.~Mereghetti, \emph{{CP Violation in Higgs-Gauge Interactions: From Tabletop
  Experiments to the LHC}},
  \href{http://dx.doi.org/10.1103/PhysRevLett.123.051801}{\emph{Phys. Rev.
  Lett.} {\bf 123} (2019) 051801},
  [\href{https://arxiv.org/abs/1903.03625}{{\tt 1903.03625}}].

\bibitem{Roussy:2022cmp}
T.~S. Roussy et~al., \emph{{An improved bound on the electron\textquoteright{}s
  electric dipole moment}},
  \href{http://dx.doi.org/10.1126/science.adg4084}{\emph{Science} {\bf 381}
  (2023) adg4084}, [\href{https://arxiv.org/abs/2212.11841}{{\tt 2212.11841}}].

\bibitem{Ethier:2021ydt}
J.~J. Ethier, R.~Gomez-Ambrosio, G.~Magni and J.~Rojo, \emph{{SMEFT analysis of
  vector boson scattering and diboson data from the LHC Run II}},
  \href{http://dx.doi.org/10.1140/epjc/s10052-021-09347-7}{\emph{Eur. Phys. J.
  C} {\bf 81} (2021) 560}, [\href{https://arxiv.org/abs/2101.03180}{{\tt
  2101.03180}}].

\bibitem{Ethier:2021bye}
{\scshape SMEFiT} collaboration, J.~J. Ethier, G.~Magni, F.~Maltoni,
  L.~Mantani, E.~R. Nocera, J.~Rojo et~al., \emph{{Combined SMEFT
  interpretation of Higgs, diboson, and top quark data from the LHC}},
  \href{http://dx.doi.org/10.1007/JHEP11(2021)089}{\emph{JHEP} {\bf 11} (2021)
  089}, [\href{https://arxiv.org/abs/2105.00006}{{\tt 2105.00006}}].

\bibitem{ATLAS:2022tnm}
{\scshape ATLAS} collaboration, \emph{{Measurement of the properties of Higgs
  boson production at $\sqrt{s} = 13$ TeV in the $H\to\gamma\gamma$ channel
  using $139$ fb$^{-1}$ of $pp$ collision data with the ATLAS experiment}},
  \href{https://arxiv.org/abs/2207.00348}{{\tt 2207.00348}}.

\bibitem{ATLAS:2022vkf}
{\scshape ATLAS} collaboration, \emph{{A detailed map of Higgs boson
  interactions by the ATLAS experiment ten years after the discovery}},
  \href{http://dx.doi.org/10.1038/s41586-022-04893-w}{\emph{Nature} {\bf 607}
  (2022) 52--59}, [\href{https://arxiv.org/abs/2207.00092}{{\tt 2207.00092}}].

\bibitem{CMS:2022dwd}
{\scshape CMS} collaboration, A.~Tumasyan et~al., \emph{{A portrait of the
  Higgs boson by the CMS experiment ten years after the discovery}},
  \href{http://dx.doi.org/10.1038/s41586-022-04892-x}{\emph{Nature} {\bf 607}
  (2022) 60--68}, [\href{https://arxiv.org/abs/2207.00043}{{\tt 2207.00043}}].

\bibitem{Melia:2011tj}
T.~Melia, P.~Nason, R.~Rontsch and G.~Zanderighi, \emph{{W+W-, WZ and ZZ
  production in the POWHEG BOX}},
  \href{http://dx.doi.org/10.1007/JHEP11(2011)078}{\emph{JHEP} {\bf 11} (2011)
  078}, [\href{https://arxiv.org/abs/1107.5051}{{\tt 1107.5051}}].

\bibitem{CMS:2021icx}
{\scshape CMS} collaboration, A.~Tumasyan et~al., \emph{{Measurement of the
  inclusive and differential WZ production cross sections, polarization angles,
  and triple gauge couplings in pp collisions at $ \sqrt{s} $ = 13 TeV}},
  \href{http://dx.doi.org/10.1007/JHEP07(2022)032}{\emph{JHEP} {\bf 07} (2022)
  032}, [\href{https://arxiv.org/abs/2110.11231}{{\tt 2110.11231}}].

\bibitem{Corbett:2023yhk}
T.~Corbett and A.~Martin, \emph{{Higgs associated production with a vector
  decaying to two fermions in the geoSMEFT}},
  \href{https://arxiv.org/abs/2306.00053}{{\tt 2306.00053}}.

\bibitem{Cirigliano:2009wk}
V.~Cirigliano, J.~Jenkins and M.~Gonzalez-Alonso, \emph{{Semileptonic decays of
  light quarks beyond the Standard Model}},
  \href{http://dx.doi.org/10.1016/j.nuclphysb.2009.12.020}{\emph{Nucl. Phys. B}
  {\bf 830} (2010) 95--115}, [\href{https://arxiv.org/abs/0908.1754}{{\tt
  0908.1754}}].

\bibitem{Janot:2019oyi}
P.~Janot and S.~Jadach, \emph{{Improved Bhabha cross section at LEP and the
  number of light neutrino species}},
  \href{http://dx.doi.org/10.1016/j.physletb.2020.135319}{\emph{Phys. Lett. B}
  {\bf 803} (2020) 135319}, [\href{https://arxiv.org/abs/1912.02067}{{\tt
  1912.02067}}].

\bibitem{CMS:2014mgj}
{\scshape CMS} collaboration, V.~Khachatryan et~al., \emph{{Measurement of the
  t-channel single-top-quark production cross section and of the $\mid V_{tb}
  \mid$ CKM matrix element in pp collisions at $\sqrt{s}$= 8 TeV}},
  \href{http://dx.doi.org/10.1007/JHEP06(2014)090}{\emph{JHEP} {\bf 06} (2014)
  090}, [\href{https://arxiv.org/abs/1403.7366}{{\tt 1403.7366}}].

\bibitem{ATLAS:2018gqq}
{\scshape ATLAS} collaboration, \emph{{Measurement of the effective leptonic
  weak mixing angle using electron and muon pairs from $Z$-boson decay in the
  ATLAS experiment at $\sqrt s = 8$ TeV}}, .

\bibitem{Breso-Pla2021Mar}
V.~Bres{\ifmmode\acute{o}\else\'{o}\fi}-Pla, A.~Falkowski and
  M.~Gonz{\ifmmode\acute{a}\else\'{a}\fi}lez-Alonso, \emph{{$A_{FB}$ in the
  SMEFT: precision Z physics at the LHC}},
  \href{http://dx.doi.org/10.1007/JHEP08(2021)021}{\emph{arXiv} (Mar., 2021) },
  [\href{https://arxiv.org/abs/2103.12074}{{\tt 2103.12074}}].

\bibitem{D0:2011baz}
{\scshape D0} collaboration, V.~M. Abazov et~al., \emph{{Measurement of
  $\sin^2\theta_{\rm eff}^{\ell}$ and $Z$-light quark couplings using the
  forward-backward charge asymmetry in $p\bar{p} \to Z/\gamma^{*} \to
  e^{+}e^{-}$ events with ${\cal L}=5.0$ fb$^{-1}$ at $\sqrt{s}=1.96$ TeV}},
  \href{http://dx.doi.org/10.1103/PhysRevD.84.012007}{\emph{Phys. Rev. D} {\bf
  84} (2011) 012007}, [\href{https://arxiv.org/abs/1104.4590}{{\tt
  1104.4590}}].

\bibitem{Baur2001Aug}
U.~Baur, O.~Brein, W.~Hollik, C.~Schappacher and D.~Wackeroth,
  \emph{{Electroweak Radiative Corrections to Neutral-Current Drell-Yan
  Processes at Hadron Colliders}},
  \href{http://dx.doi.org/10.1103/PhysRevD.65.033007}{\emph{arXiv} (Aug., 2001)
  }, [\href{https://arxiv.org/abs/hep-ph/0108274}{{\tt hep-ph/0108274}}].

\bibitem{SLD:2000jop}
{\scshape SLD} collaboration, K.~Abe et~al., \emph{{First direct measurement of
  the parity violating coupling of the Z0 to the s quark}},
  \href{http://dx.doi.org/10.1103/PhysRevLett.85.5059}{\emph{Phys. Rev. Lett.}
  {\bf 85} (2000) 5059--5063},
  [\href{https://arxiv.org/abs/hep-ex/0006019}{{\tt hep-ex/0006019}}].

\bibitem{Johnson:1963zza}
C.~H. Johnson, F.~Pleasonton and T.~A. Carlson, \emph{{Precision Measurement of
  the Recoil Energy Spectrum from the Decay of He-6}},
  \href{http://dx.doi.org/10.1103/PhysRev.132.1149}{\emph{Phys. Rev.} {\bf 132}
  (1963) 1149--1165}.

\bibitem{Adelberger:1999ud}
{\scshape ISOLDE} collaboration, E.~G. Adelberger, C.~Ortiz, A.~Garcia, H.~E.
  Swanson, M.~Beck, O.~Tengblad et~al., \emph{{Positron neutrino correlation in
  the 0+ ---\ensuremath{>} 0+ decay of Ar-32}},
  \href{http://dx.doi.org/10.1103/PhysRevLett.83.1299}{\emph{Phys. Rev. Lett.}
  {\bf 83} (1999) 1299--1302},
  [\href{https://arxiv.org/abs/nucl-ex/9903002}{{\tt nucl-ex/9903002}}].

\bibitem{Gorelov:2004hv}
A.~Gorelov et~al., \emph{{Scalar interaction limits from the beta-ne
  correlation of trapped radioactive atoms}},
  \href{http://dx.doi.org/10.1103/PhysRevLett.94.142501}{\emph{Phys. Rev.
  Lett.} {\bf 94} (2005) 142501},
  [\href{https://arxiv.org/abs/nucl-ex/0412032}{{\tt nucl-ex/0412032}}].

\bibitem{Wauters:2010gh}
F.~Wauters et~al., \emph{{Precision measurements of the $^{60}$Co
  $\beta$-asymmetry parameter in search for tensor currents in weak
  interactions}},
  \href{http://dx.doi.org/10.1103/PhysRevC.82.055502}{\emph{Phys. Rev. C} {\bf
  82} (2010) 055502}, [\href{https://arxiv.org/abs/1005.5034}{{\tt
  1005.5034}}].

\bibitem{Soti:2014xua}
G.~Soti et~al., \emph{{Measurement of the $\beta$-asymmetry parameter of
  $^{67}$Cu in search for tensor-type currents in the weak interaction}},
  \href{http://dx.doi.org/10.1103/PhysRevC.90.035502}{\emph{Phys. Rev. C} {\bf
  90} (2014) 035502}, [\href{https://arxiv.org/abs/1409.1824}{{\tt
  1409.1824}}].

\bibitem{Wauters:2009jw}
F.~Wauters, I.~Kraev, M.~Tandecki, E.~Traykov, S.~Van~Gorp and N.~Severijns,
  \emph{{Beta asymmetry parameter in the decay of In-114}},
  \href{http://dx.doi.org/10.1103/PhysRevC.80.062501}{\emph{Phys. Rev. C} {\bf
  80} (2009) 062501}, [\href{https://arxiv.org/abs/0901.0081}{{\tt
  0901.0081}}].

\bibitem{Carnoy:1991jd}
A.~S. Carnoy, J.~Deutsch, T.~A. Girard and R.~Prieels, \emph{{Limits on
  nonstandard weak currents from the polarization of O-14 and C-10 decay
  positrons}}, \href{http://dx.doi.org/10.1103/PhysRevC.43.2825}{\emph{Phys.
  Rev. C} {\bf 43} (1991) 2825--2834}.

\bibitem{Wichers:1986es}
V.~A. Wichers, T.~R. Hageman, J.~Van~Klinken, H.~W. Wilschut and D.~Atkinson,
  \emph{{Bounds on Right-handed Currents From Nuclear Beta Decay}},
  \href{http://dx.doi.org/10.1103/PhysRevLett.58.1821}{\emph{Phys. Rev. Lett.}
  {\bf 58} (1987) 1821--1824}.

\bibitem{DescotesGenon:2005pw}
S.~Descotes-Genon and B.~Moussallam, \emph{{Radiative corrections in weak
  semi-leptonic processes at low energy: A Two-step matching determination}},
  \href{http://dx.doi.org/10.1140/epjc/s2005-02316-8}{\emph{Eur. Phys. J. C}
  {\bf 42} (2005) 403--417}, [\href{https://arxiv.org/abs/hep-ph/0505077}{{\tt
  hep-ph/0505077}}].

\bibitem{QCDSFUKQCD:2010cha}
{\scshape QCDSF/UKQCD} collaboration, M.~Gockeler, P.~Hagler, R.~Horsley,
  Y.~Nakamura, D.~Pleiter, P.~E.~L. Rakow et~al., \emph{{Baryon Axial Charges
  and Momentum Fractions with $N_f$ = 2+1 Dynamical Fermions}},
  \href{http://dx.doi.org/10.22323/1.105.0163}{\emph{PoS} {\bf LATTICE2010}
  (2010) 163}, [\href{https://arxiv.org/abs/1102.3407}{{\tt 1102.3407}}].

\bibitem{FlaviaNetWorkingGrouponKaonDecays:2010lot}
{\scshape FlaviaNet Working Group on Kaon Decays} collaboration, M.~Antonelli
  et~al., \emph{{An Evaluation of $|V_{us}|$ and precise tests of the Standard
  Model from world data on leptonic and semileptonic kaon decays}},
  \href{http://dx.doi.org/10.1140/epjc/s10052-010-1406-3}{\emph{Eur. Phys. J.
  C} {\bf 69} (2010) 399--424}, [\href{https://arxiv.org/abs/1005.2323}{{\tt
  1005.2323}}].

\bibitem{Baum:2011rm}
I.~Baum, V.~Lubicz, G.~Martinelli, L.~Orifici and S.~Simula, \emph{{Matrix
  elements of the electromagnetic operator between kaon and pion states}},
  \href{http://dx.doi.org/10.1103/PhysRevD.84.074503}{\emph{Phys. Rev. D} {\bf
  84} (2011) 074503}, [\href{https://arxiv.org/abs/1108.1021}{{\tt
  1108.1021}}].

\bibitem{PiENu:2015seu}
{\scshape PiENu} collaboration, A.~Aguilar-Arevalo et~al., \emph{{Improved
  Measurement of the $\pi \to \textrm{e} \nu$ Branching Ratio}},
  \href{http://dx.doi.org/10.1103/PhysRevLett.115.071801}{\emph{Phys. Rev.
  Lett.} {\bf 115} (2015) 071801},
  [\href{https://arxiv.org/abs/1506.05845}{{\tt 1506.05845}}].

\bibitem{Seng:2021nar}
C.-Y. Seng, D.~Galviz, W.~J. Marciano and U.-G. Mei\ss{}ner, \emph{{Update on
  |Vus| and |Vus/Vud| from semileptonic kaon and pion decays}},
  \href{http://dx.doi.org/10.1103/PhysRevD.105.013005}{\emph{Phys. Rev. D} {\bf
  105} (2022) 013005}, [\href{https://arxiv.org/abs/2107.14708}{{\tt
  2107.14708}}].

\bibitem{Moulson:2017ive}
M.~Moulson, \emph{{Experimental determination of $V_{us}$ from kaon decays}},
  \href{http://dx.doi.org/10.22323/1.291.0033}{\emph{PoS} {\bf CKM2016} (2017)
  033}, [\href{https://arxiv.org/abs/1704.04104}{{\tt 1704.04104}}].

\bibitem{Yushchenko:2003xz}
O.~P. Yushchenko et~al., \emph{{High statistic study of the K-
  ---\ensuremath{>} pi0 mu- nu decay}},
  \href{http://dx.doi.org/10.1016/j.physletb.2003.12.002}{\emph{Phys. Lett. B}
  {\bf 581} (2004) 31--38}, [\href{https://arxiv.org/abs/hep-ex/0312004}{{\tt
  hep-ex/0312004}}].

\bibitem{ATLAS:2020zms}
{\scshape ATLAS} collaboration, G.~Aad et~al., \emph{{Search for heavy Higgs
  bosons decaying into two tau leptons with the ATLAS detector using $pp$
  collisions at $\sqrt{s}=13$ TeV}},
  \href{http://dx.doi.org/10.1103/PhysRevLett.125.051801}{\emph{Phys. Rev.
  Lett.} {\bf 125} (2020) 051801},
  [\href{https://arxiv.org/abs/2002.12223}{{\tt 2002.12223}}].

\bibitem{CMS:2021ctt}
{\scshape CMS} collaboration, A.~M. Sirunyan et~al., \emph{{Search for resonant
  and nonresonant new phenomena in high-mass dilepton final states at $
  \sqrt{s} $ = 13 TeV}},
  \href{http://dx.doi.org/10.1007/JHEP07(2021)208}{\emph{JHEP} {\bf 07} (2021)
  208}, [\href{https://arxiv.org/abs/2103.02708}{{\tt 2103.02708}}].

\bibitem{ATLAS:2021bjk}
{\scshape ATLAS} collaboration, \emph{{Search for high-mass resonances in final
  states with a tau lepton and missing transverse momentum with the ATLAS
  detector}}, .

\bibitem{ATLAS:2019lsy}
{\scshape ATLAS} collaboration, G.~Aad et~al., \emph{{Search for a heavy
  charged boson in events with a charged lepton and missing transverse momentum
  from $pp$ collisions at $\sqrt{s} = 13$ TeV with the ATLAS detector}},
  \href{http://dx.doi.org/10.1103/PhysRevD.100.052013}{\emph{Phys. Rev. D} {\bf
  100} (2019) 052013}, [\href{https://arxiv.org/abs/1906.05609}{{\tt
  1906.05609}}].

\bibitem{CMS:2022fsw}
{\scshape CMS} collaboration, A.~Tumasyan et~al., \emph{{Search for heavy
  resonances and quantum black holes in e\ensuremath{\mu}, e\ensuremath{\tau},
  and \ensuremath{\mu}\ensuremath{\tau} final states in proton-proton
  collisions at $ \sqrt{s} $ = 13 TeV}},
  \href{http://dx.doi.org/10.1007/JHEP05(2023)227}{\emph{JHEP} {\bf 05} (2023)
  227}, [\href{https://arxiv.org/abs/2205.06709}{{\tt 2205.06709}}].

\bibitem{Straub:2018kue}
D.~M. Straub, \emph{{flavio: a Python package for flavour and precision
  phenomenology in the Standard Model and beyond}},
  \href{https://arxiv.org/abs/1810.08132}{{\tt 1810.08132}}.

\bibitem{Aebischer:2018bkb}
J.~Aebischer, J.~Kumar and D.~M. Straub, \emph{{Wilson: a Python package for
  the running and matching of Wilson coefficients above and below the
  electroweak scale}},
  \href{http://dx.doi.org/10.1140/epjc/s10052-018-6492-7}{\emph{Eur. Phys. J.
  C} {\bf 78} (2018) 1026}, [\href{https://arxiv.org/abs/1804.05033}{{\tt
  1804.05033}}].

\bibitem{Aoki:2019cca}
{\scshape Flavour Lattice Averaging Group} collaboration, S.~Aoki et~al.,
  \emph{{FLAG Review 2019: Flavour Lattice Averaging Group (FLAG)}},
  \href{http://dx.doi.org/10.1140/epjc/s10052-019-7354-7}{\emph{Eur. Phys. J.
  C} {\bf 80} (2020) 113}, [\href{https://arxiv.org/abs/1902.08191}{{\tt
  1902.08191}}].

\bibitem{Duraisamy:2016gsd}
M.~Duraisamy, S.~Sahoo and R.~Mohanta, \emph{{Rare semileptonic $B \to
  K(\pi)l_i^- l_j^+$ decay in a vector leptoquark model}},
  \href{http://dx.doi.org/10.1103/PhysRevD.95.035022}{\emph{Phys. Rev. D} {\bf
  95} (2017) 035022}, [\href{https://arxiv.org/abs/1610.00902}{{\tt
  1610.00902}}].

\bibitem{Zhang:2014rja}
C.~Zhang, \emph{{Effective field theory approach to top-quark decay at
  next-to-leading order in QCD}},
  \href{http://dx.doi.org/10.1103/PhysRevD.90.014008}{\emph{Phys. Rev. D} {\bf
  90} (2014) 014008}, [\href{https://arxiv.org/abs/1404.1264}{{\tt
  1404.1264}}].

\bibitem{Durieux:2014xla}
G.~Durieux, F.~Maltoni and C.~Zhang, \emph{{Global approach to top-quark
  flavor-changing interactions}},
  \href{http://dx.doi.org/10.1103/PhysRevD.91.074017}{\emph{Phys. Rev. D} {\bf
  91} (2015) 074017}, [\href{https://arxiv.org/abs/1412.7166}{{\tt
  1412.7166}}].

\bibitem{ATLAS:2023qzr}
{\scshape ATLAS} collaboration, G.~Aad et~al., \emph{{Search for
  flavor-changing neutral-current couplings between the top quark and the Z
  boson with proton-proton collisions at s=13\,\,TeV with the ATLAS detector}},
  \href{http://dx.doi.org/10.1103/PhysRevD.108.032019}{\emph{Phys. Rev. D} {\bf
  108} (2023) 032019}, [\href{https://arxiv.org/abs/2301.11605}{{\tt
  2301.11605}}].

\end{thebibliography}\endgroup

\end{document}